\documentclass[a4paper,12pt]{amsart}
\usepackage[sfdefault=cmbr,OMLmathsans]{isomath}  
\usepackage{upgreek}  

\usepackage{tensor}  

\usepackage{wrapfig}

\usepackage{etex}
\usepackage{graphicx}      
\usepackage[latin1]{inputenc}
\usepackage[T1]{fontenc}
\usepackage{fixltx2e}
\usepackage{longtable}
\usepackage{float}
\usepackage{wrapfig}
\usepackage{rotating}
\usepackage[normalem]{ulem}
\usepackage{amsmath}
\usepackage{textcomp}
\usepackage{marvosym}
\usepackage{wasysym}
\usepackage{amssymb}
\usepackage{hyperref}
\usepackage{color}
\usepackage{listings}
\usepackage[final, colorinlistoftodos]{todonotes}
\usepackage{pst-all}
\usepackage{ulem}

\newcommand{\figures}{./figures}
\newcommand{\figuresDoubleSpringSplit}{./figuresDoubleSpringSplit}

\renewcommand{\vec}[1]{{\bf #1}}
\newcommand{\vek}[1]{\mathchoice{\displaystyle\boldsymbol{#1}}
{\textstyle\boldsymbol{#1}}{\scriptstyle\boldsymbol{#1}}
{\scriptscriptstyle\boldsymbol{#1}}}

\newcommand{\ten}[1]{{\mathbb #1}}
\newcommand \flux[1]{f_#1}		

\def \pot{\mathcal P}
\begin{document}

\title[Smoothness and Balance in Explicit Cosimulation]{Treating Smoothness and Balance during Data Exchange in Explicit Simulator Coupling or Cosimulation } 


\author{Dirk Scharff, Thilo Moshagen, Jaroslav Vond\v{r}ejc} 

\maketitle

\begin{abstract}                
Cosimulation methods allow combination of  simulation tools of physical systems running in parallel to act as a single simulation environment for a
 big system. As data is passed  across subsystem boundaries instead of solving the system as one single equation system, it is not  ensured that systemwide balances are fulfilled. If the exchanged data is a flow of a conserved quantitiy, approximation errors can accumulate and make simulation results inaccurate. The problem of approximation errors is typically adressed with extrapolation of exchanged data. Nevertheless balance errors occur as extrapolation  is approximation. This  problem can be handled with balance correction methods which compensate these errors by adding corrections for the balances to the signal in next coupling time step. This work aims at combining extrapolation of exchanged data and balance correction in a way that the exchanged signal not only remains smooth, meaning the existence of continuous derivatives, 
 but  even in a way reducing the derivatives, in 
 order to avoid unphysical dynamics caused by the coupling. 
 To this end, suitable switch and hat functions are constructed and applied to the problem.
\end{abstract}



\section{Introduction}
\label{sec-1}

Engineers are increasingly relying on numerical simulation techniques. Models and 
simulation tools for various physical problems have come into existence in the past 
decades. The desire to simulate a system that consists of well described and treated 
subsystems by using appropriate solvers for each subsystem and letting them exchange the data that forms the mutual 
influence is immanent. \\ 
The situation usually is described by two coupled differential-algebraic systems $S_1$ and $S_2$ that together form a system $S$:
\begin{align}
S_1: \quad \nonumber \\ 
\dot{\vek x}_ 1 &= \vek f_1(\vek x_1,\vek x_2,\vek z_1, \vek z_2)\\
0 &= \vek g_1(\vek x_1, \vek x_2, \vek z_1, \vek z_2) \\
S_2: \nonumber \quad\\
\dot{\vek x}_2  &= \vek f_2(\vek x_1,\vek x_2,\vek z_1, \vek z_2)\\
0 &= \vek g_2(\vek x_1, \vek x_2, \vek z_1, \vek z_2). 
\end{align}
The $(\vek x_1,\vek x_2)$ are the differential states of $S$, their splitting into $\vek{ x}_i$ determines the subsystems $S_i$ together with the choices of the $\vek{z}_i$.
In \emph{Co-Simulation} the immediate mutual influence of subsystems 
 is  replaced by exchanging data at fixed time points 
 and subsystems
are solved separately and parallely but using the received parameter:
\begin{align}
S_1: \nonumber \quad\\
\dot{\vek x}_ 1 &= \vek f_1(\vek x_1, \vek z_1, \vek u_2)\label{split1} \\ 
0 &= \vek g_1(\vek x_1,  \vek z_1, \vek u_2) \label{splitAlgebraic1} \\ 
S_2:\quad \nonumber \\
\dot{\vek x}_2 &= \vek f_2(\vek x_2,\vek z_2, \vek u_1)\\
0 &= \vek g_2(\vek x_2, \vek z_2,  \vek u_1)  \label{splitLast}
\end{align}
where $\vek u_i$ are given by coupling conditions that have to be fulfilled at exchange times $T_k$ 
\begin{align}
\vek 0 &=\vek h_1(\vek x_1, \vek z_1, \vek u_1)\label{coupling1}  \\ 
\vek 0 &=\vek h_2(\vek x_2, \vek z_2, \vek u_2)  \label{coupling2}
\end{align}
and are not dependent on subsystem $i$'s states any more, so are mere parameters between exchange time steps.\\
Full row rank of $d_{\vek z_i} \vek g_i$ can be assumed, such that the differential-algebraic systems are of index 1. This description of the setting is widespread (\cite{ArnoldGuenther2001}). 
With the $\vek h_i$ being solved for $\vek u_i$ inside the $S_i$ ( let solvability be given), for systems with more than two subsystems ist is more convienient to write  output variable $\vec y_i$  instead of $\vek u_i$   and now redefine $\vek u_j$ as the input of $S_j$, consisting of some components of the outputs $(\vec y_i)_{i=1..n} $\cite{ArnoldClaussSchierz2013}. This structure is defined as kind of a standard for connecting simulators for cosimulation  by the \emph{Functional Mockup Interface} Standard \cite{FMI}. It defines clearly what information a subsystems implementation provides. From chapter 2 on, we use this notation. Mind input $\vek u_j$ is then indexed with its subsystems index.\\
In Co-Simulation the variables establishing the mutual influence of subsystems 
 are exchanged at fixed time points.
This results in continuous variables being approximated by piecewise constant
  extrapolation, as shown in the following picture: \\
\begin{figure}[h!tb]
\begin{center}
\resizebox{0.5 \textwidth}{!}{
\includegraphics{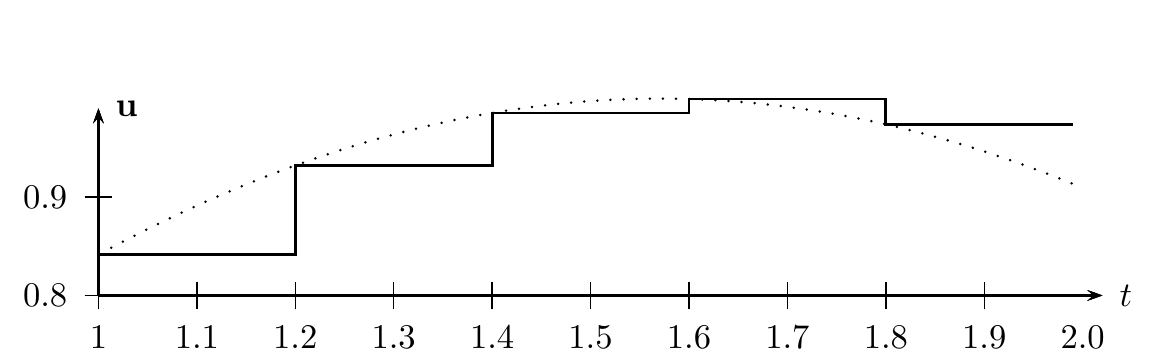}
}\end{center}
\caption{Constant extrapolation of an input signal}
\end{figure}

If one does not want to iterate on those inputs by restarting the simulations using the newly calculated inputs, one just proceeds to the next timestep. \\
This gives the calculations an explicite component, the mutual influence is now not immediate any more, inducing the typical stability problems, besides the
  approximation errors.\\
But for good reasons, co-simulation is a widely used method: 
It allows to put  separate submodels, for each of which a solver exists, together into one system  and simulate that system by simulating each subsystem with its   specialised solver - examination of mutual influence becomes possible without rewriting everything into one system, and   simulation speed benefits from the parallel calculation of the submodels. \\

The following fields of work on explicite co-simulation can be named to be the ones of most interest:
\begin{enumerate}
\item Improvement of the approximation  of the exchanged data will most often improve simulation results \cite{Busch2012}. This  is usually done by \emph{higher-order extrapolation} of exchanged data, as shown in this plot, where  the function plotted with dots is linearly extrapolated:\\ 
\begin{figure}[h!tb]
\begin{center}
\resizebox{0.5 \textwidth}{!}{
\includegraphics{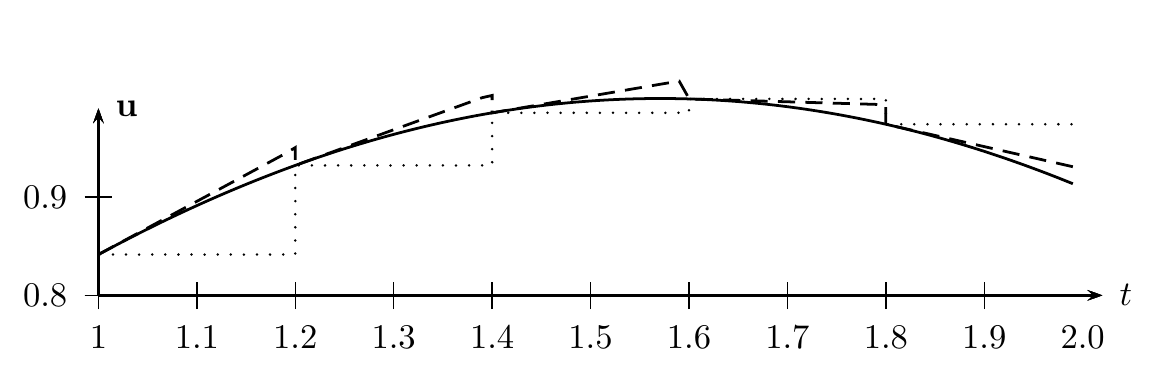}
}\end{center}
\caption{Linear extrapolation of an input signal}
\end{figure}
%

\item When the mutual influence between subsystems consists of flow of conserved quantities like mass or energy, it turns out that the improvement of the approximation of this influence by extrapolation of past data is not sufficient to establish the conservation of those quantities with the necessary accuracy.
The error that arises from the error in exchange adds up over time and
 becomes obvious (and lethal to simulation results many times). 
 In a cooling cycle example (\cite[Section 6.3]{KosselDiss}),  a gain of 1.25\% in coolant mass occurs when simulating a common situation.
\begin{figure}[h!tb]
\begin{center}
\resizebox{0.5 \textwidth}{!}{
\includegraphics{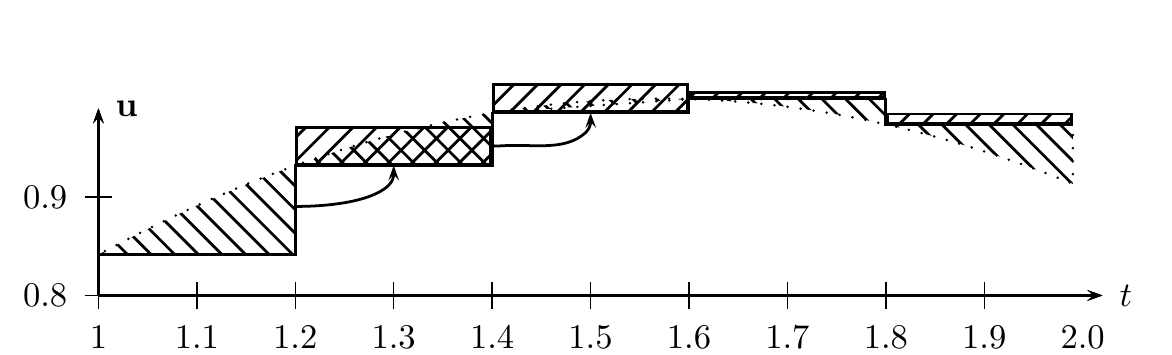}
}
\end{center}
\caption{Constant extrapolation of an input signal, balance error and its recontribution}
\end{figure}
It has been tried  to meet this challenge by   passing the amount of exchanged quantity for the past timestep along with the actual flow on to the receiving system, where then the error that has just been  commited is calculated  and added to the current flow to compensate the past error. For well damped example problems in fluid circles this method has fulfilled the expectations \cite{Scharff12}. It has been labelled \emph{balance correction}. 
\item There is good reason to prevent jumps in exchanged data by \emph{smoothing}. Higher order extrapolation polynomials cannot make extrapolated  data at the end of the exchange timestep match the newly given value.
\end{enumerate}
While under all circumstances it is desirable to guess the influence
information at the subsystem boundaries from past data as well as
possible, balance correction techniques bear the profound problem that
for establishing conservation \emph{ a posteriori} over the whole time
an instantanious error in the exchanged data has to be
accepted. More precisely, balance correction means making an error in the amount of a quantity $\int^{t_i+1}_{t_i}u(\tau)d\tau$  that is exchanged between subsystems  during the interval $[t_i,t_{i+1})$,   $u$ being the exchanged signal,      
for the purpose of lowering the error in states $\vek x$ that would be caused by this accumulated and now persistent error in amount of $\int u d\tau$. This paper reveals the extend of this problem and proposes a way  to reduce the time delay.\\
 Of course, also an  error in the derivatives of the exchanged data is made, which might cause dynamics in the receiving system and its neighbours. The
presented work prepares cosimulation to control the error in the
derivatives by construction and use of suitable functions for smoothing
during switching and adding of correction terms.

\section{Preliminaries}
\subsection{Notation}
\label{sec-2}
Let information
exchange times start with $t_0$ and end\footnote{Exchange at calculations end could be relevant as data determines algebraic variables at end of simulation, so $t_n$ is also the end of simulation.} with $t_n$ 
and time intervals of exchange $\left[ t_{i-1}, t_i\right) $ be indexed
from 1 to $n$, such that $\Delta_i x := x(t_i)-x(t_{i-1})$ for any variable $x$. 

%
Input variables in general are denoted by $\vec u$, algebraic and differential states by $\vec z$ and $\vec x$, respectively. Lower indices indicate vector indices, here the subsystem the components belong to (a slight change from indexing $\vec u$ in eq. \eqref{split1}-\eqref{coupling2}), upper indices indicate the restriction of the function to the time interval, e.g. $\vek u_i^j$ denotes the function $\vek u_i: [t_{j-1},t_j]\mapsto \mathbb R^{n_i}$. 

Within this paper, concepts  are extended from flows of conserved quantities to general input treatment. Conserved variables are not distinguished in notation, conservation is mentioned in text if given. Input variables  $\vec u$ extrapolation is denoted by  $\operatorname{Ext}\vec u$,  sometimes referred to as \emph{hold function}, thus
a guess of $\vec u^i$ calculated from the known $\vec u^j$, $j<i$, while
 $\overline{  \vec u}^j$  be the signal as it is used
for calculation constructed on $\left[ t_{j-1}, t_j\right) $, here a smooth combination of  $
 { \operatorname{Ext}\vec u_{i}^{j-1}}$ and  $\operatorname{Ext}\vec u_{i}^{j}$, referred to as the \emph{smooth extrapolation}. 
 Let  
$\Delta E_{j} = \int_{t_{j-1}}^{t_j}\vek u_i dt  - \int_{t_{j-1}}^{t_j} \overline{ \vek u_i} dt$ be the error of the exchanged quantity itself.
 

\subsection{Choice of exchanged Variables}
The direction in which the variables are exchanged cannot be chosen arbitrarily. Careful consideration is needed to not place contradictorily constraints on the models. If i.e. at one of the model boundaries $S_1$ exchanges the value of a differential state, it must only export and not receive this value -- otherwise this state is turned into a parameter and extrapolated instead of solved, or one would overwrite an integration result and reinitialize the solver at each exchange. The  choice is to receive arguments of the derivative  of the state and send the state value itself \cite{Scharff12}, \cite{KosselDiss}. Often, the physics of coupling is described by a flow driven by a potential, where the flux depends on the potential by $f\sim \nabla_{\vek x} \pot $, or in non-spatial context   $f = \operatorname{const}\left[\pot_{S_2}- \pot_{S_1}\right]$.  Then exchange is commonly implemented in the following way: System $S_1$ passes the value of the flow-determining potential $\pot$ on to $S_2$, whereas the latter calculates the flux $f $ and passes it to $S_1$.

\subsection{Detailed description of Techniques in coupling Simulation Software}
\label{sec-2-1}
After the  brief overview in the introduction on the issues that appear when coupling simulation software and means to treat them was given. Those fields are described more precisely here, giving citations. 

\subsubsection{Determination of consistent Inputs}
As $\vek u_i$ fulfils $\vek 0 =\vek h_i(\vek x_i, \vek z_i, \vek u_i)$, where $\vek z_i$ results from  $\vek 0 =\vek g_i(\vek x_i, \vek z_i, \vek u_j)$ it in general depends on $\vek u_j$. So the coupling equations \eqref{coupling1} \eqref{coupling2} $(\vec h_i)_{i=1..n}=\vek 0$ are a coupled system in fact.
 The $\vek 0 =\vek h_i$ can be solved with respect to the $\vek u_i$ one after the other  only if the directed graph of  influence of outputs on each other contains neither bidirectional dependencies nor loops. Otherwise, the  coupling conditions remain not fully fulfilled, or one solves them iteratively. Such
  methods are the Interface-Jacobian based methods,  e.g. given in \cite{KueblerSchiehlen2000},  which determine the $\vec u_i$ with respect to $\vec h_i = 0$ and $\vec g_i = 0$ with the help of a newton solver at exchange times, which means full consistency at those times. But not all  tools simulating the subsystems provide the residual of  $\vek g_i=0$.

\subsubsection{Iterative coupled Methods}
 If restarting the timestep is an option, iterations can be performed and instead of using $\operatorname{Ext}(\vek u_i)$, the time-dependant numerical solution for $\vec u_i$ from the old iteration can be used.
 Schemes are distinguished by the exchange directions of input,
 as the waveform iteration or the Gauss-Seidel scheme, which can exploit
  the directions of influences between the subsystems (\cite{MiekkalaNevanlinna87}, \cite{ArnoldGuenther2001}, \cite{Busch2012}).
  Stability and accuracy, also in terms of balance, such can be augmented, of course at computational cost due to the restarts.\\ 
In \cite{SicklingerBelskyEngelmannElmqvist14},  a Newton method is used to solve the coupling equations  \eqref{coupling1}-\eqref{coupling2}, but different from \cite{KueblerSchiehlen2000} not only $(\vec h_i(\vek x_i(T_k), \vek z_i, \vek u_i))_{i=1..n}=\vek 0$ at the end of the timestep, but   $(\vec h_i(\vek x_i(\vek u_i), \vek z_i, \vek u_i))_{i=1..n}=\vek 0$. This requires repeated solving of the subsystems alone for Jacobian evaluation.
 
\subsubsection{Increasing the extrapolation order of inputs} To improve approximation $  \operatorname{Ext}\vec u^i$ has been calculated as an extrapolation polynomial of past
   $\vec u^{j}$, $j<i$. 
In his dissertation \cite{Busch2012} Busch examines Lagrangian and Hermite extrapolation polynomials for exchanged quantities using a system concatenated from two coupled spring-mass oscillators, thus a linear  second order ODE with four degrees of freedom,  as model problem.

\subsubsection{Investigations of convergence}
Busch examines  the effect of the approximation order in exchanged data on global convergence and judges the stability of the method for his problem, assuming exact integration of each subsystem. 
 Convergence to exact solution in this case is limited to $O(h^2)$ for piecewise constant extrapolation, while it is limited to $O(h^3)$ for degree one Hermite polynomial extrapolation.\\
As Busch examines a system of ordinary differential equations without algebraic parts and given explicitely, he feels no need to consider smoothing. 
Balance errors that occur in his problem remain undiscussed. 
Besides the aforementioned examination \cite{Busch2012} which is limited to linear problems,  in \cite{ArnoldGuenther2001} there is a more general examination, treating iterative coupling schemes. Mainly, a fixed point argument is used here.
\cite{MiekkalaNevanlinna87}

\subsubsection{Smooth switching between old and new input }
The smoothing of exchanged data is originally motivated by the need to  provide a starting value close to the solution during searching for solutions of the algebraic  part of the differential-algebraic equation system. 
\\
Another reason for smoothing the inputs can be that it enables the calculation of derivatives resp. difference quotients from them. Although this is unfortunate, such needs sometimes occur in practice.\\
In  \cite{KosselDiss}, two halves  of a cooling cycle are co-simulated, the mass flow $\flux{m}$ being exchanged, so a component of $\vek u$.  The smooth input $\overline{  \vek{ u}}_{i}^{j}$ is concatenated as a convex sum of $\operatorname{Ext}{ \vek{ u}}_{i}^{j-1}$ and $\operatorname{Ext} \vek{ u}_{i}^{j}$ weighted
with a sufficiently smooth, here degree 5 polynomial, function switching from 0 at $T_{j-1}$ to 1 at $T_{j}$.
Besides from stabilizing the Newton solver as intended, Kossels work gives an impression on balance errors: During some simulations balance error of about 1.25\% in coolant mass is observed.\\
It is obvious
that by this smoothed switching 
the balance error
 $\Delta E_{i}^j$ possibly increases compared to a
unsmoothed switching procedure.\\ 
Another work that considers smoothing is the paper \cite{WellsHasanLucas}.

\subsubsection{Balance Correction}\label{sec:BC}
Classical balance errors of extrapolations is  $\Delta E_{i}^k = 
\int_{T_{k-1}}^{T_k}\vek u_{i} - \overline{  (\vec u)}_{i}^k(t) dt$,  
where $\vek u_{i}$ is a flux of a conserved quantity, but it is defined for arbitrary quantities. 
 Negative and positive contributions from the 
 intervals partly compensate each other, and 
as in a typical simulation the system often ends up in a stationary state similar to the one at begin, thus graphs of quantities and their derivatives tend to end at values where they started, the conserved quantities balance error at the end of a simulation may be small. This does not imply it is small during all intervals of the simulation. As a typical such simulation situation think of an automotive driving cycle (\cite{KosselDiss}): Using piecewise constant extrapolation of $f$, there  is a gap in mass after the phase of rising system velocity has passed - this loss remains uncompensated during the (significant) phase of elevated speed.\\
Scharff, motivated by the loss in balance stated in \cite{KosselDiss} as a collateral result, proposed in \cite{Scharff12} that the errors 
\begin{equation}\label{eq:BC}
\Delta E_i^{j} = \int_{T_{j-1}}^{T_j}\vek u_i dt  - \int_{T_{j-1}}^{T_j} \overline{ \vek u}_i dt
\end{equation} 
are
added in the time step $j+1$.  
The correction that is applied in  the $j$-th interval is then
$\Delta E_i^{j-1}\phi_j (t)$,
where   the function $\phi_j$ is scaled such that its integral is 1 and  may be constant but when one wants to  preserve smoothness, it is a smooth function smoothly vanishing
at the boundaries of the time interval $j$. 
In spite of the
lack of strictness and the increase of errors in derivatives of the
exchanged quantities, this method enables coupling of simulations that
would be impossible without - Kossels example was recalculated successfully using balance correction. 

\section{Smooth Balancing and Stability}
\def\fac{0.5}
\def\e{2.718282}
\def\balanceErr{\fac*(\e^\rL * ((\rM - \rL)^3 /6 + (\rM - \rL)^4 /24))}
\def\hatFunc{\e^(-1/(1 - (( x -\rMid)/0.2501)^2))/(0.25*0.6)}
\def\switchErrPrev{\fac*0.5*(\rR-\rM)*( (\e^\rL)	
		-(\e^\rLL + \e^\rLL*(\rL-\rLL)))}
\def\switchErr{\fac*0.5*(\rR-\rM)*( (\e^\rM)
		 -(\e^\rL + \e^\rL*(\rM-\rL)))}

\begin{figure}[h!bt]
\center{\includegraphics[ width=0.7\textwidth, trim={0 0cm  0 2.3cm},clip]{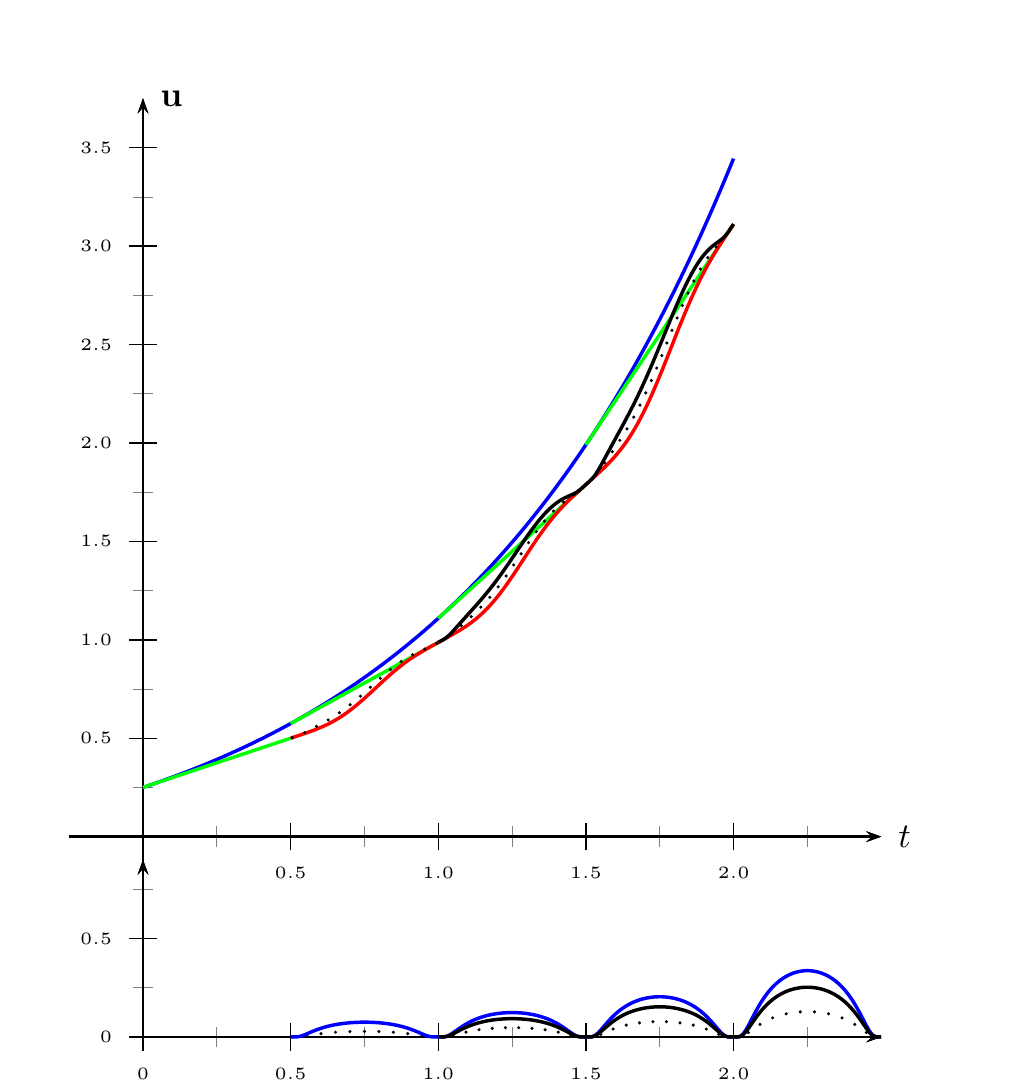}}

\caption{\label{ClassicalBalanceCorrection}  An input signal and ways it might be treated by the receiving system: green: linearly interpolated, red: smoothed with s-shaped $\psi$, black: error from previous time step added to smooth signal. In  the lower plot, the contributions themselves are plotted: 
 Black dotted: error of linear extrapolation from previous timestep, blue: balance error due to smooth (slow) switching in actual timestep, black: sum of balance error from previous and switching error from previous step. It is obvious that switching error cannot be neglected, and that derivatives of the input signal oscillate.}
\end{figure}
As a first generalisation, this paper  uses that also \emph{nonconserved} quantities can be balanced over short times. It makes sense to apply for example an amount of force that has been omitted due to extrapolation error in the next timestep.\\

If the algebraic equations $\vek g_i=\vek 0$ are solvable w.r.t. $\vek z_i$, the systems can be solved by state space method, this is determining $\vek z_i$ with a Newton solver at each evaluation of the ODE. Such system is then solved as an ODE, why in the following we consider only ODE. The right scheme below gives the cosimulation scheme for the coupled problem on the left: \\[11pt] 
\begin{tabular}{p{0.2\textwidth} | p{0.2\textwidth}} \label{cosimScheme}
\center{$S_1$} & \center{$S_2$}\tabularnewline
\multicolumn{2}{ c }{System States}\tabularnewline
\center{$\vek x_1  $} 		&\center{$\vek x_2 $} \tabularnewline
\multicolumn{2}{ c }{Outputs}\tabularnewline
\center{$\vek y_1  $} 		&\center{$\vek y_2 $} \tabularnewline
\multicolumn{2}{ c }{Inputs}\\
\center{$\vek u_1=\vek y_2$}	& \center{$\vek u_2 = \vek y_1$}\tabularnewline
\multicolumn{2}{ c }{Equations}\tabularnewline
\center{$\dot {\vek x_1} = \vek f_1(\vek x_1, \vek u_1)	$\\} & \center{$\dot{ \vek x_2} = \vek f_2(\vek x_2, \vek u_2)$}	
\end{tabular}
\begin{tabular}{p{0.25\textwidth} | p{0.25\textwidth}} 
\center{$S_1$} & \center{$S_2$}\tabularnewline
\multicolumn{2}{ c }{System States}\tabularnewline
\center{$\vek x_1  $} 		&\center{$\vek x_2 $} \tabularnewline
\multicolumn{2}{ c }{Outputs}\tabularnewline
\center{$\vek y_1  $} 		&\center{$\vek y_2 $} \tabularnewline
\multicolumn{2}{ c }{Inputs}\\
\center{$\vek u_1=\vek y_2$}	& \center{$\vek u_2 = \vek y_1$}\tabularnewline
\multicolumn{2}{ c }{Equations}\tabularnewline
\center{$\dot {\vek x_1} = \vek f_1(\vek x_1, \operatorname{Ext}(\vek u_1))	$} & \center{$\dot{ \vek x_2} = \vek f_2(\vek x_2, \operatorname{Ext}(\vek u_2))$}	
\end{tabular}\\[11pt]
Each method described in the preceding section provides relief for a certain problem that arises when applying the standard cosimulation scheme:
 In a now  widened field of physical problems, this paper names simulation settings in which the problems 
balance and stability,
smoothness and subsequent oscillations,
and both of them arise, 
and
combines balance correction and smoothing for solving them,
 examines the results and suggests an early refeed of early-known balance error contributions to tackle the delaying effect  of smooth switching. It makes suggestions for further reducing the exchange-induced oscillations in receiving systems.\\
Smooth switching as in \cite{KosselDiss}  is done using an S-shaped function $\psi(t)$ for switching
between extrapolation polynomials $\operatorname{Ext}{ \vek{ u}}_{i}^{j-1}$ and $\operatorname{Ext}{ \vek{ u}}_{i}^{j}$ such that the smoothed input is
\begin{equation}\label{smoothSwitching}
\overline{ \vek u_i}^j(t)= (1-\psi(t)) \operatorname{Ext}{ \vek{ u}}_{i}^{j-1}(t) + \psi(t)  \operatorname{Ext}{ \vek{ u}}_{i}^{j}(t).
\end{equation}
The ODE in above scheme then is $\dot {\vek x_1} = \vek f_1(\vek x_1, \overline{\vek u_1})$.
If balance correction contributions $\Delta E_i^{j-1}\phi_j (t)$ as in Section \ref{sec:BC} are added to such smooth $\overline{ \vek{ u}_{i}}^{j}$, the  function $\phi(t)$ should not be constant but a hat such that it is smooth on the boundaries of the interval, i.e. that $\partial_t^{(n)} \phi_j(t) = 0$ at $t_j$ and $t_{j-1}$ for some, better for all $n$, in order not to spoil the smoothness of the $\overline{ \vek{ u}_{i}}$. Implementations of $\phi$ and $\psi$ are presented in Section \ref{sec:HatsSwitches}. A sketch of the input signals used during combination of methods is given in figure  \ref{ClassicalBalanceCorrection}. 

For solving the ODE inside each subsystem, in this contribution pythons \emph{vode} and \emph{zvode} are used,  methods relying on  implicit Adams method and backward differentiation formulas (BDF). For both methods, a stepwidth can be found that conserves stability of linear problem, and the internal step size control does so, so subsystems methods behave like stable methods.

\subsection{Spring and Mass Example and its Implications}

Consider the linear mechanical oscillator given by
\begin{equation}\label{eq:springMass}
\dot{ \vec{ x}}
= \ten A\vec{x} 
= \begin{pmatrix}
0 & 1\\
-\frac{c}{m} & \left(-\frac{d}{m}\right)
 \end{pmatrix} \vec x, 
 \qquad \vec{x} = 
 \begin{pmatrix}
x\\
\dot x
 \end{pmatrix}
\end{equation}
with mass $m=1$, spring constant $c=1$, and in which the damping constant $d$ shall vanish.\\ 
None of its states is a conserved quantity in the view of this system.

 \subsubsection{Cosimulation} 
The system is treated as subsystems in the cosimulation scheme given in Table \ref{CosimSchemesSpringMass}, an implementation of the standard scheme given in the beginning of Section  \ref{cosimScheme}.
\\
\begin{table}[h!tb]
\begin{tabular}{c|c}
Spring  & Mass \tabularnewline
\multicolumn{2}{ c }{System States}\tabularnewline
 $ x_1 := s =  x $ 		&  $ x_2 := v = \dot x $ \tabularnewline
\multicolumn{2}{ c }{Outputs}\tabularnewline
$ y_1:=F =-cx  $		&$ y_2:= v = \dot x $ \tabularnewline
    &\tabularnewline
\multicolumn{2}{ c }{Inputs}\\
$  u_1:=  y_2$	& $ u_2 :=  y_1$\tabularnewline
\multicolumn{2}{ c }{Equations}\tabularnewline
\parbox[t]{0.22\textwidth}{ 
$\dot { x_1} = \operatorname{Ext}( u_1)$\\$ = v $}
& \parbox[t]{0.22\textwidth}{
$\dot{  x_2} = -\frac{\operatorname{Ext}(u_2) }{m}  $ $= -\frac{F}{m}$}
\end{tabular}
\begin{tabular}{c|c}
Spring & Mass\tabularnewline
\multicolumn{2}{ c }{System States}\tabularnewline
 	$\dots$	&  $\dots$\tabularnewline
\multicolumn{2}{ c }{Outputs}\tabularnewline
\parbox[t]{0.21\textwidth}{$ \vek y_1:=(f,\dot f)$\\ $=(-cx, -cv) $ } 		&\parbox[t]{0.21\textwidth}{ $\vek y_2:=( v,a)$ \\$= (\dot x, \dot f/m )$} \tabularnewline
\multicolumn{2}{ c }{Inputs}\\
$\vek  u_1:= \vek y_2$	& $\vek u_2 := \vek y_1$\tabularnewline
\multicolumn{2}{ c }{Equations}\tabularnewline
 $\vdots$&$\vdots$	\tabularnewline
 &\tabularnewline
\end{tabular}
\caption{\label{CosimSchemesSpringMass}Cosimulation Schemes for the spring-mass system, left constant, right linear extrapolation}
\end{table}

\begin{wrapfigure}{l}{0.2\textwidth}
\includegraphics{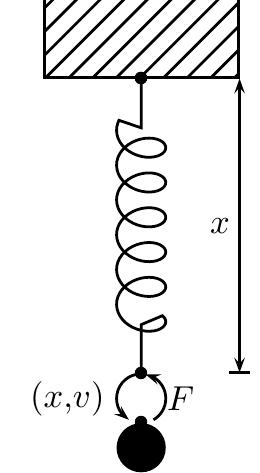}
\end{wrapfigure}
Output of the spring is the force $F =-cx$, 
that of the mass is the displacement $s = x$, and these quantities are again non-conserved. Note this problem is the simplest splittable problem possible. It has the least number of states that can be splitted, the splitted systems depend only on input, and in the most simple way, which is linear. \\
\begin{figure}
\includegraphics[ width=0.45\textwidth]{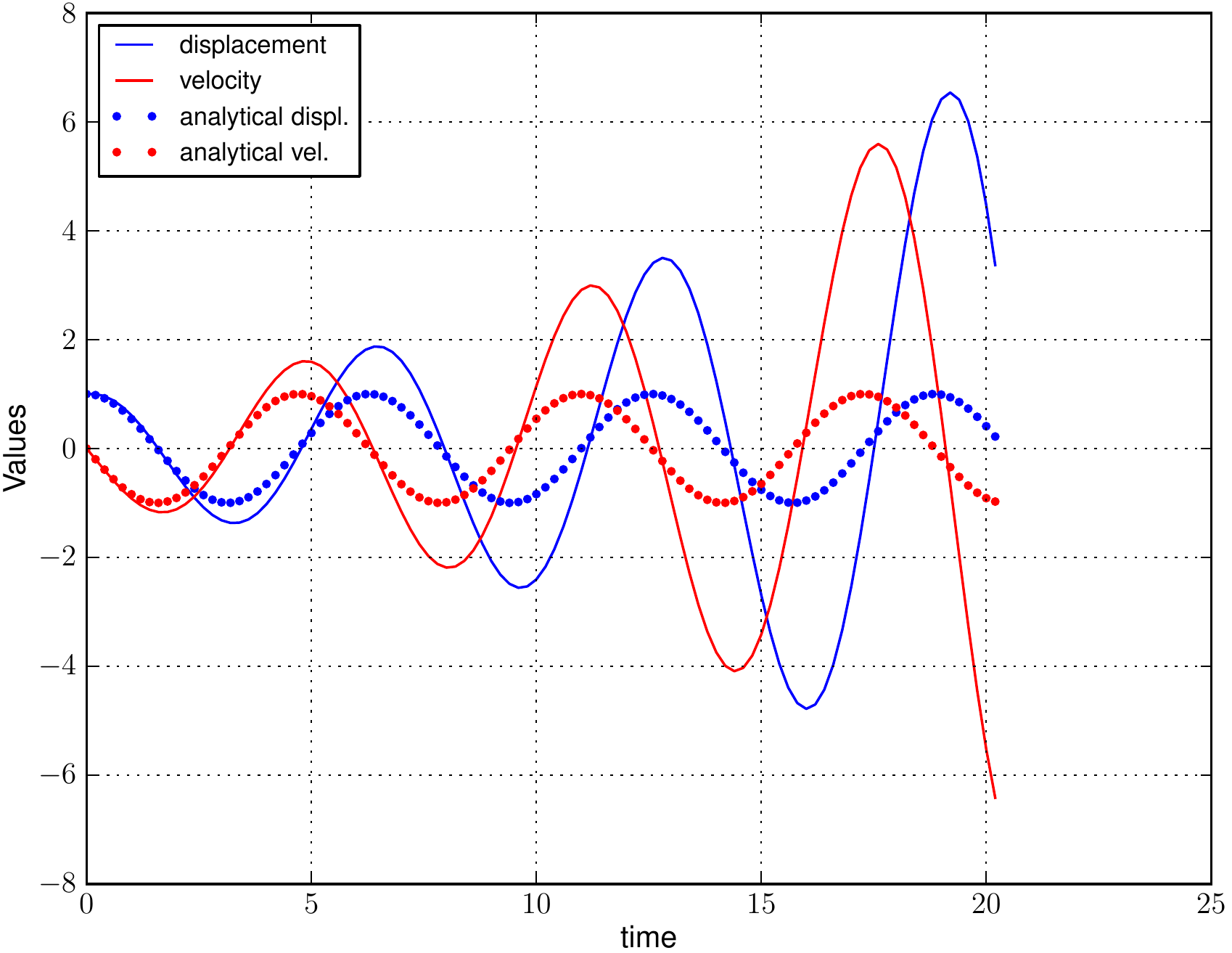}
\includegraphics[ width=0.45\textwidth]{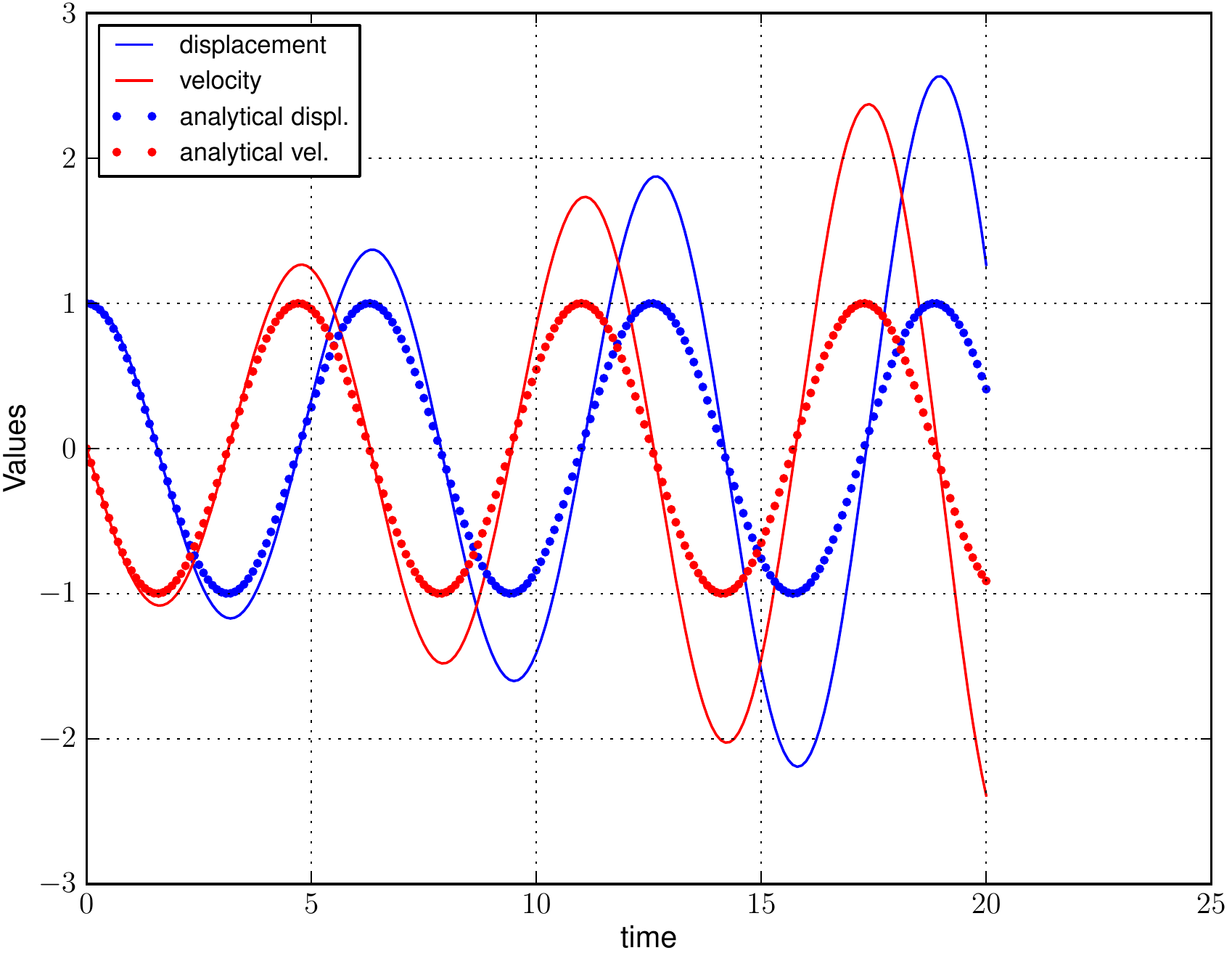}\\
\includegraphics[ width=0.45\textwidth]{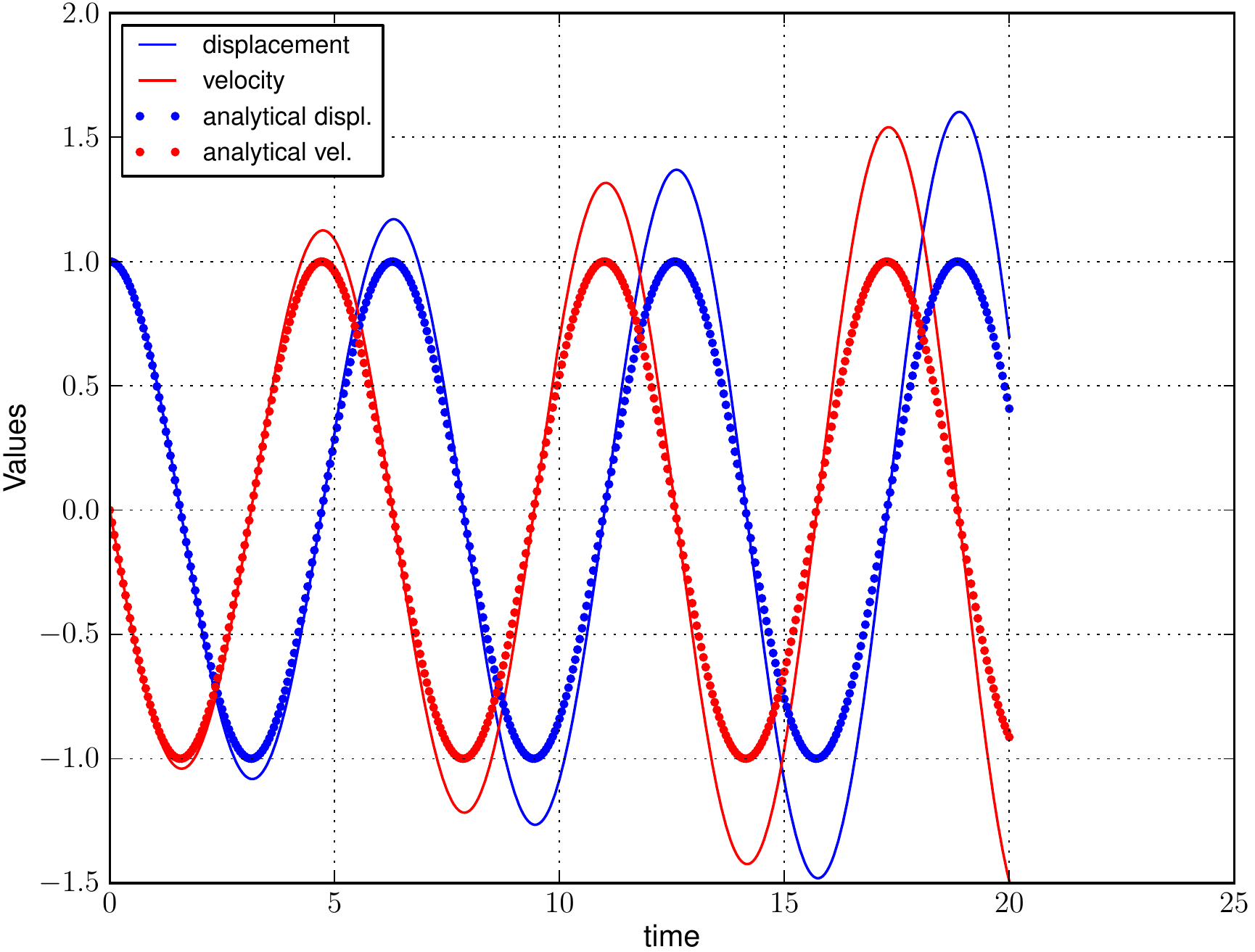}
\includegraphics[ width=0.45\textwidth]{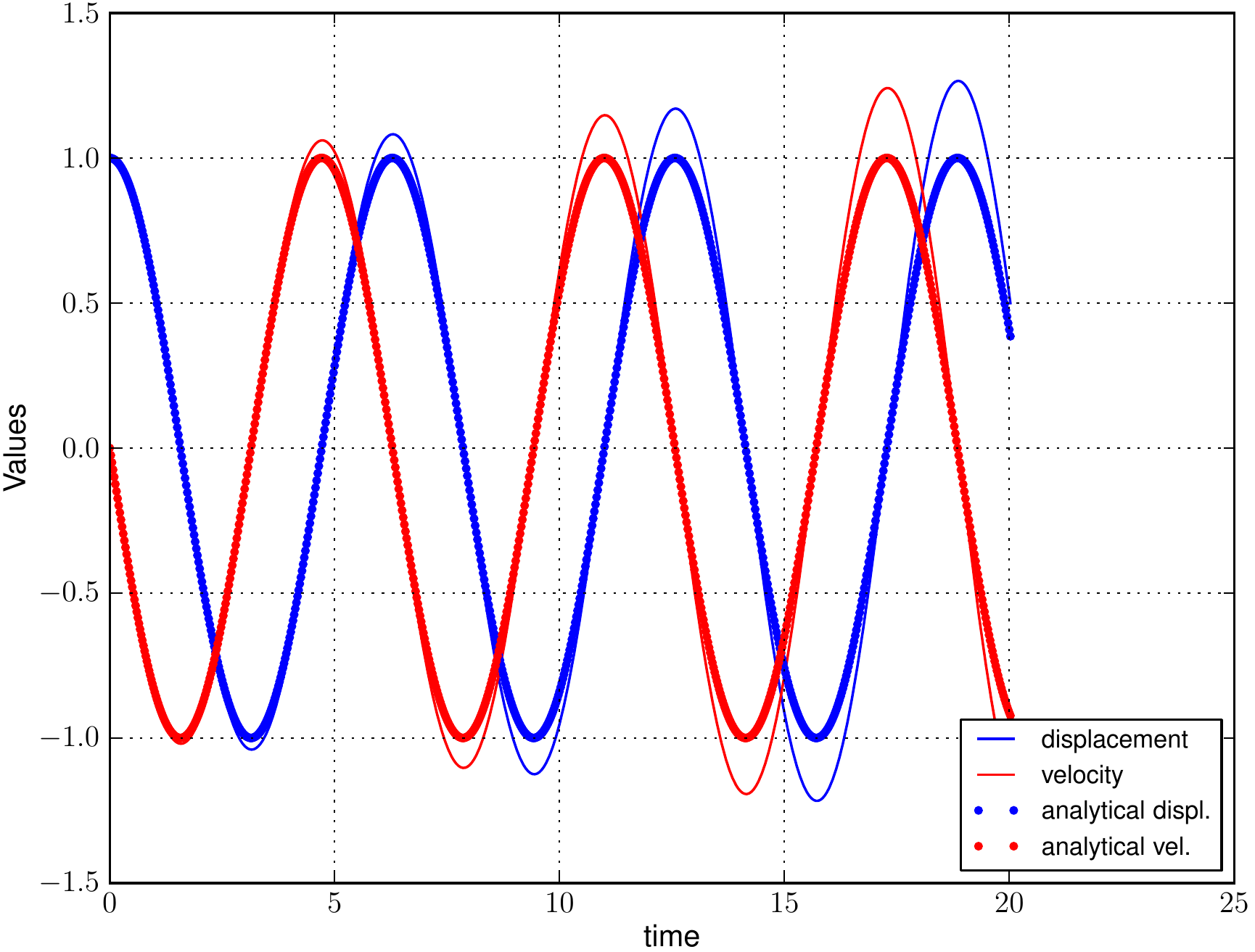}\\
\caption{\label{fig:ConstCosimConvergence} Simulation of the system \eqref{eq:springMass} in the cosimulation scheme with constant extrapolation, varying the exchange step size $H$.  Upper row, left: $H = 0.2$, right:  $H = 0.1$, lower row: left: $H = 0.05$, right:  $H = 0.025$. Convergence of order $H$ is given, but there is no stability for any step size in sight.}
\end{figure}
\begin{figure}
\includegraphics[ width=0.45\textwidth]{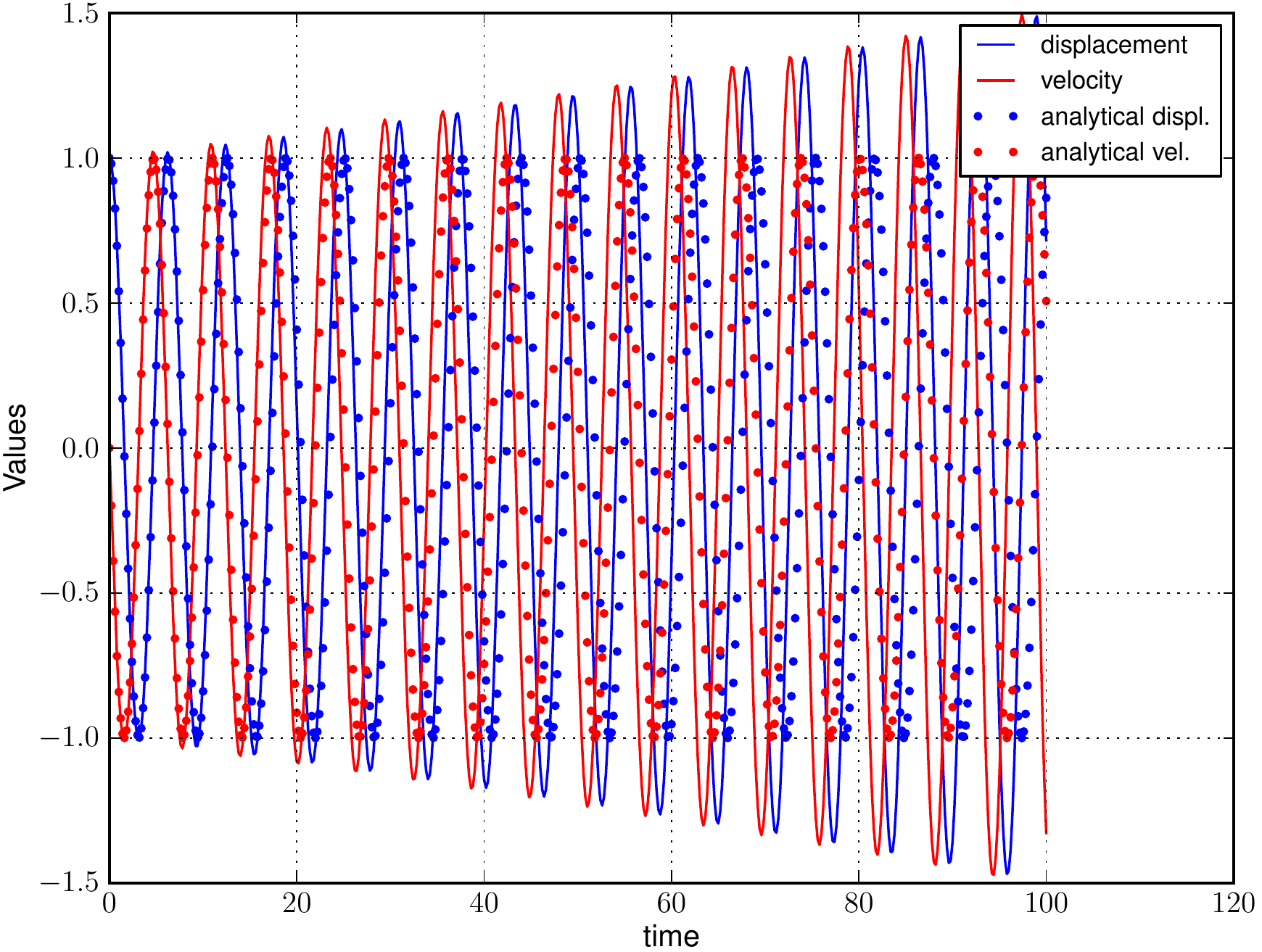}
\includegraphics[ width=0.45\textwidth]{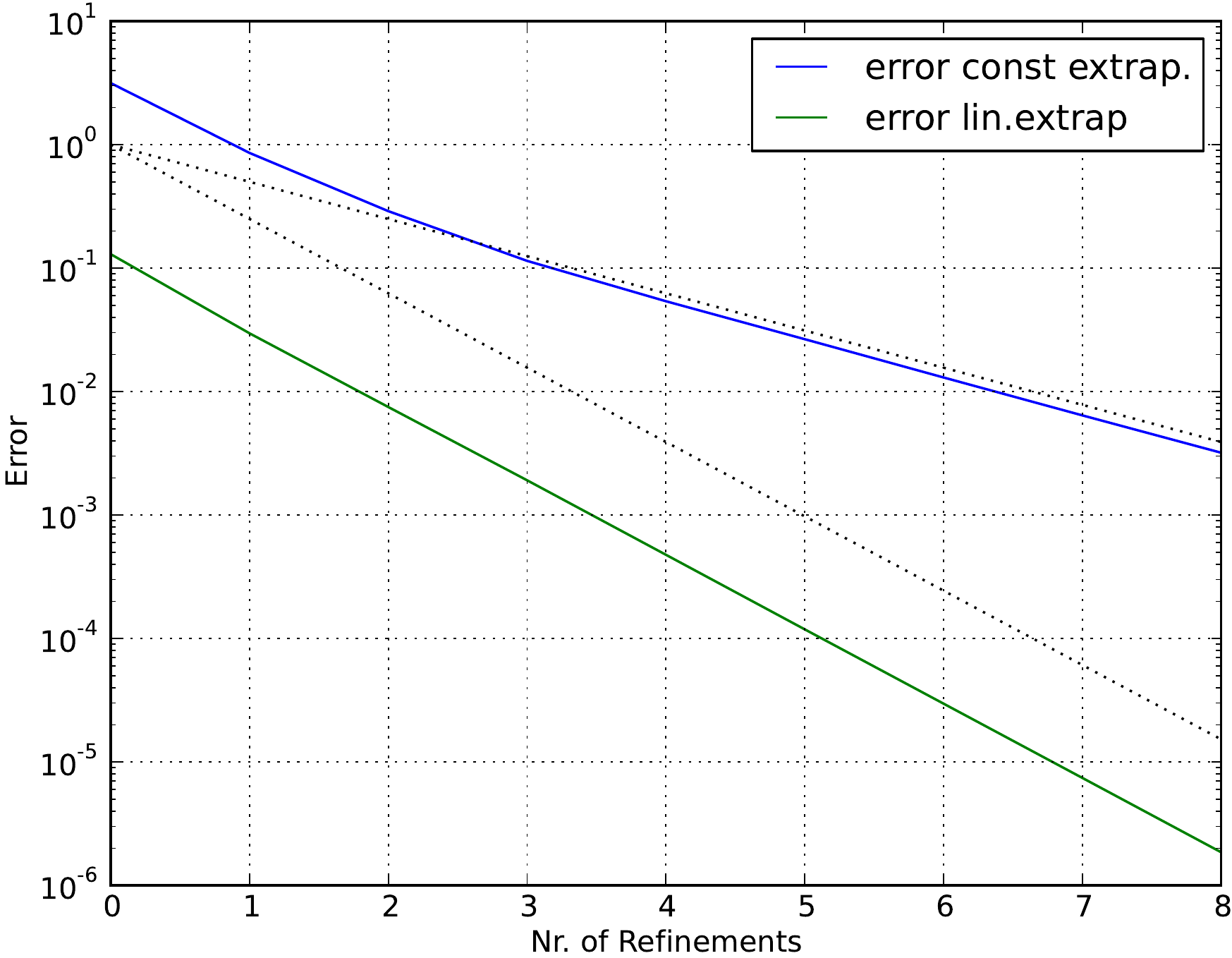}\\
\caption{\label{fig:LinCosimConvergence} Simulation of the system \eqref{eq:springMass} in the cosimulation scheme with linear extrapolation, 
exchange step size $H = 0.2$. Simulation is obviously instable. 
Right plot: The cosimulation method using constant extrapolation converges of order $H$, using linear extrapolation, with order $H^2$.
}
\end{figure}
\begin{figure}
\includegraphics[ width=0.45\textwidth]{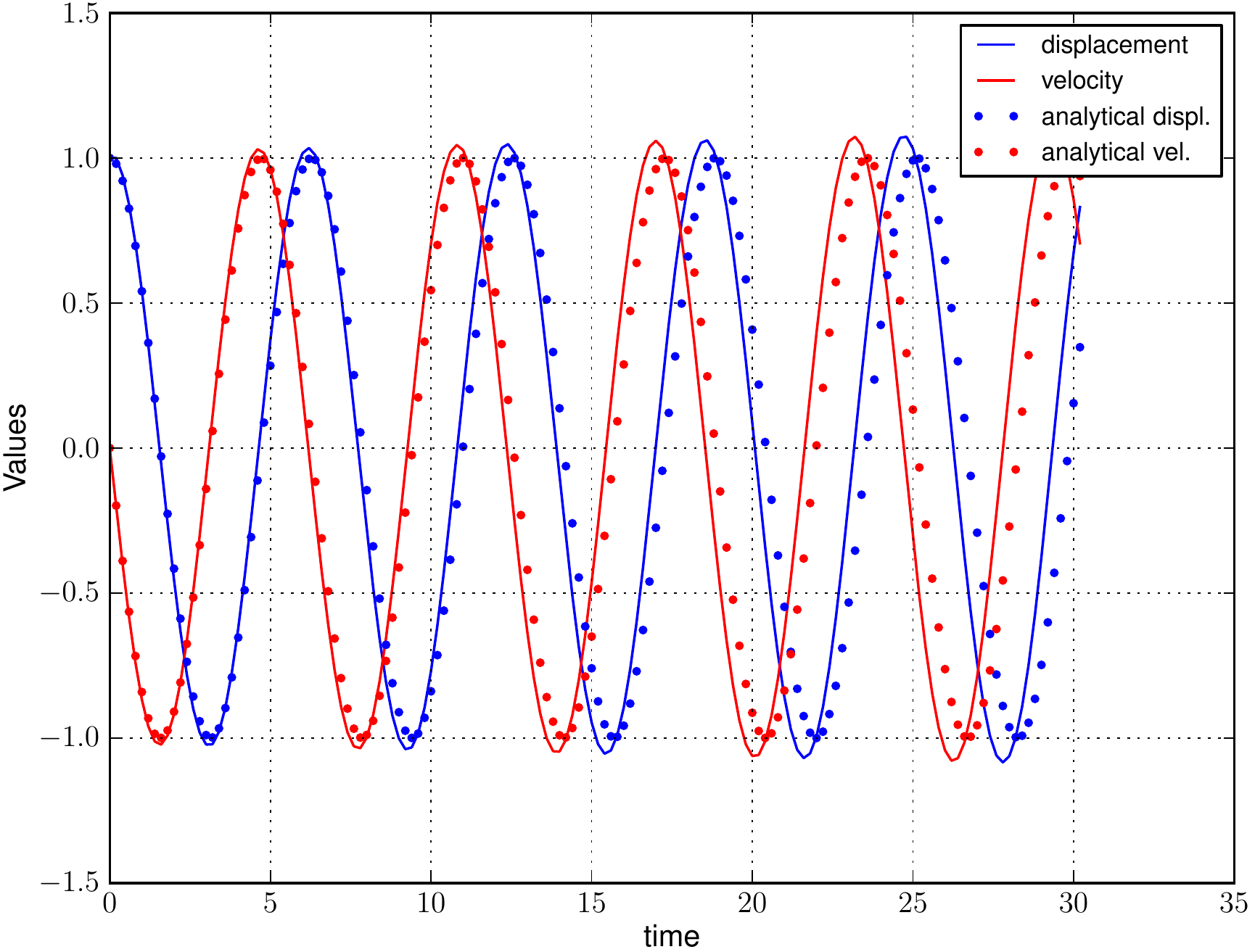}
\includegraphics[ width=0.45\textwidth]{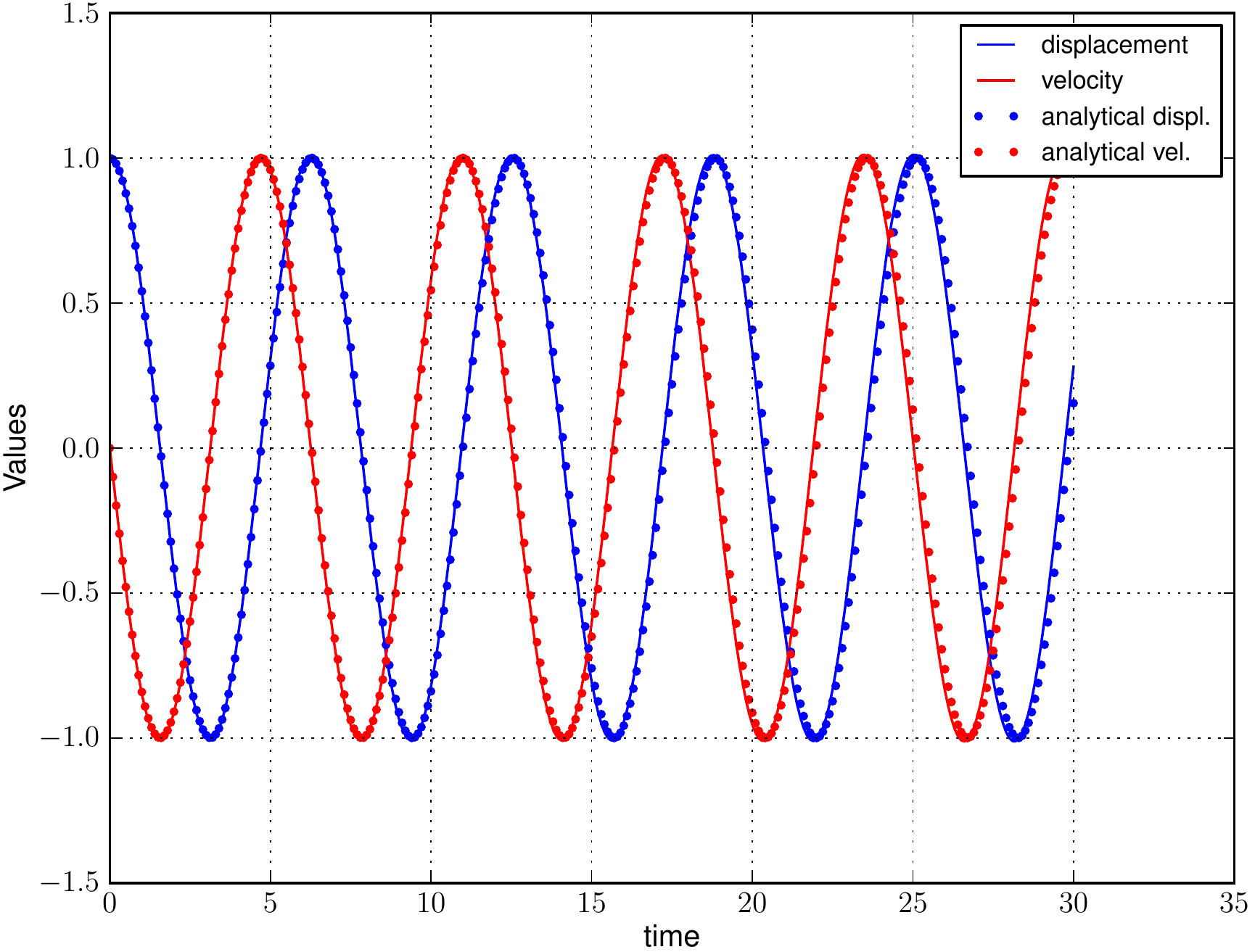}\\
\caption{\label{fig:ConstCosimBCConvergence} Simulation of the system \eqref{eq:springMass} in the cosimulation scheme with constant extrapolation and balance correction, varying the exchange step size $H$.   Left: $H = 0.2$, right:  $H = 0.1$.  The step size $H = 0.1$ is now stable, but the instability of the $H = 0.2$ calculation reveals that the scheme is not made stable by the balance correction.  }
\end{figure}
\begin{figure}
\includegraphics[ width=0.45\textwidth]{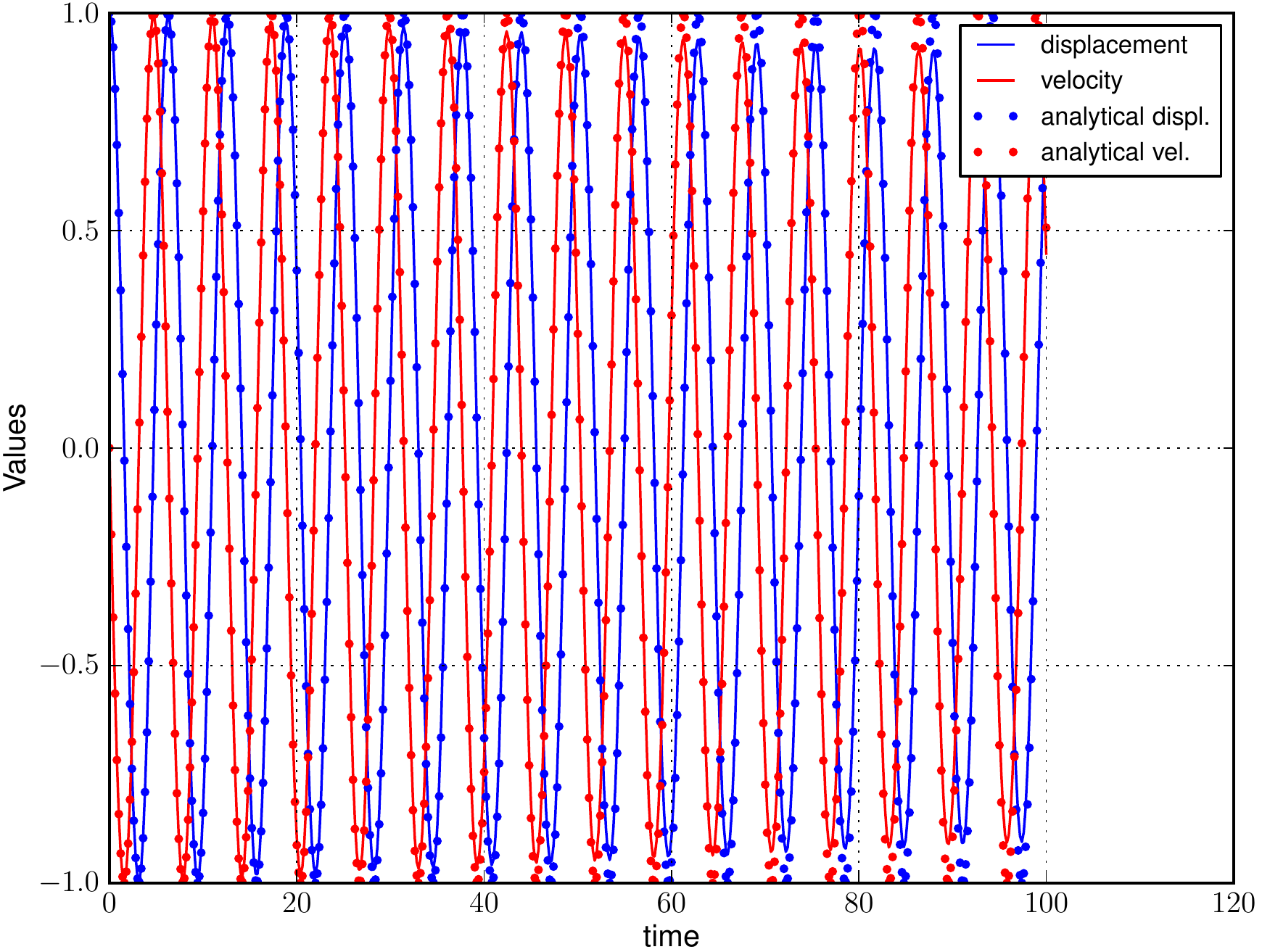}
\includegraphics[ width=0.45\textwidth]{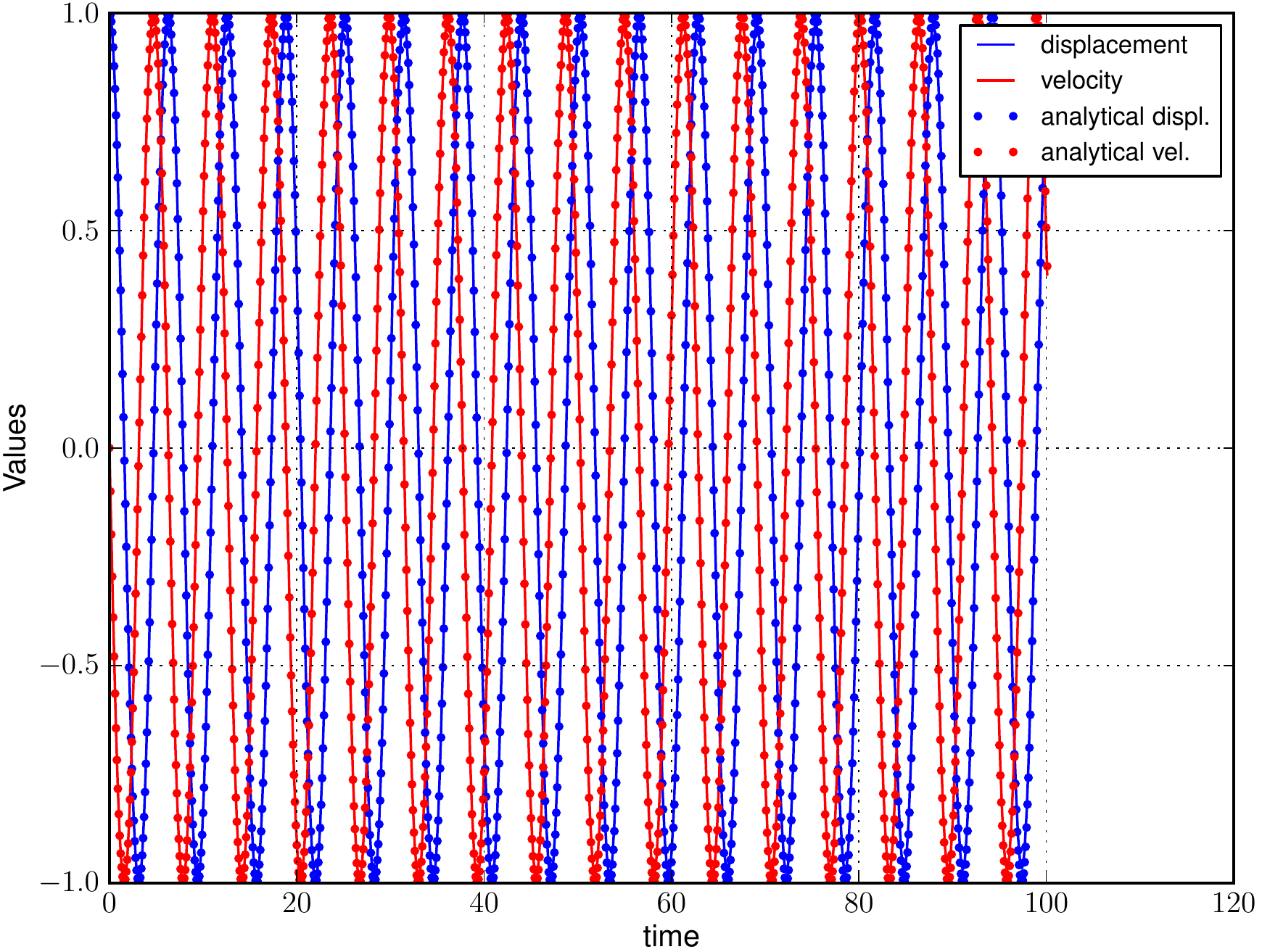}\\
\caption{\label{fig:LinCosimBCConvergence} Simulation of the system \eqref{eq:springMass} in the cosimulation scheme with linear extrapolation and balance correction, varying the exchange step size $H$.   Left: $H = 0.2$, right:  $H = 0.1$.  The step size $H = 0.1$ is now stable, as well as the $H = 0.2$-calculation, but no conservation is established for the latter.  }
\end{figure}
Whereas the system is easily simulated in whole using an A-stable method, when simulated with the cosimulation scheme as above, 
the system picks up energy (figures \ref{fig:ConstCosimConvergence} and \ref{fig:LinCosimConvergence}) as the exchanged quantities force and displacement are factors of it and the approximation error of those factors cause an error (here: rise) in the systems energy  as will be treated in section \ref{factors}. \\
The balance correction algorithm ( see Section \ref{sec:BC}) delivers a better result (figures \ref{fig:ConstCosimBCConvergence} and \ref{fig:LinCosimBCConvergence}), but as most of the quantity missing in one exchange step is refeeded with delay, the energy flux from the spring to the mass in cosimulation is still higher than in the exact solution. As a conclusion, balancing the flows improves, but does not enforce stability, and it
motivates a reclassification of the balance errors.

\subsection{Conclusion: Reclassification of Balance Errors}

\subsubsection{ Errors in balances while exchanging conserved Quantities}
These errors are defined as those  that can be calculated as   $\Delta E_i^{j} = \int_{t_{j-1}}^{t_j}\vek u_i dt  - \int_{t_{j-1}}^{t_j} \overline{ \vek u_i} dt$, 
thus    those resulting directly from extrapolation errors in flow of the conserved quantity.
 They for example occur the mass of a fluid in a cycle like in \cite{KosselDiss} and are treated in \cite{Scharff12} as the original balance errors, before one tried to balance non-conserved quantities like in this paper.\\

\subsubsection{Errors in balances while exchanging  factors of conserved quantities}\label{factors}
If an exchanged quantity influences a conserved quantity, the balance of that quantity is disturbed by the approximation error of an input as described above, even if the input quantity is nonconserved. Furthermore, as the conserved quantity depends on other factors, its imbalance may persist even if the balance of the exchanged quantity is reestablished, e.g. by errors compensating each other. For all we know, this problem has not been described so far. Yet, in our case studies, it becomes apparent and it will be subject of our future work.

\section{Effects of discontinuous Inputs: Method-induced Oscillations}
 \subsection{Spring-Mass system on moving ground}
\label{sec:MovingGround}
To examine the effect of discontinuity of input due to extrapolation on subsequent subsystems, consider a spring on moving soil.\\  
To relate it to the cosimulation context, this system can be seen as two coupled spring and mass systems in its limiting case, in which the first mass is very big, and so is the stiffness of the first spring, but still such  that the frequency  $f=\omega/2\pi$, $\omega_{1,2}=\lambda_{1,2}\sim \pm\sqrt{-\frac{c}{m}}$ is well bigger than the frequency of the second system.   So the system 
\begin{equation}\label{eq:DoubleSpringMass}
\frac{d}{dt}
\begin{pmatrix}
x_1\\
\dot x_1\\
x_2\\
\dot x_2
 \end{pmatrix}
=
\frac{d}{dt}
\vec x
=\begin{pmatrix}
0	& 1	& 0	& 0  \\
-\frac{c_1+c_2}{m_1}	& -\frac{d_1+d_2}{m_1}	& \frac{c_2}{m_1} 	& \frac{d_2}{m_1} \\
 0 	& 0		& 0 & 1 \\
 \frac{c_2}{m_2}		& \frac{d_2}{m_2} & -\frac{c_2}{m_2}   & -\frac{d_2}{m_2} 
\end{pmatrix}
\vek x
+\begin{pmatrix}
0\\
\frac{1}{m_1}(c_1l_{0,1}-c_2l_{0,2})\\
0\\
\frac{1}{m_2}(c_2l_{0,2})\\
\end{pmatrix},
\end{equation}
is mimicked, for the case it is nearly decoupled by  $m_1\gg m_2$, $\frac{c_2}{m_2} >\frac{c_1}{m_1} $,  by the system
\begin{equation}\label{eq:MovingGround}
\begin{split}
\dot{ \vec{ x}}
= \ten A\vec{x} 
= \begin{pmatrix}
0 & 1\\
-\frac{c}{m} & \left(-\frac{d}{m}\right)
 \end{pmatrix} \left(\vec x - \begin{pmatrix}x_0(t)\\ 0 \end{pmatrix}\right), 
 \qquad \vec{x} = 
 \begin{pmatrix}
x\\
\dot x
 \end{pmatrix}\\
 x_0(t) = x_{t_0}\cos(\omega_1 t)
 \end{split}
\end{equation}
%

The real behavior of such a system, given that the spring is relaxed at $t=0$, is that the  mass quietly follows the ground movement, i.e. the position of the mass is approximately $x(t)\sim x_{t_0}\cos(\omega_1 t)+h_0$ without own oscillations. \\
 The position of the very heavy systems mass  $x_0(t)$, moving slowly, is then taken as the input $u$, and the exchange step width is chosen in orders of magnitude of the small spring-mass systems eigen frequency. Now the effect of the discontinuities between  piecewise extrapolation can be studied  (the simplifications of course might hide instability).
\begin{figure}[h!tb]
   \centering
\includegraphics[width=.45\textwidth]{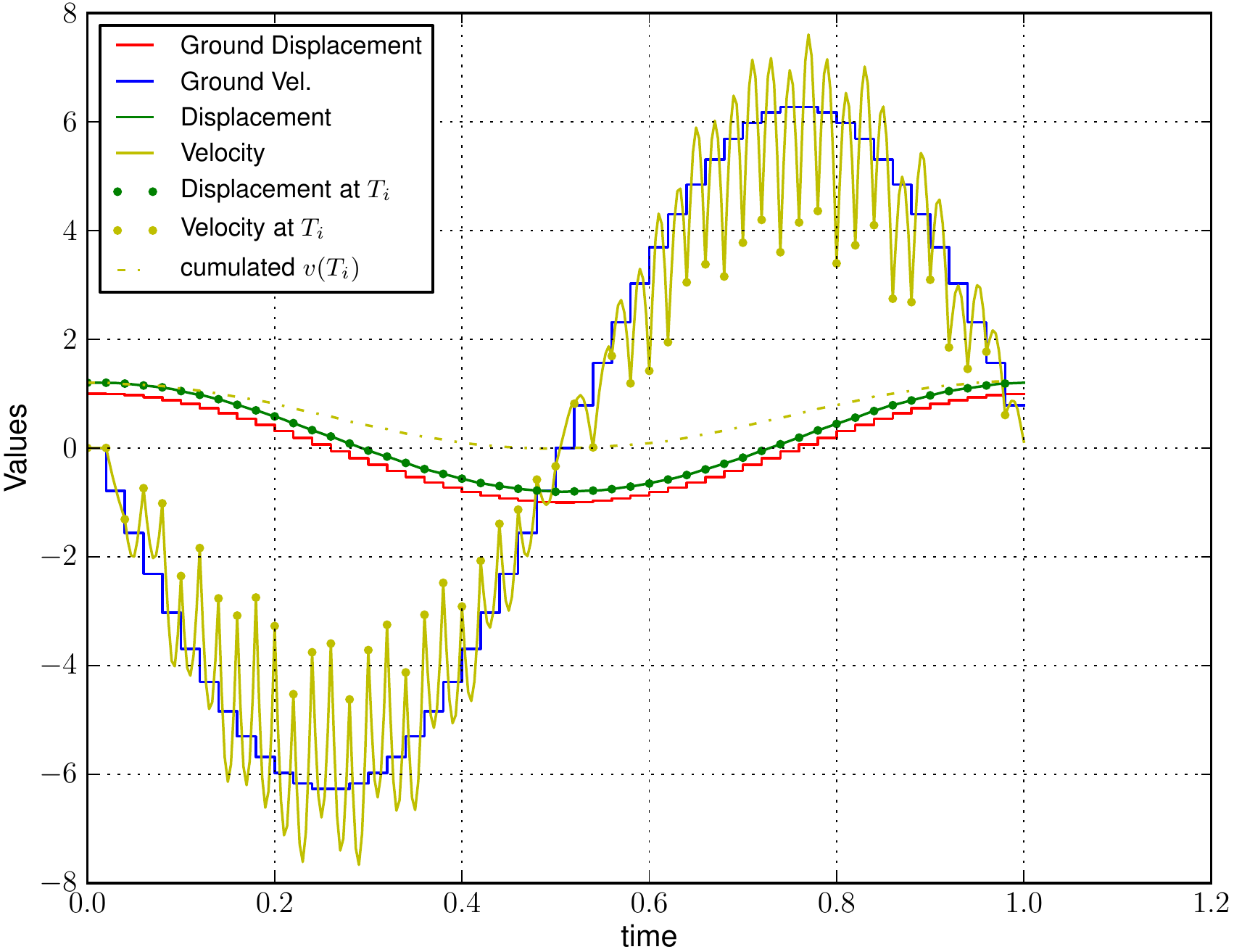} \includegraphics[width=.45\textwidth]{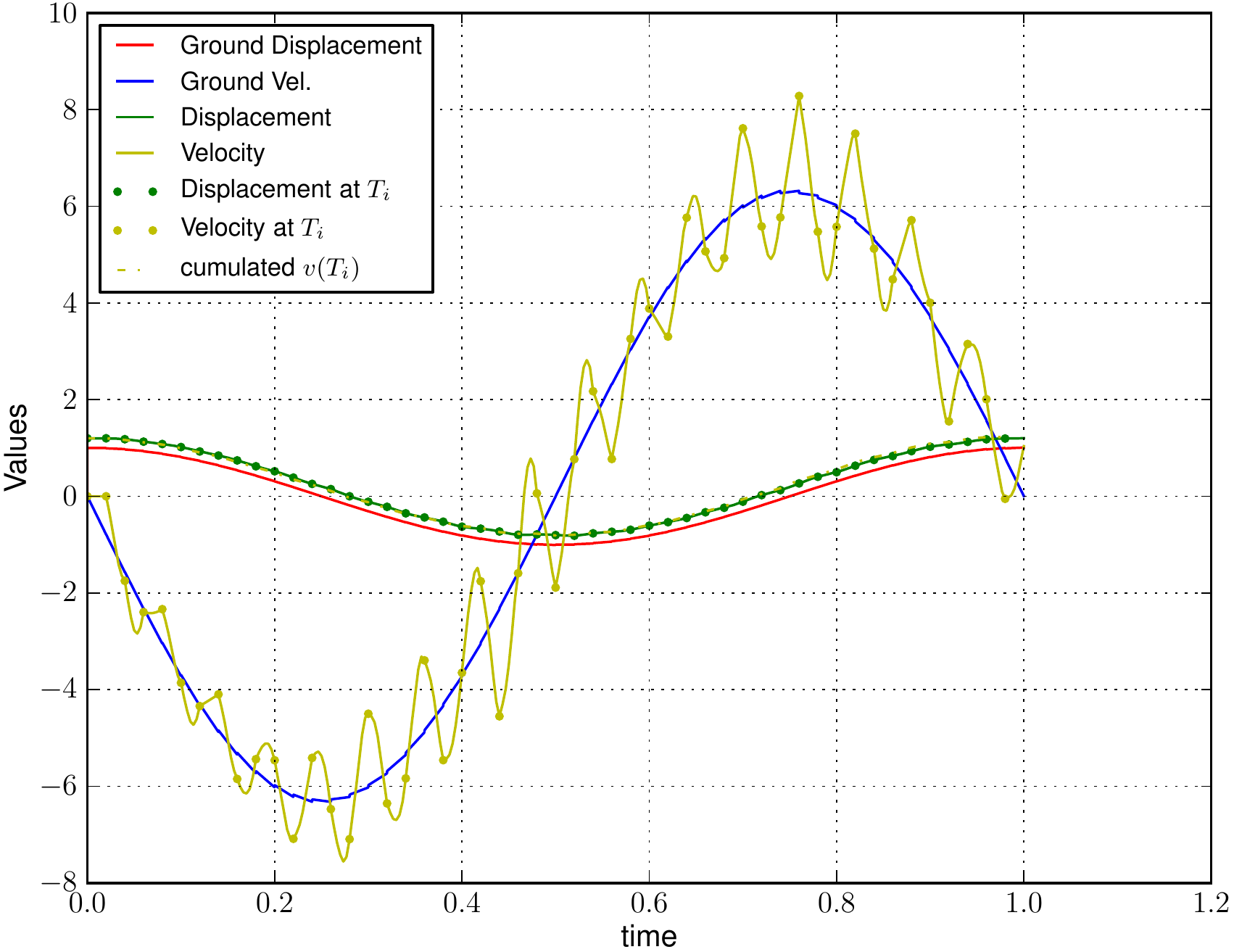} 
 \caption{The effect of extrapolation on the receiving spring mass system, using constant and linear extrapolation, respectively. Here and in the following simulations $m_2=0.0005$, $k_2=5.0$, and $\omega_1=2\pi$, corresponding to e.g. $m_1=1.0$, $k_1=4\pi^2$, see eq. \eqref{eq:DoubleSpringMass}. Red and blue curve are the extrapolants of moving soils displacement and velocity. The light green curve is the velocity of the mass, showing strong oscillations due to discontinuous input. Additionally,  the numerical integral of the velocity is plotted, using only the values at exchange times (upper, light green  dotted line) and values inside exchange interval (dark green dots on the displacement graph. This reveals a further pitfall in cosimulation. }
 \label{fig:MovingGround}
\end{figure}

\begin{figure}[h!tb]
   \centering
\includegraphics[width=.45\textwidth]{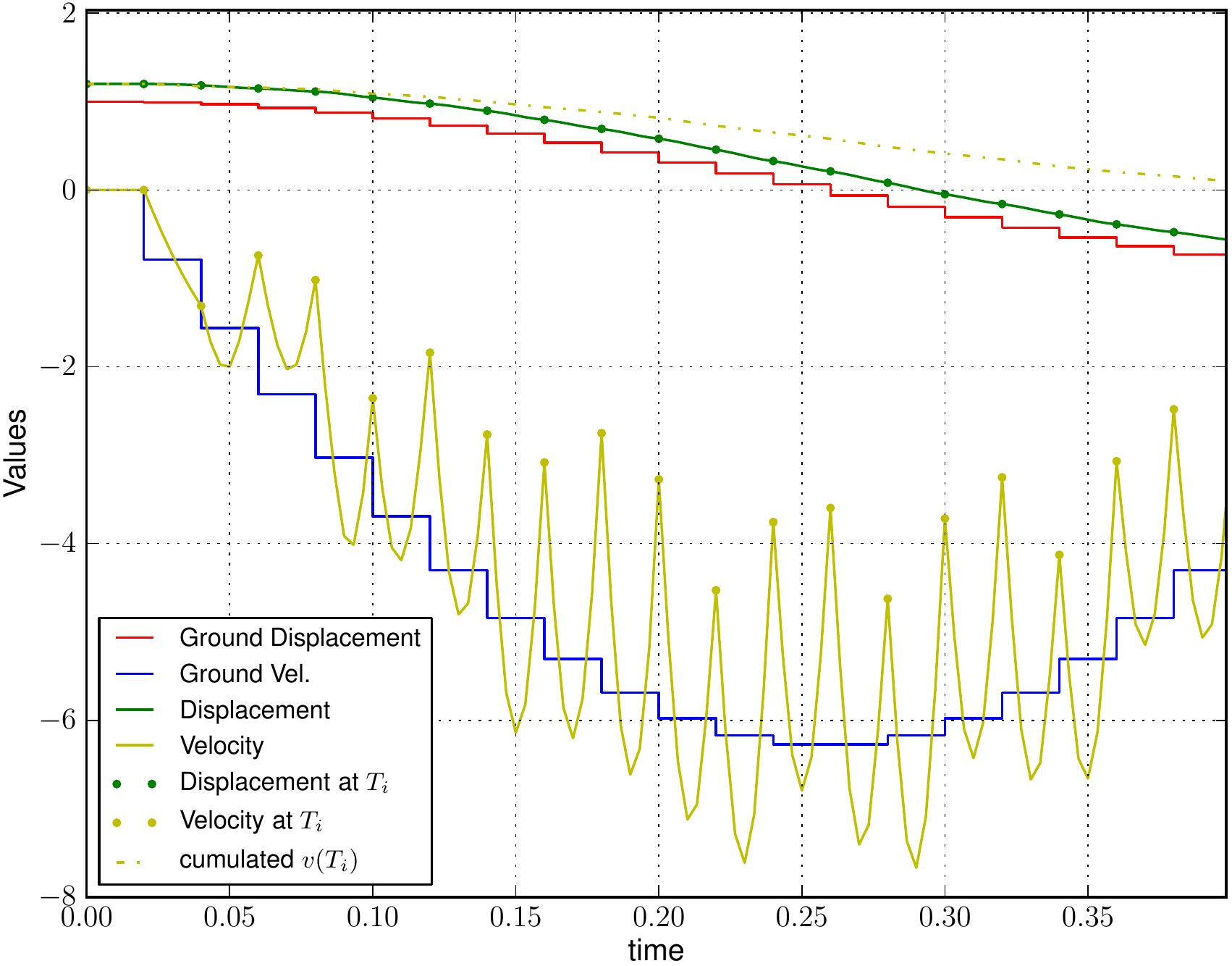} \includegraphics[width=.45\textwidth]{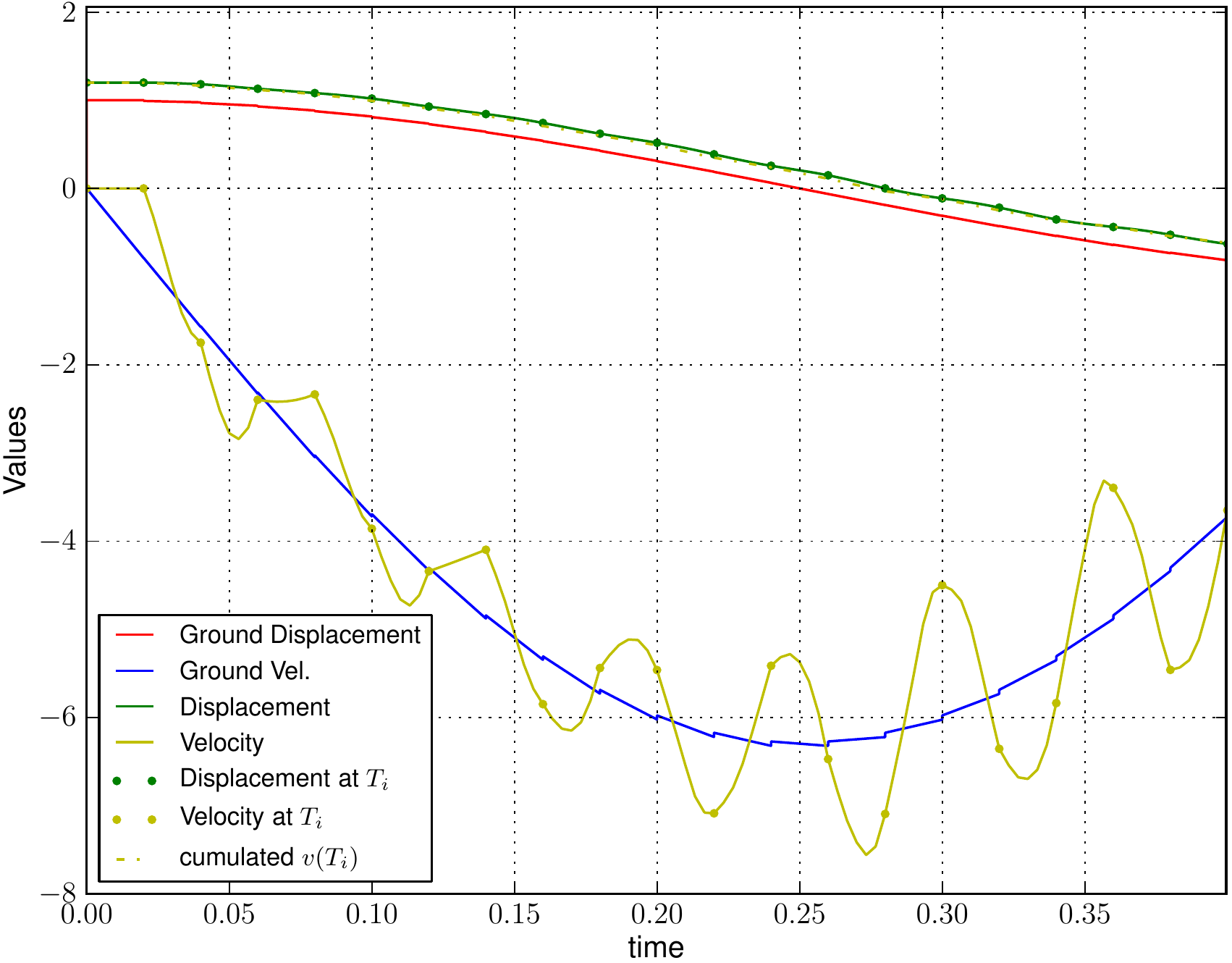} 
 \caption{First half of the previous plot \ref{fig:MovingGround}, for  details.}
 \label{fig:MovingGroundDetail}
\end{figure}
\begin{figure}[h!tb]
   \centering
\includegraphics[width=.45\textwidth]{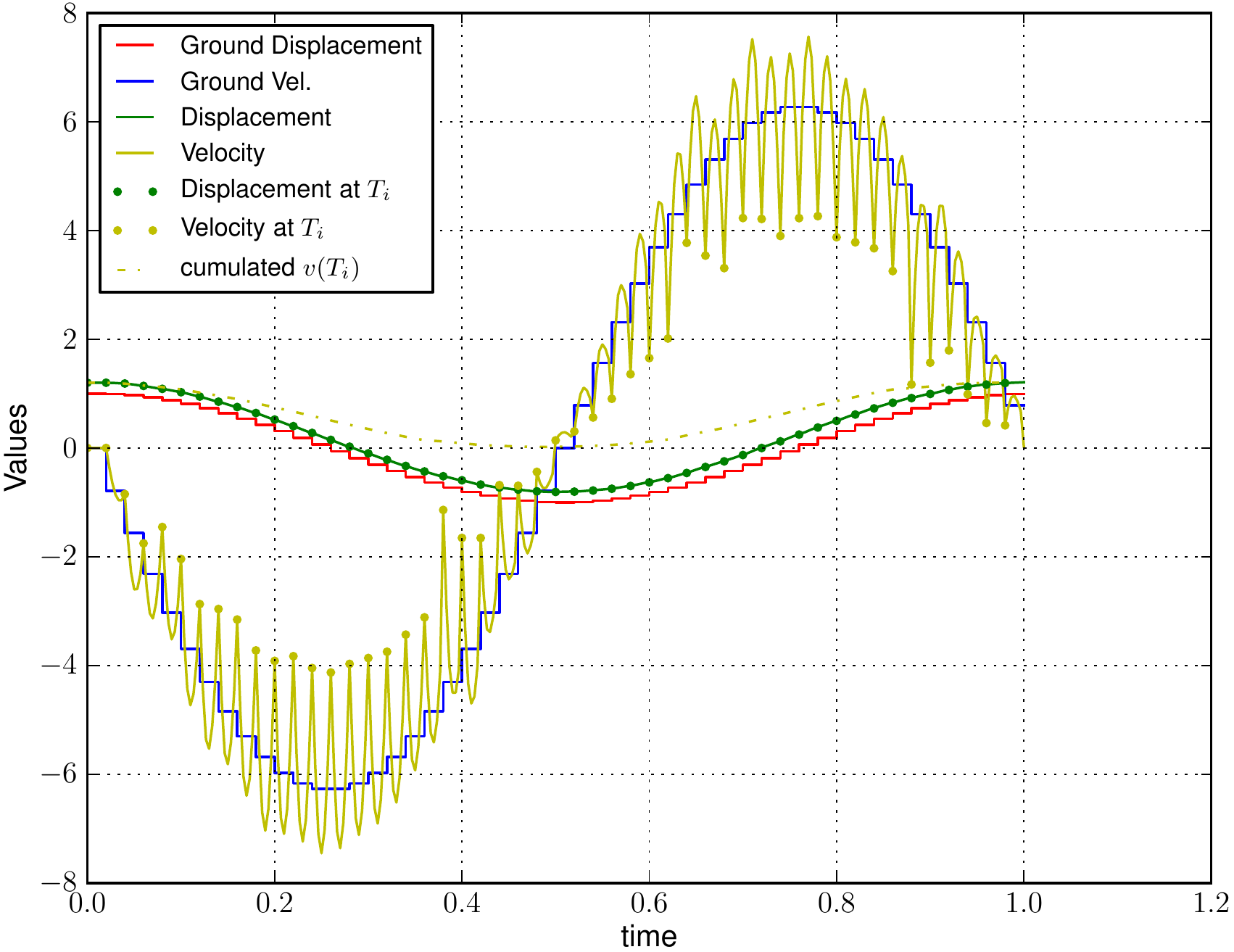} \includegraphics[width=.45\textwidth]{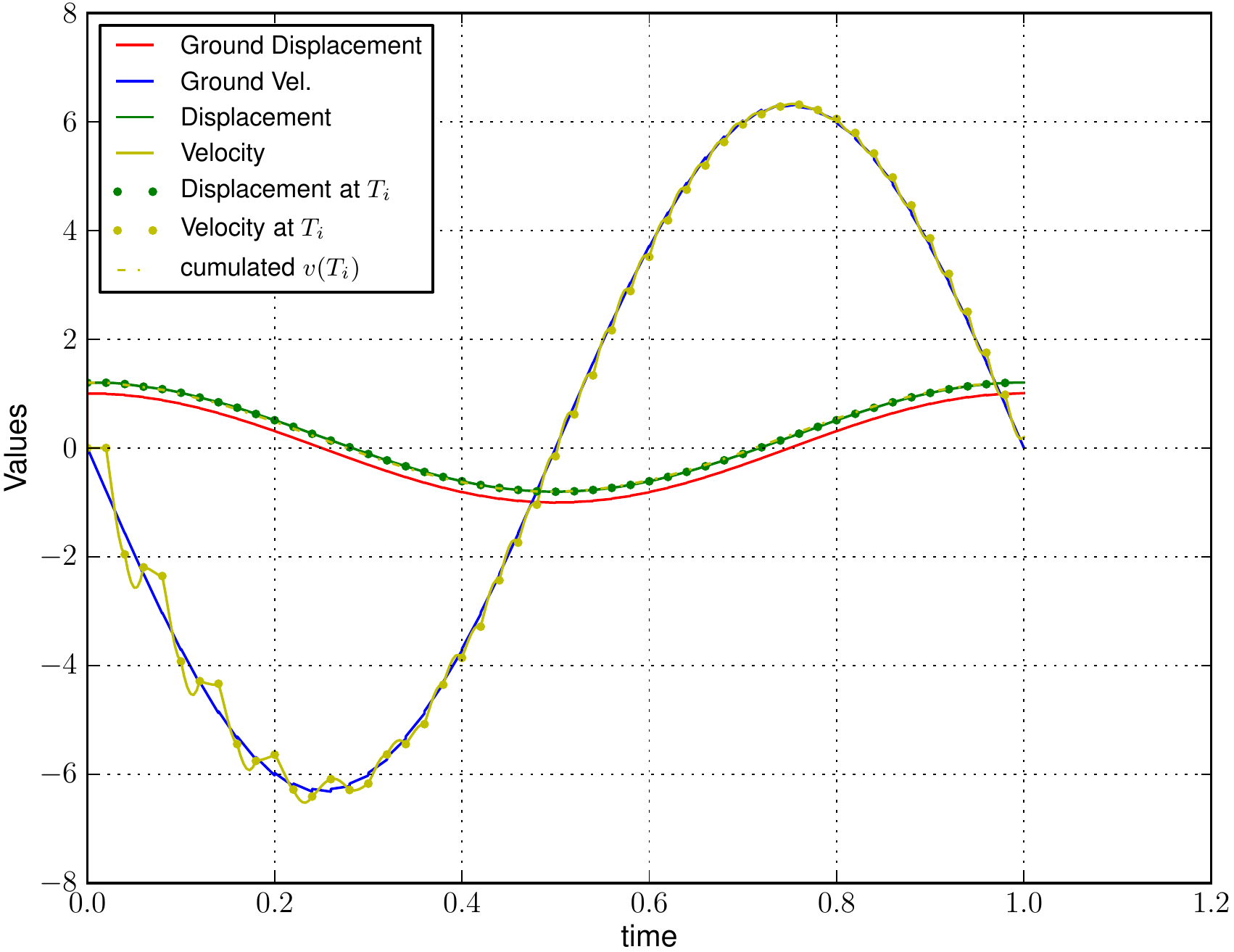} 
 \caption{The effect of extrapolation on the receiving spring mass system, using constant and linear extrapolation with Balance Correction, respectively. Plot colouring and conclusions as in fig. \ref{fig:MovingGround}. }
 \label{fig:MovingGroundBC}
\end{figure}
\subsubsection{Simulations without and with balance correction}
The figures \ref{fig:MovingGround} and \ref{fig:MovingGroundBC} show well these effects  on the subsystem. In the constant extrapolation case, the velocity inside a step shows typical bellies, which are caused by the extrapolant being too high in the beginning of each interval and  lower as the real one in the end, or vice versa.       In the linear extrapolation case, the velocity 
also shows  bellies, but they are due to the initial error caused by the constant extrapolation in first step: When $H$ is modified, that oscillations stay. Nevertheless quadratic errors in $x_0$ and thus in the force amplify these oscillations: They keep rising in the standard case (fig. \ref{fig:MovingGround} right). 
It is clear that the oscillations in velocity cause oscillations in the displacement. Both are unphysical.\\
Figure \ref{fig:MovingGroundBC} shows identical setting of the simulation, except that now balance correction is applied - it can be seen that this has the desirable effect of stabilizing the initial oscillation in the linear extrapolation calculation, but the oscillations induced by input discontinuity remain. Balance correction does not affect  smoothness.
\\
Another treacherous property of the cosimulation scheme is revealed here: If velocity values are only stored at exchange times (dots in the plot), in the constant extrapolation case the impression is made that the velocity is low. Then the displacement is inconsistent with the velocity, see  figures  \ref{fig:MovingGroundDetail}, left, and \ref{fig:MovingGroundBC}, left image.\\
These plots of simulation behavior clarify how desireable it is to provide a method that smoothens the input and the correction contributions. Considering DAE problems with disturbation index greater zero (see \cite[chapt. 3.1]{DeuflhardBornemann94}), it is clear that unsmoothed cosimulation must fail in total. \\
\subsubsection{Conclusion}Exchange step induced oscillations matter in some situations, and the methods so far discussed, even smoothed linear extrapolation with balance correction, need further improvements for settings vulnerable to exchange induced oscillations.

\subsubsection{Simulations while smoothing the input}
\begin{figure}[h!tb]
   \centering
\includegraphics[width=.45\textwidth]{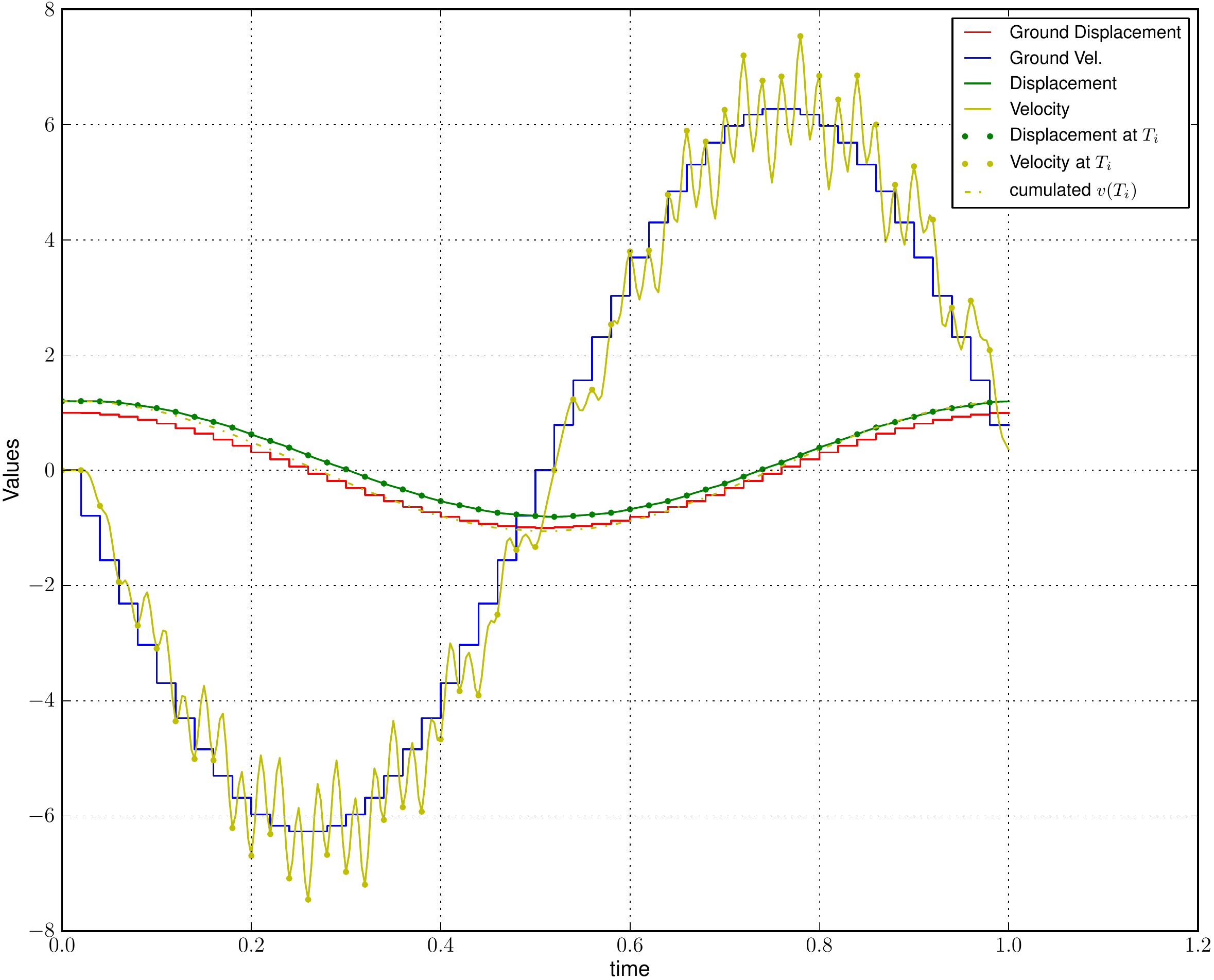} \includegraphics[width=.45\textwidth]{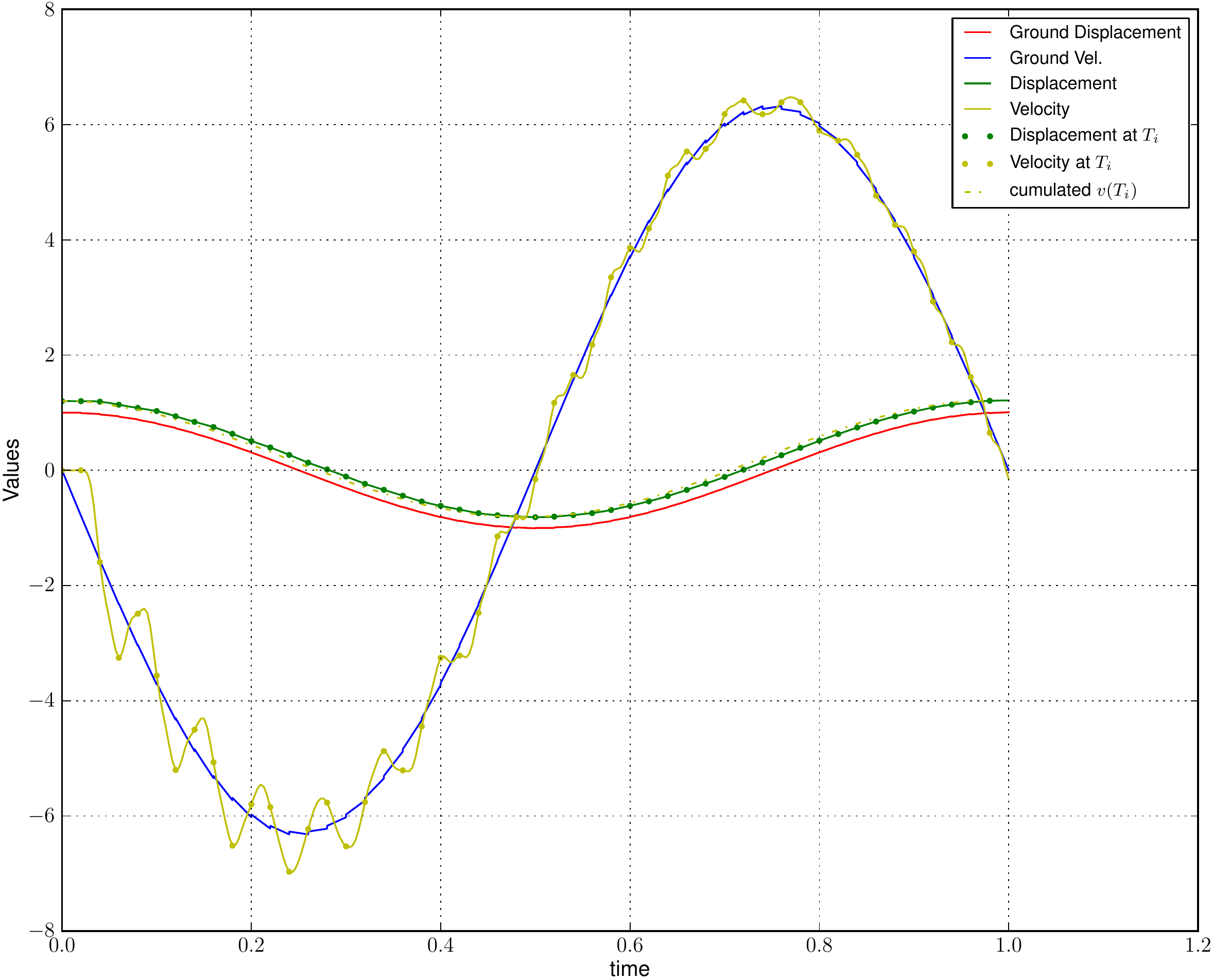} 
 \caption{The effect of extrapolation on the receiving spring mass system, using constant and linear smooth extrapolation, respectively. Plot colouring as in fig. \ref{fig:MovingGround}. The oscillations are reasonably reduced (leaving aside the initial numerical excitation) compared to nonsmoothed input (\ref{fig:MovingGround}).}
 \label{fig:MovingGroundSmooth}
\end{figure}
\begin{figure}[h!tb]
   \centering
\includegraphics[width=.45\textwidth]{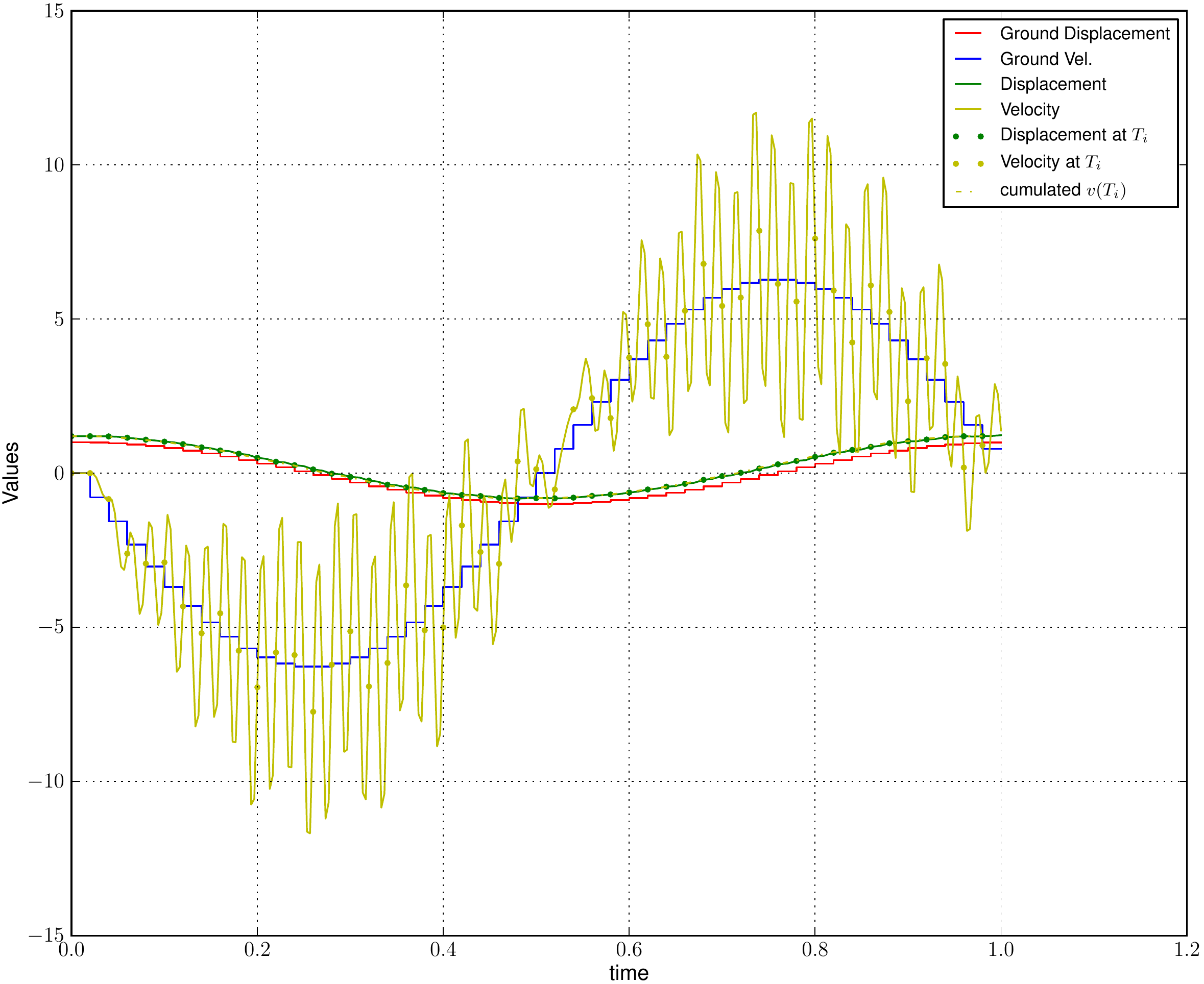} \includegraphics[width=.45\textwidth]{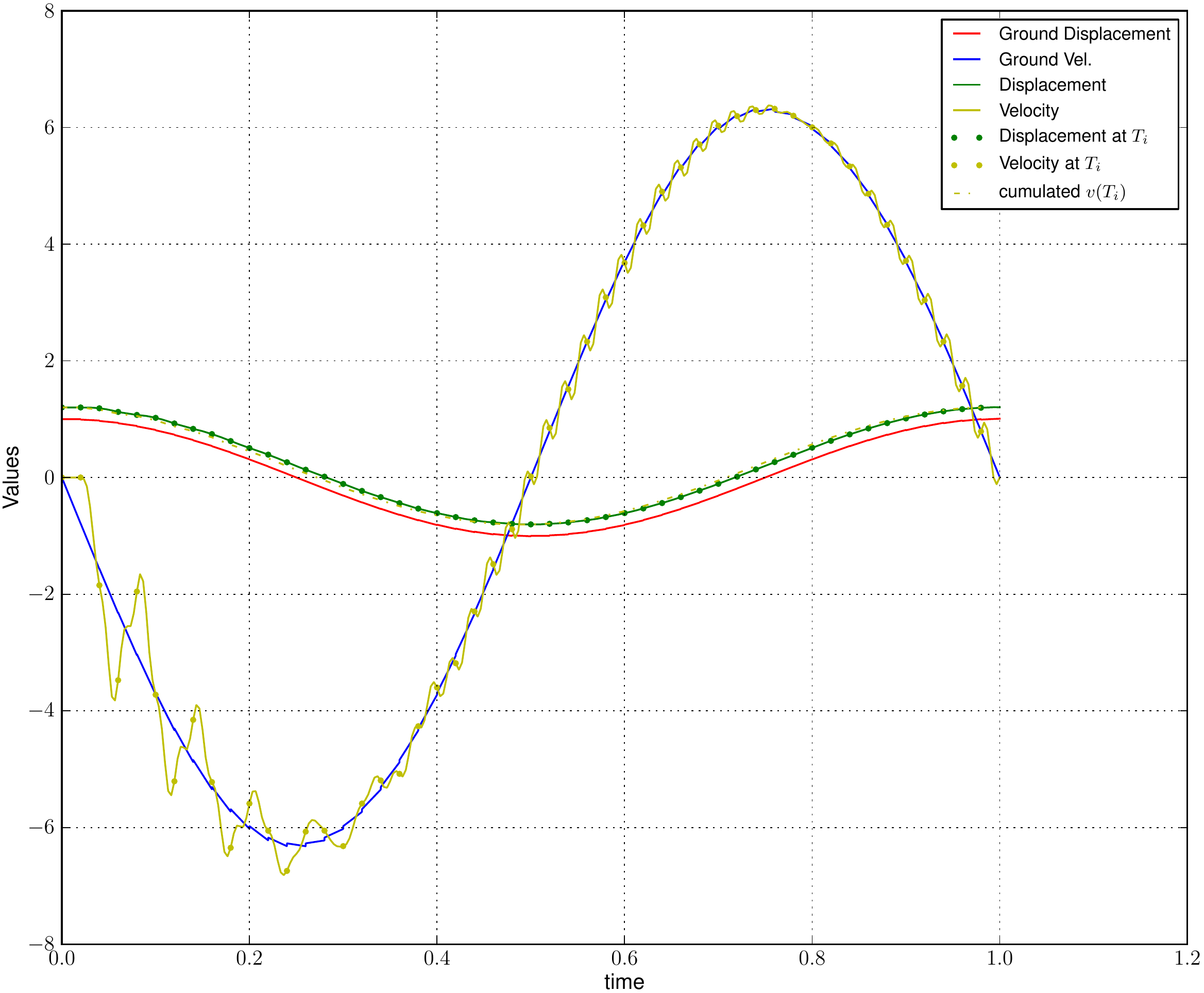} 
 \caption{The effect of extrapolation on the receiving spring mass system, using constant and linear smooth extrapolation, respectively, and balance correction. Plot colouring as in fig. \ref{fig:MovingGround}. The oscillations are amplified by the balance correction contributions, compare to (\ref{fig:MovingGround}).}
 \label{fig:MovingGroundSmoothBC}
\end{figure}
Now the effect of smoothed input on oscillations of subsequent systems is examined using the moving ground example. Figure \ref{fig:MovingGroundSmooth}
shows that if input is smoothed, input-induced oscillations are reasonably reduced  compared to nonsmoothed input situation. In the linear extrapolation case, the smoothing even damps the high-frequency oscillations.\\
If balance correction is used, again big oscillations in data exchange frequency can be observed, quite similar as in the  nonsmoothed input (\ref{fig:MovingGroundSmooth}) case and much bigger  than without those contributions. Obviously, the balancing contributions excite the spring-mass in a similar way as the discontinuities. This can be explained by the fact that switch errors contribute to the corrections, and their amplitude may increase, even if fed back with a smooth hat function.
\subsubsection{Conclusion} The problem of exchange step induced  oscillations can be relieved by smooth switching, but balance correction contributions cause high derivatives in input too, a matter that should be adressed.

\subsection{Stability effect of smooth switching and balance correction}
\begin{figure}
\includegraphics[ width=0.4\textwidth]{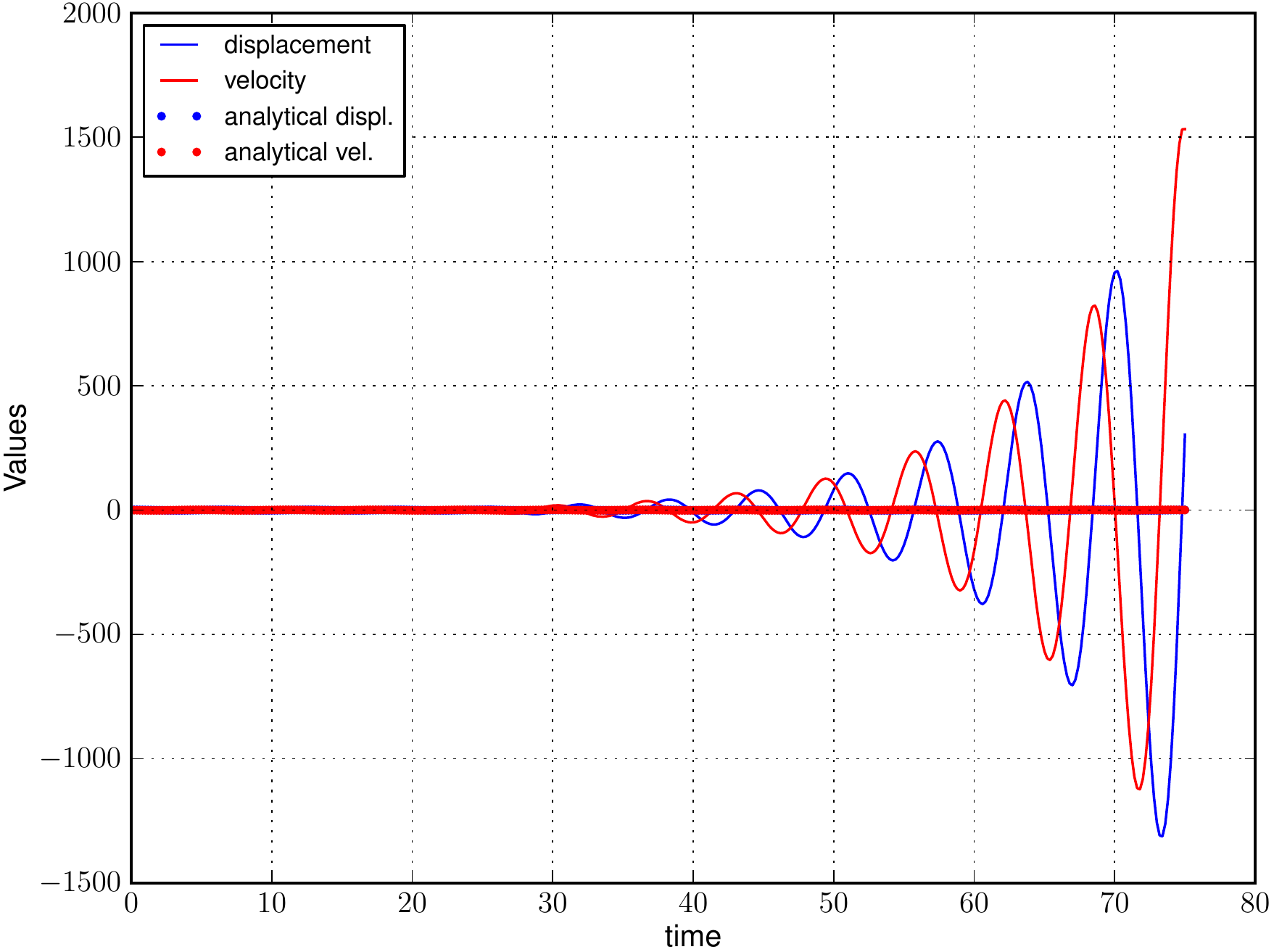}
\includegraphics[ width=0.4\textwidth]{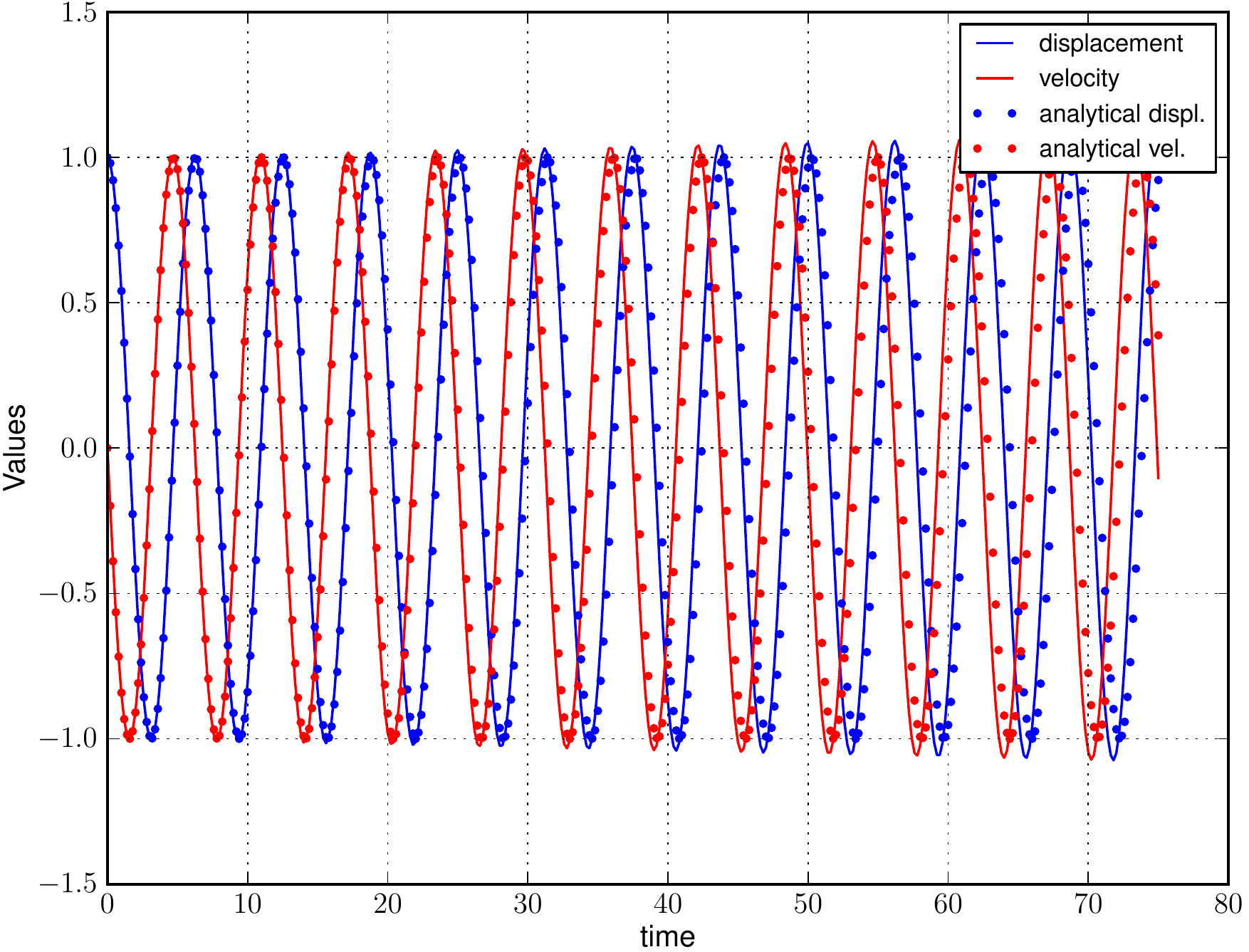}\\
\includegraphics[ width=0.4\textwidth]{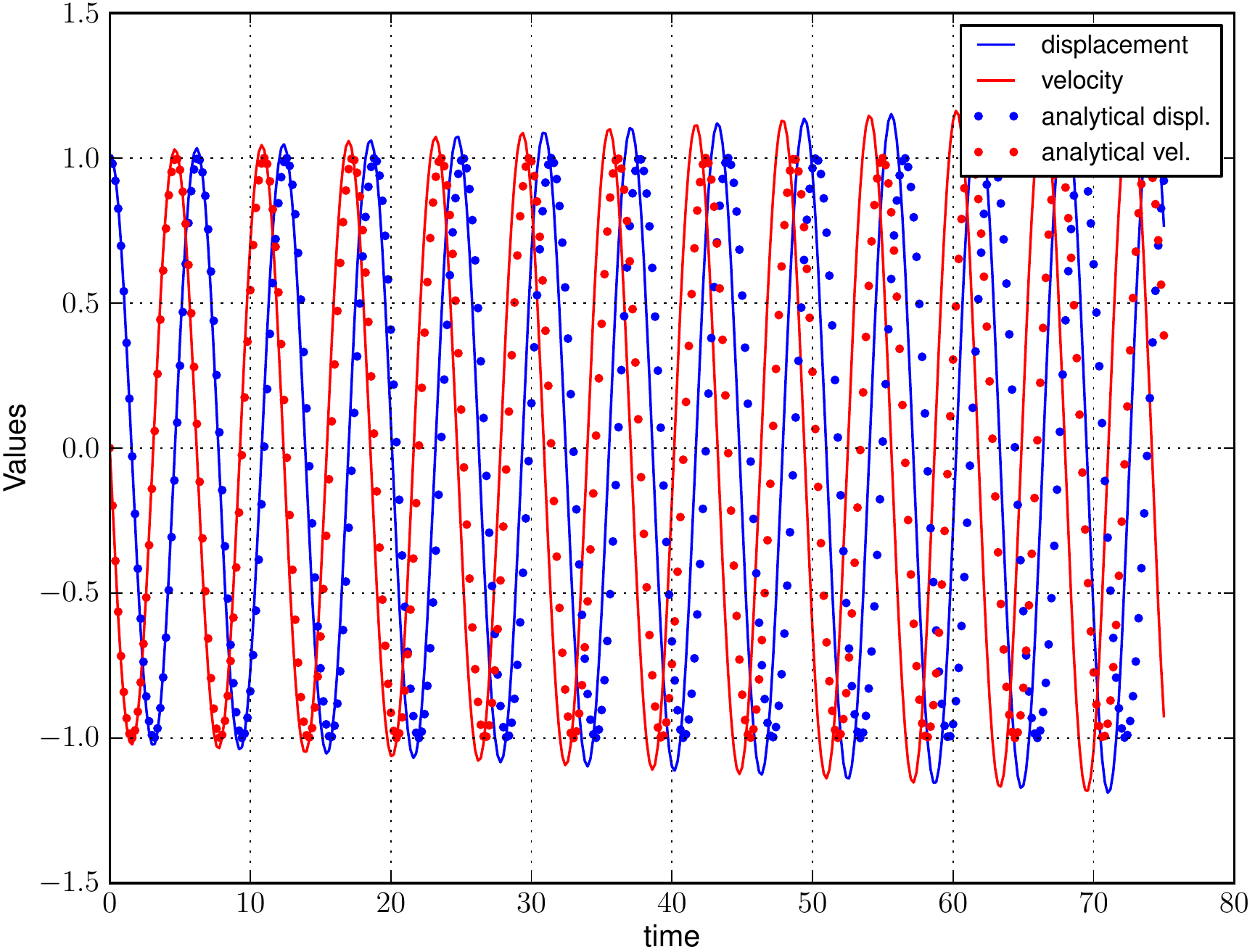}
\includegraphics[ width=0.4\textwidth]{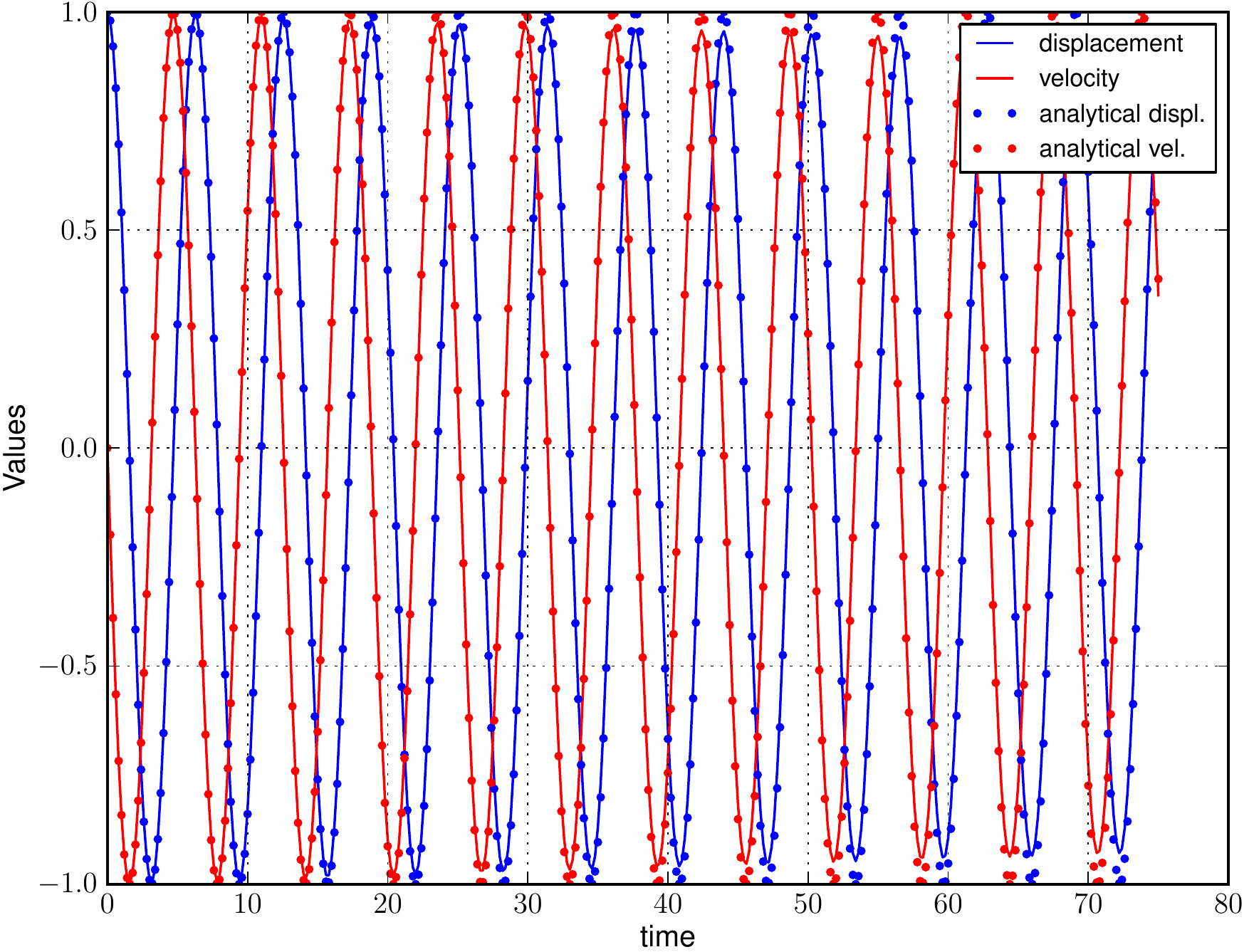}\\
\caption{\label{fig:cosimStabilityReminder} Reminder: Simulation of the system \eqref{eq:springMass} in the cosimulation scheme with constant (left) and linear (right) extrapolation, upper plots without, lower with Balance correction.   $H = 0.2$.  }
\end{figure}
\begin{figure}
\includegraphics[ width=0.45\textwidth]{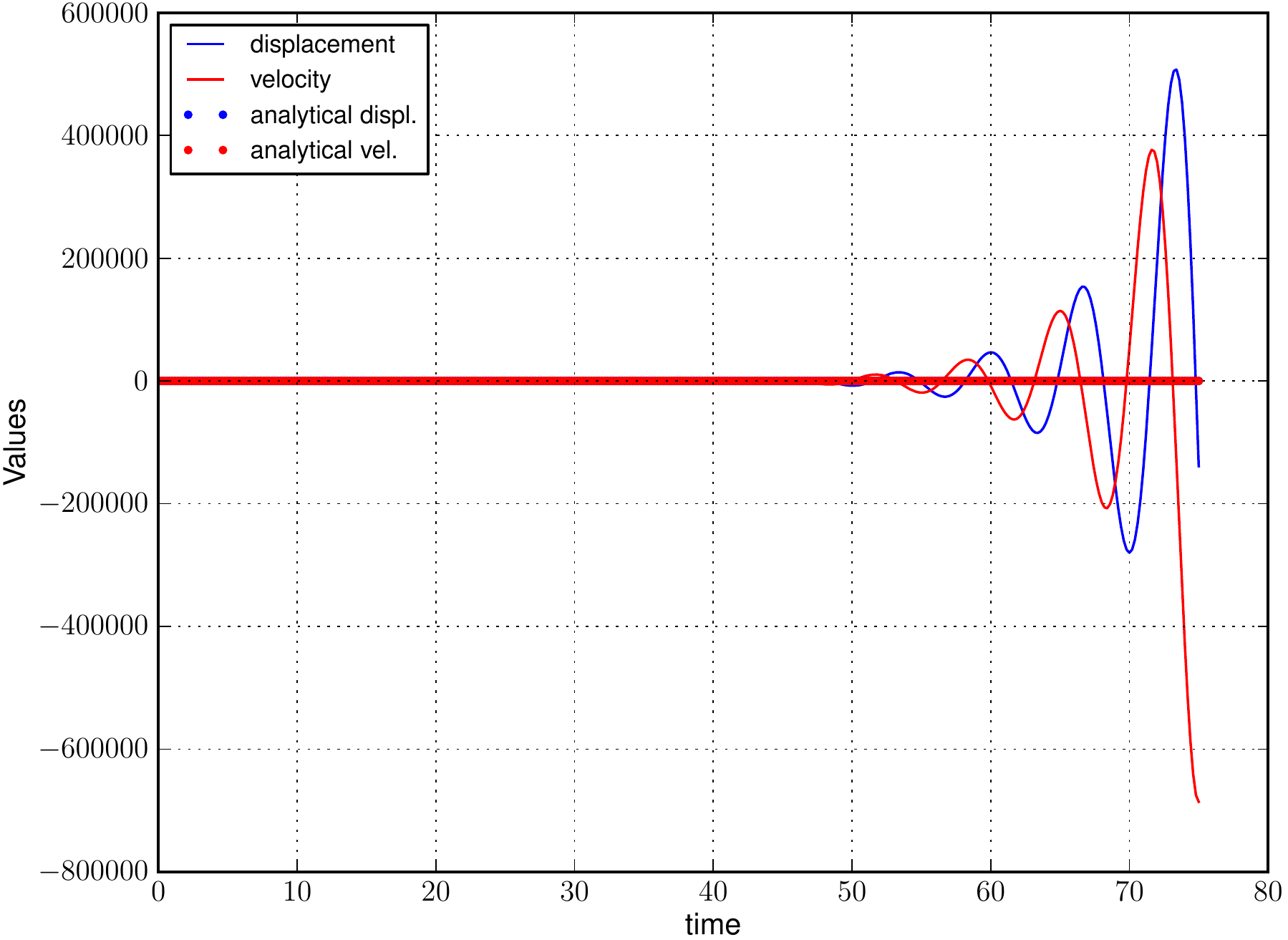}
\includegraphics[ width=0.45\textwidth]{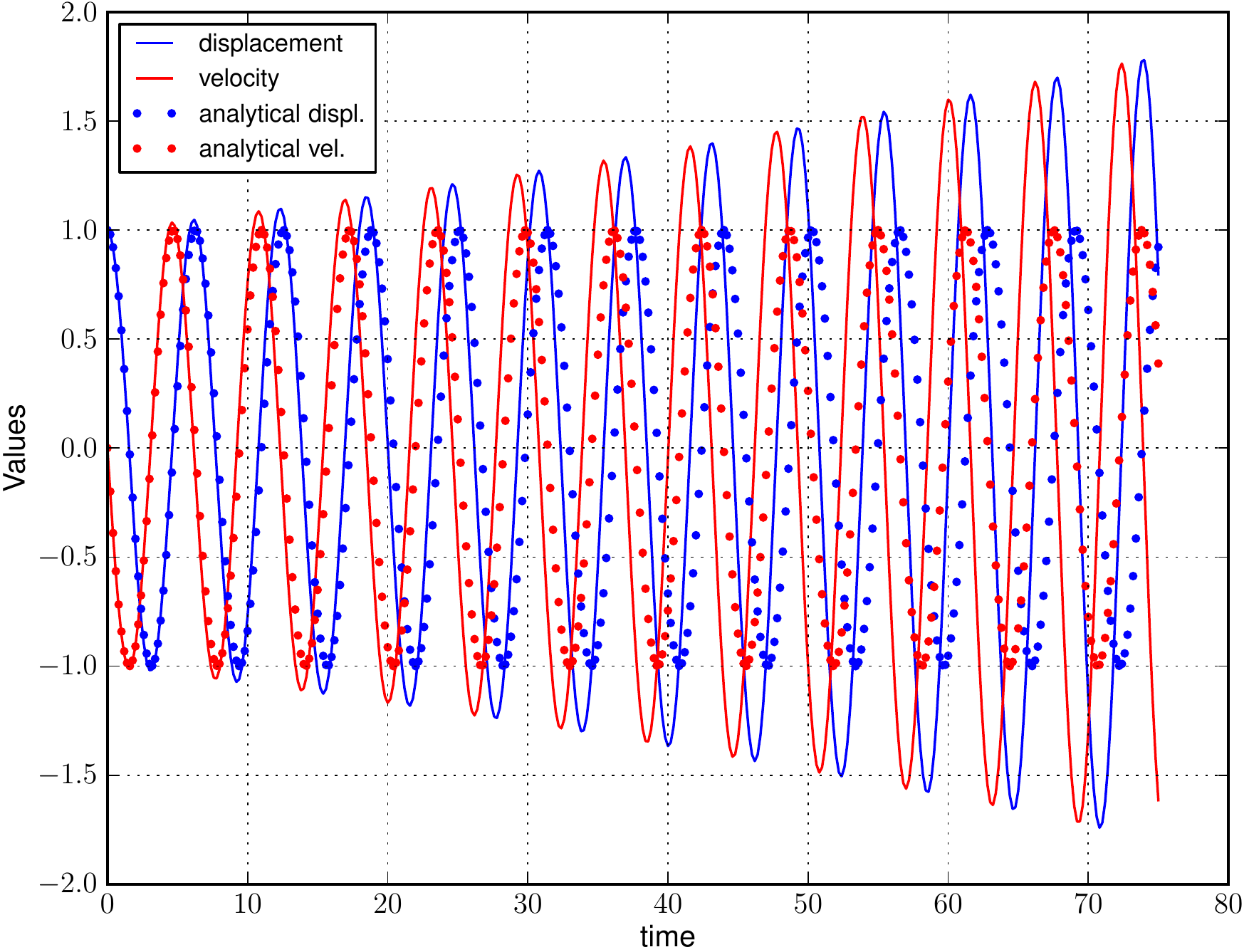}\\
\includegraphics[ width=0.45\textwidth]{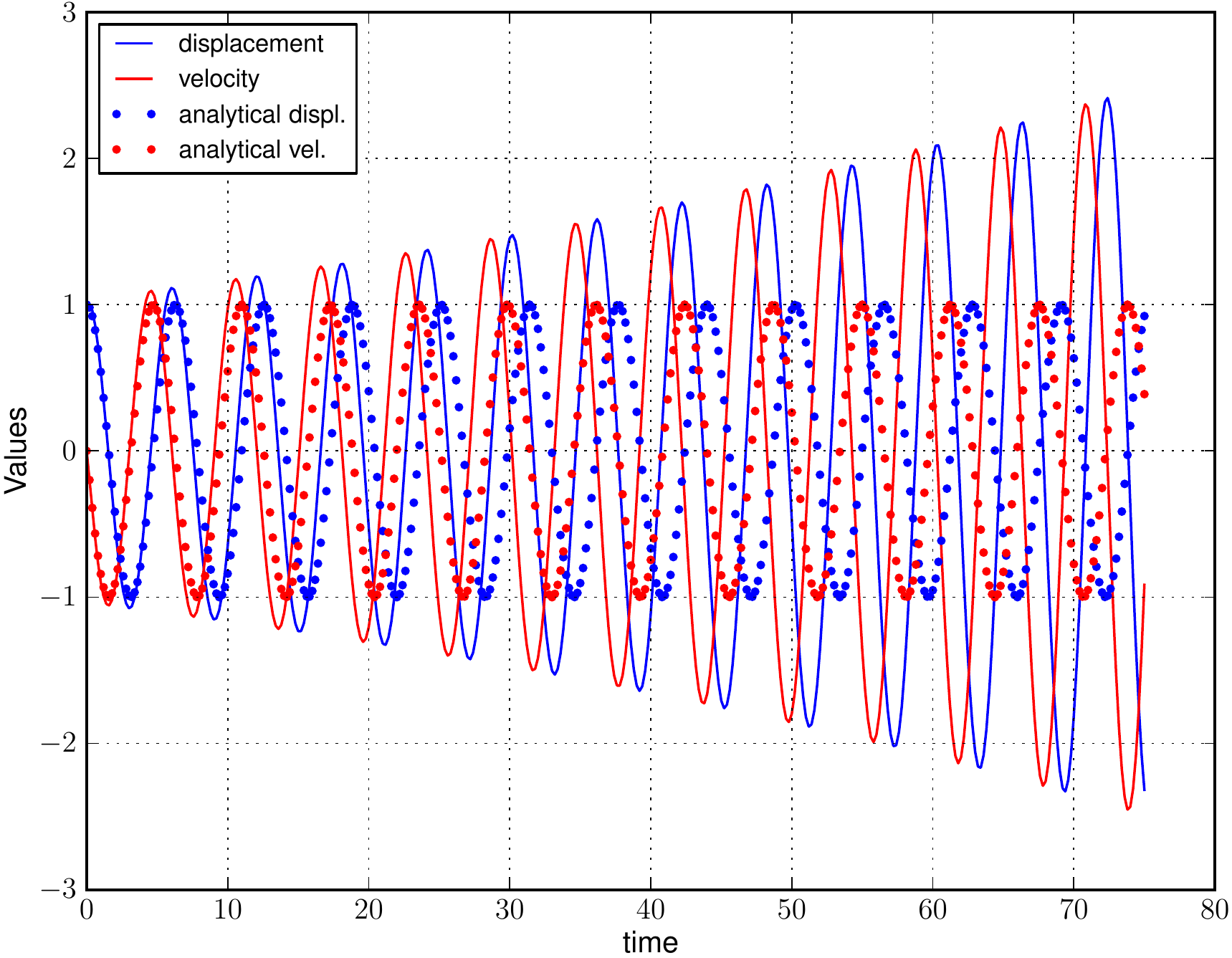}
\includegraphics[ width=0.45\textwidth]{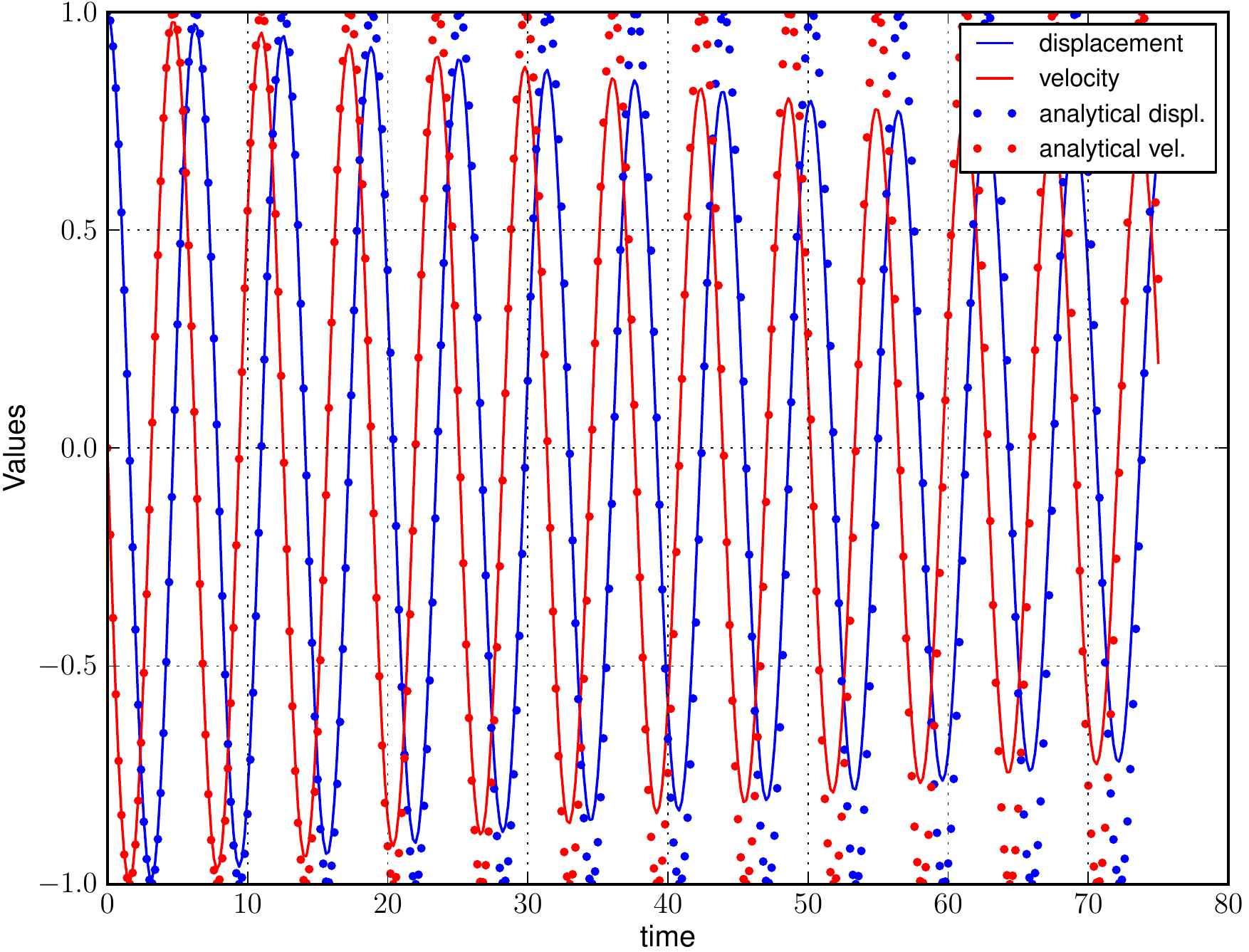}\\
\caption{\label{fig:smoothCosimStability} Simulation of the system \eqref{eq:springMass} in the cosimulation scheme with smoothing according to \eqref{smoothSwitching} and constant (left) and linear (right) extrapolation, upper plots without, lower with balance correction.   $H = 0.2$. Smoothing, as predicted, is harmful to stability, compare to fig.  \ref{fig:cosimStabilityReminder}. }
\end{figure}

\begin{figure}
\includegraphics[ width=0.45\textwidth]{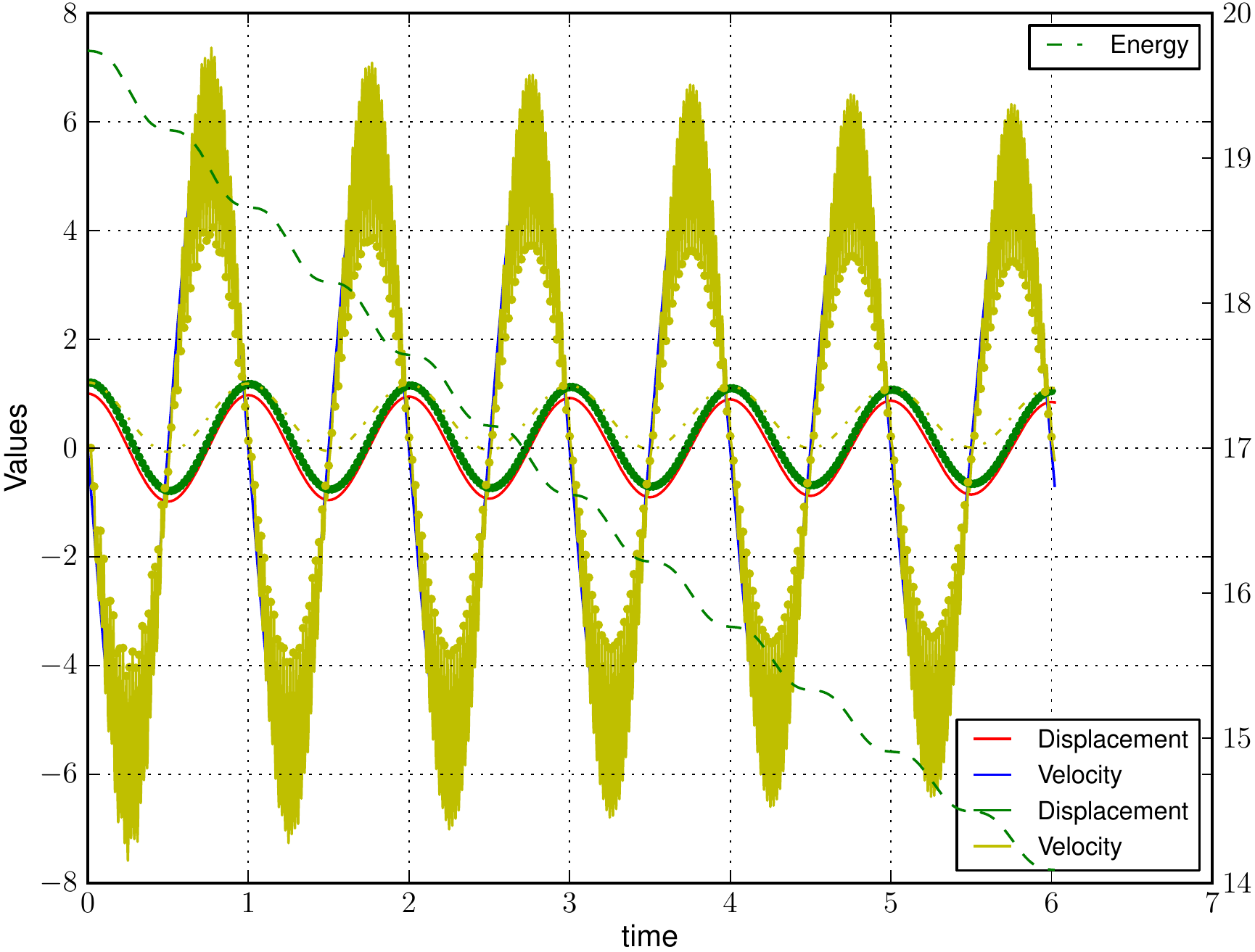}
\includegraphics[ width=0.45\textwidth]{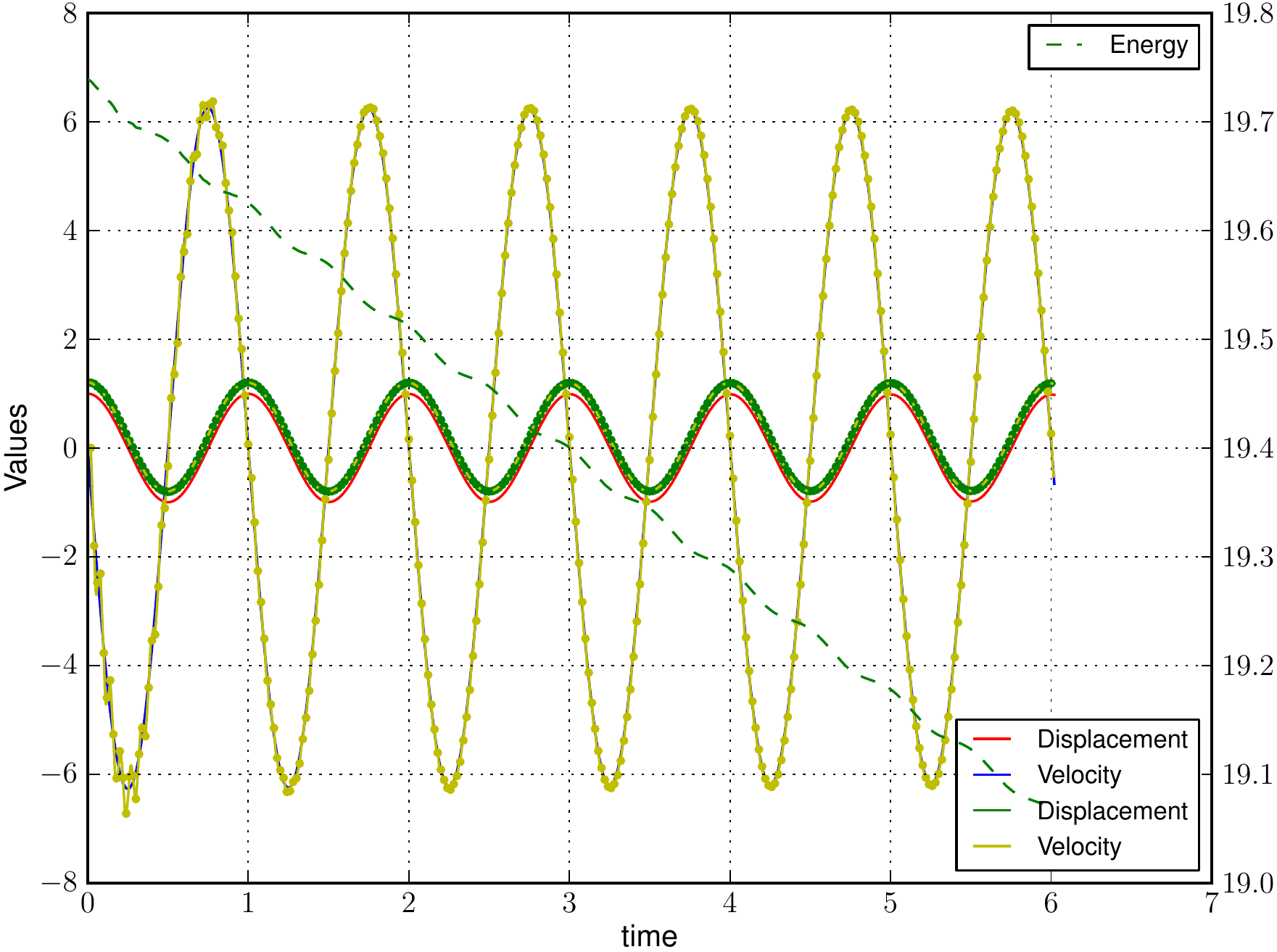}\\
\includegraphics[ width=0.45\textwidth]{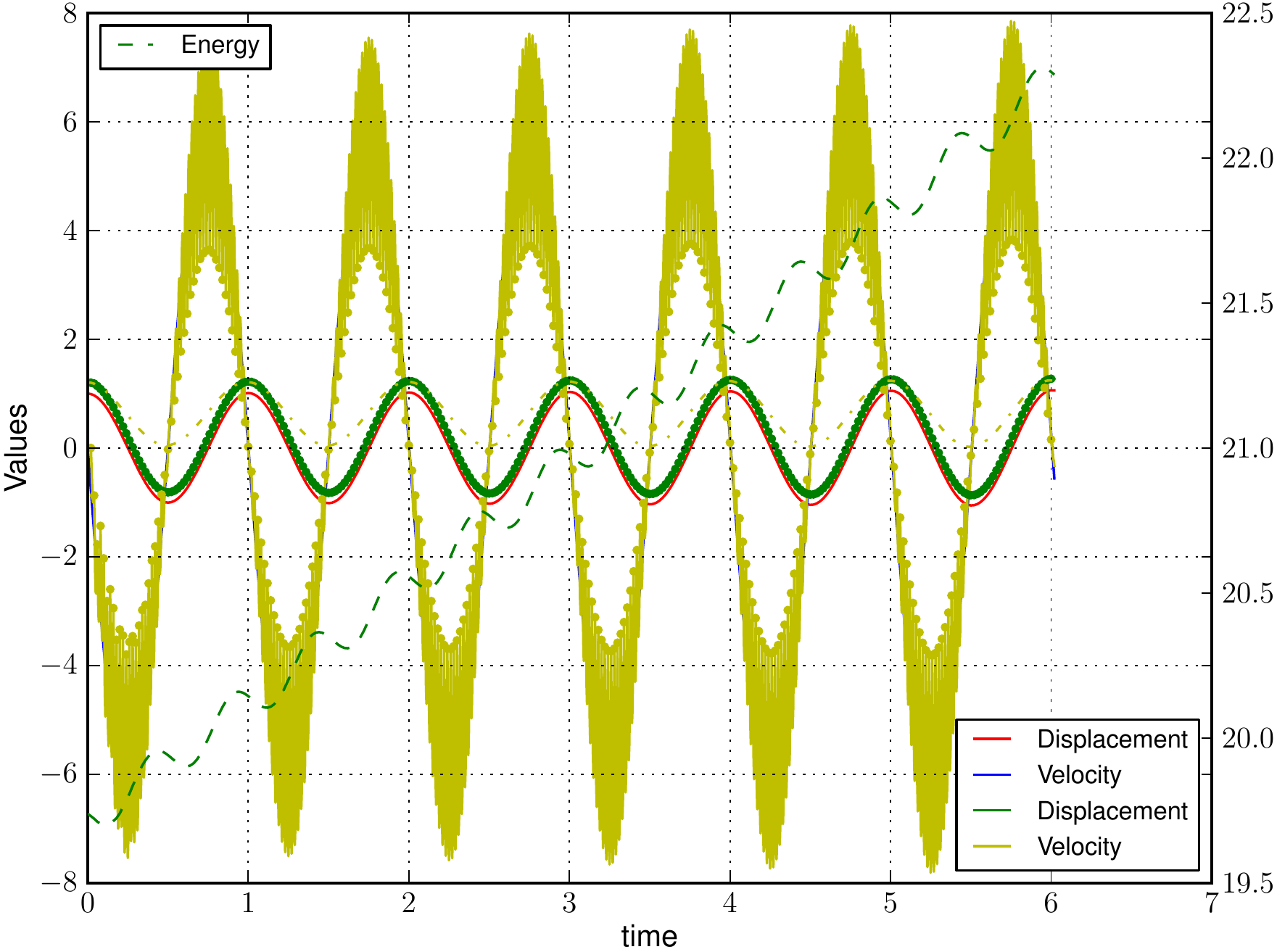}
\includegraphics[ width=0.45\textwidth]{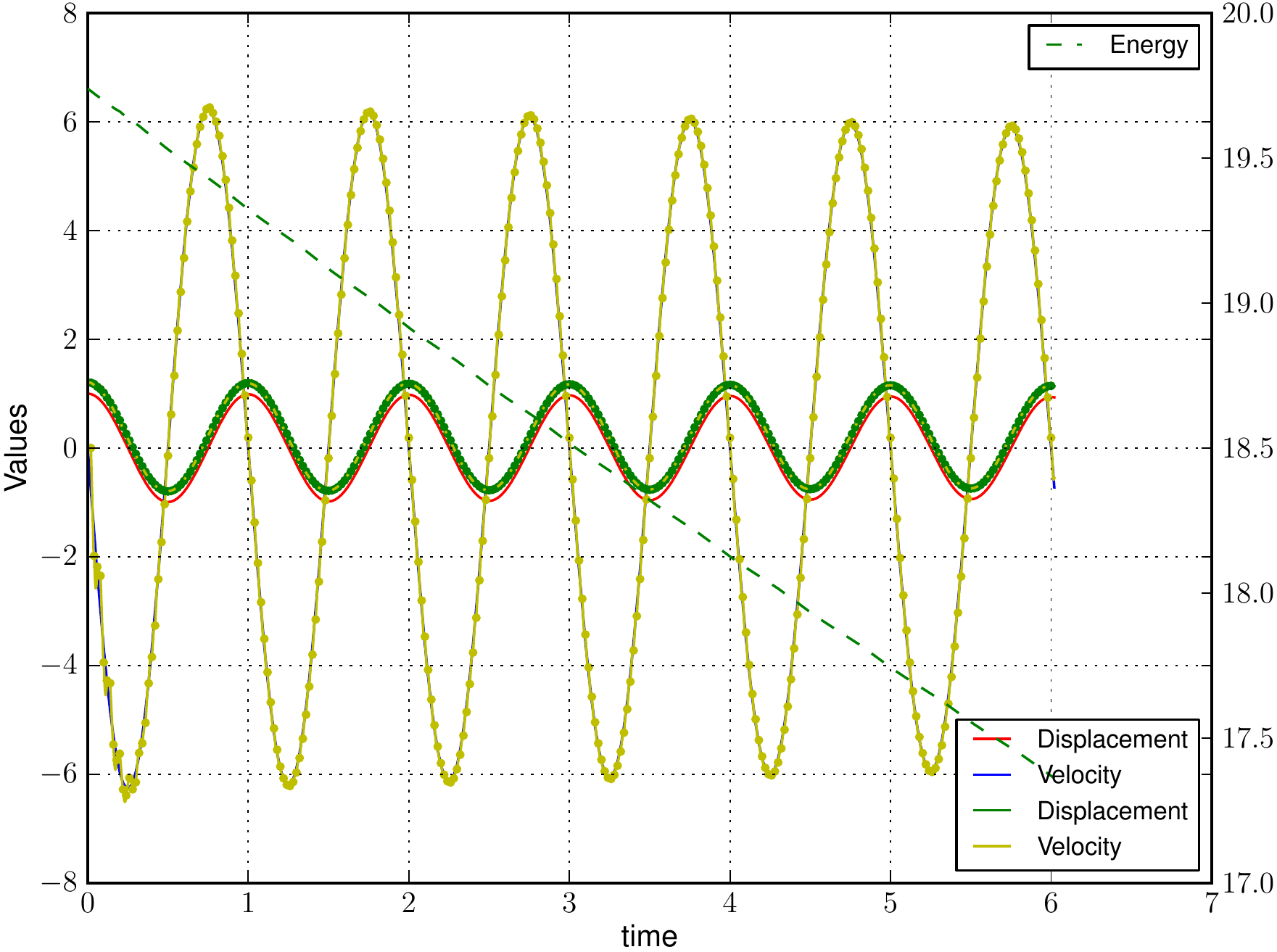}\\
\caption{\label{fig:cosimStabilityDoubleSpringMass}  Simulation of the system \eqref{eq:DoubleSpringMass} in the cosimulation scheme with constant (left) and linear (right) unsmoothed extrapolation, upper plots without, lower with balance correction.   $H = 0.02$. Balance correction does not improve energy conservation in the linear case.  }
\end{figure}
\begin{figure}
\includegraphics[ width=0.45\textwidth]{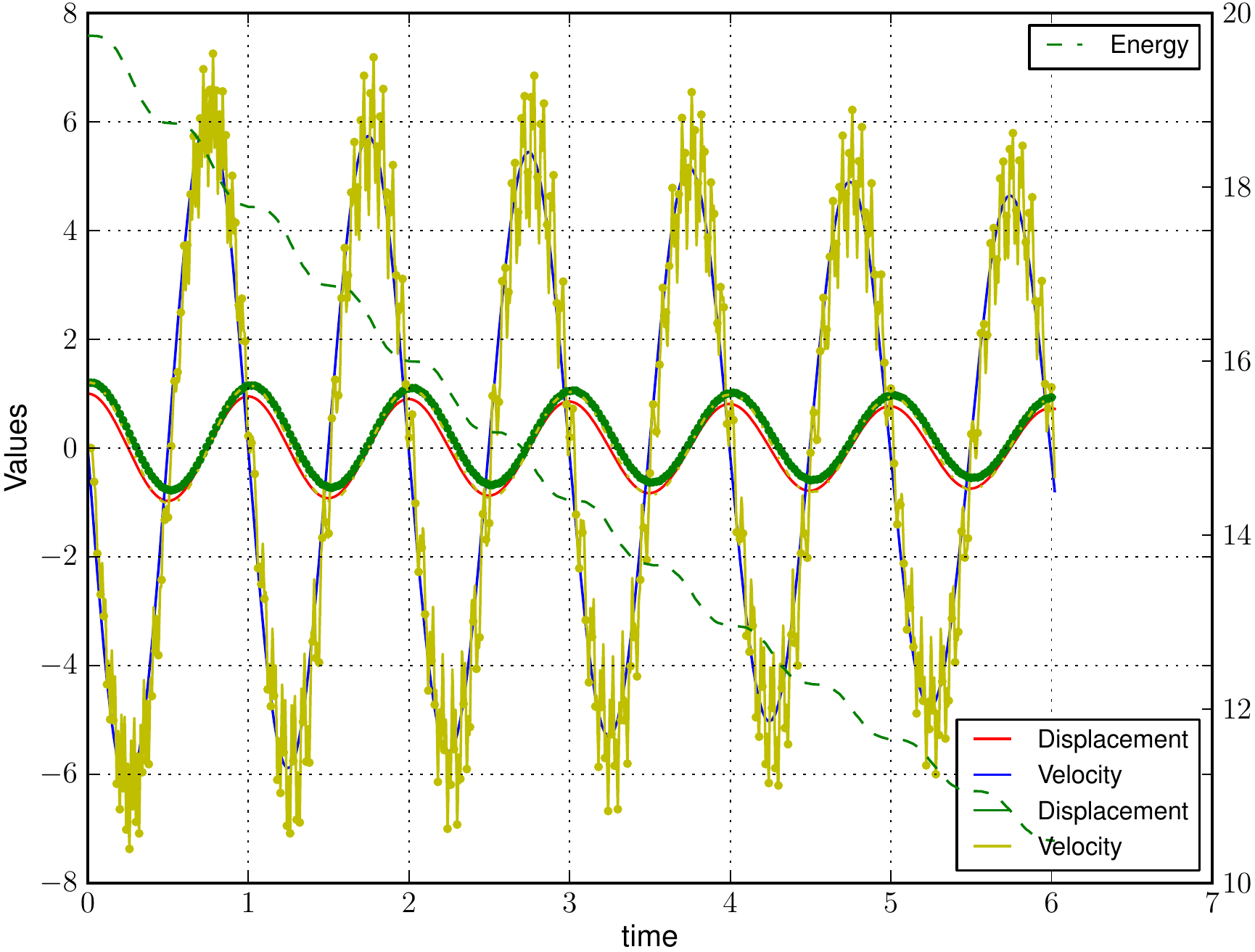}
\includegraphics[ width=0.45\textwidth]{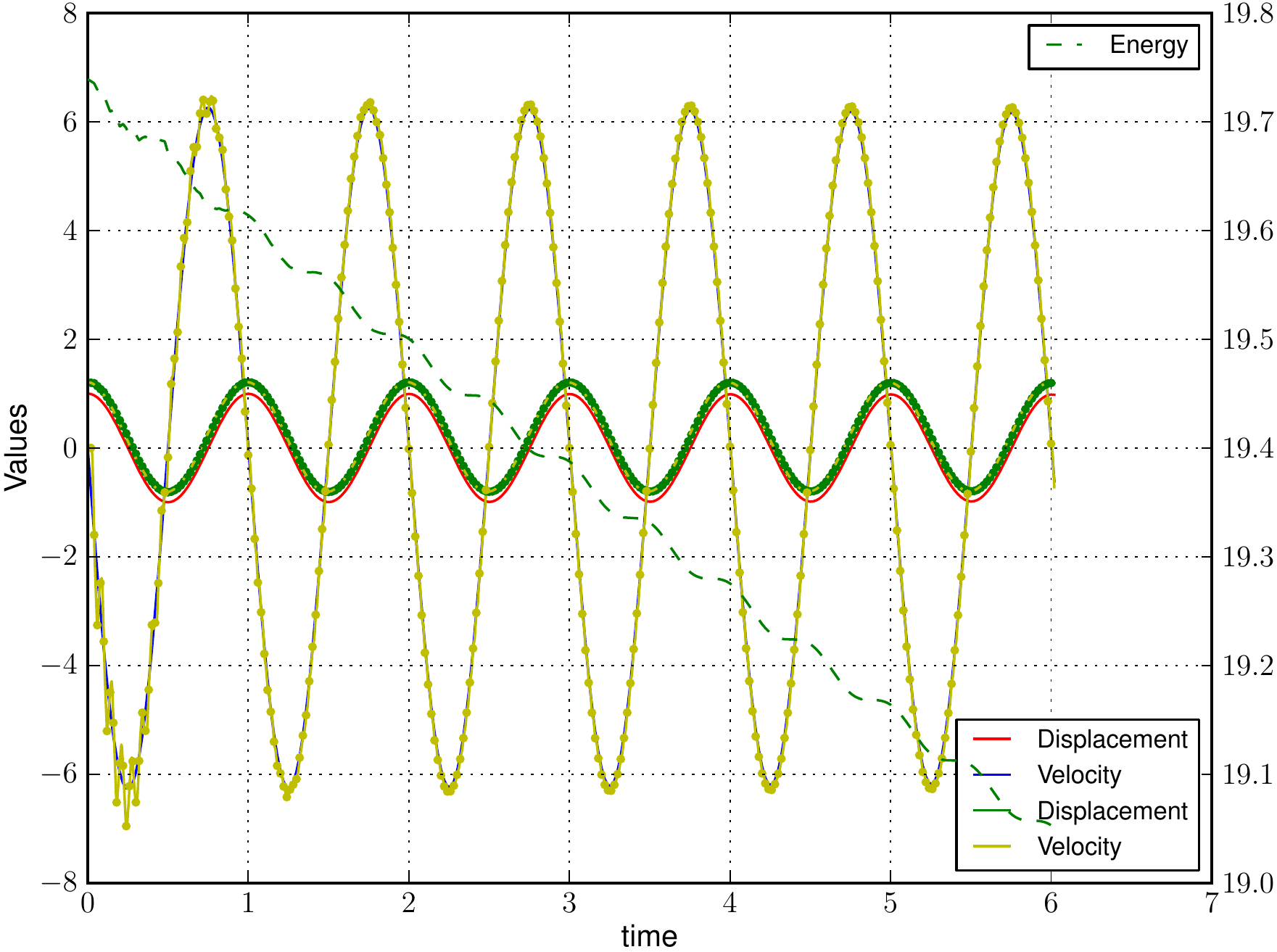}\\
\includegraphics[ width=0.45\textwidth]{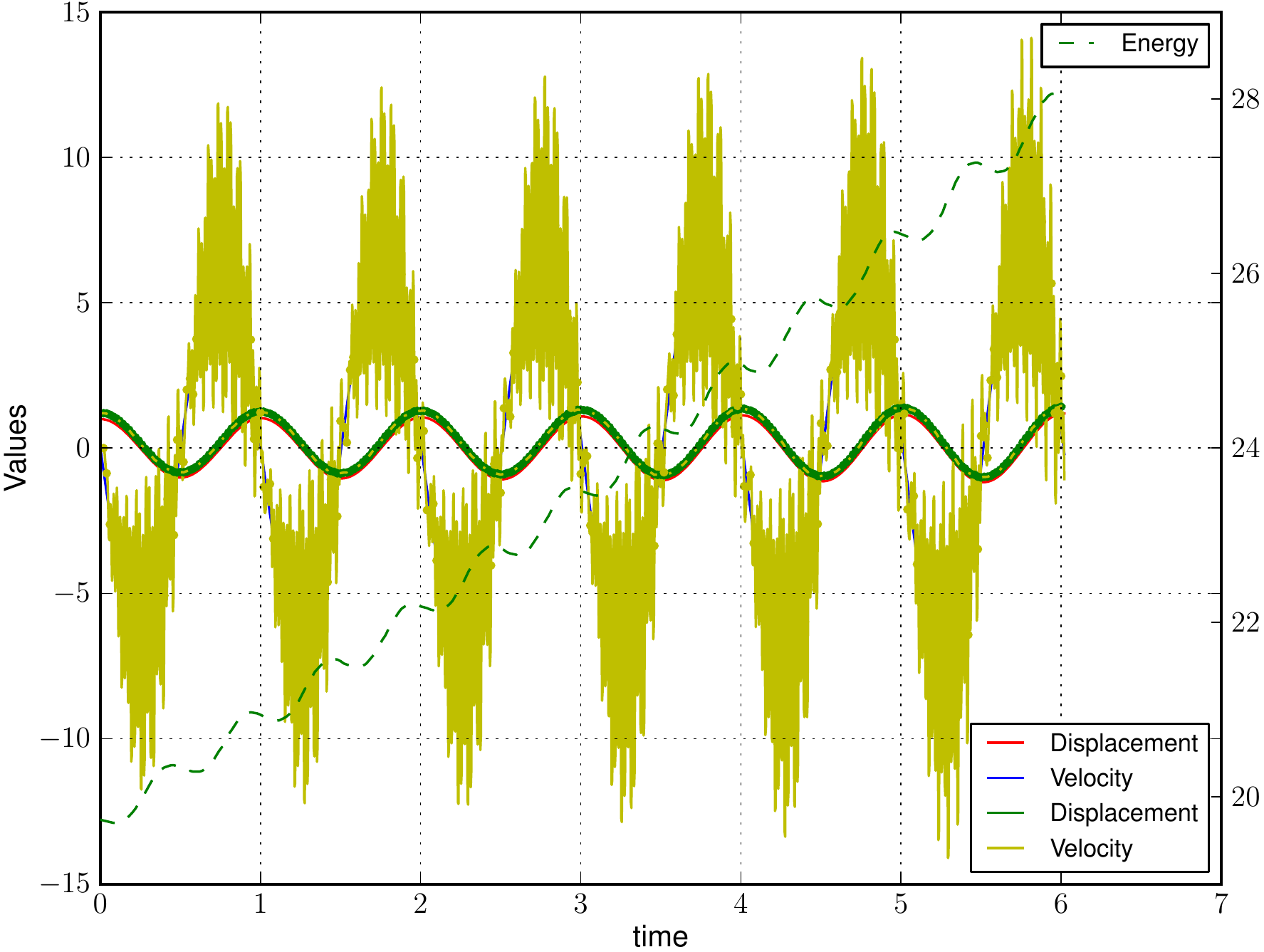}
\includegraphics[ width=0.45\textwidth]{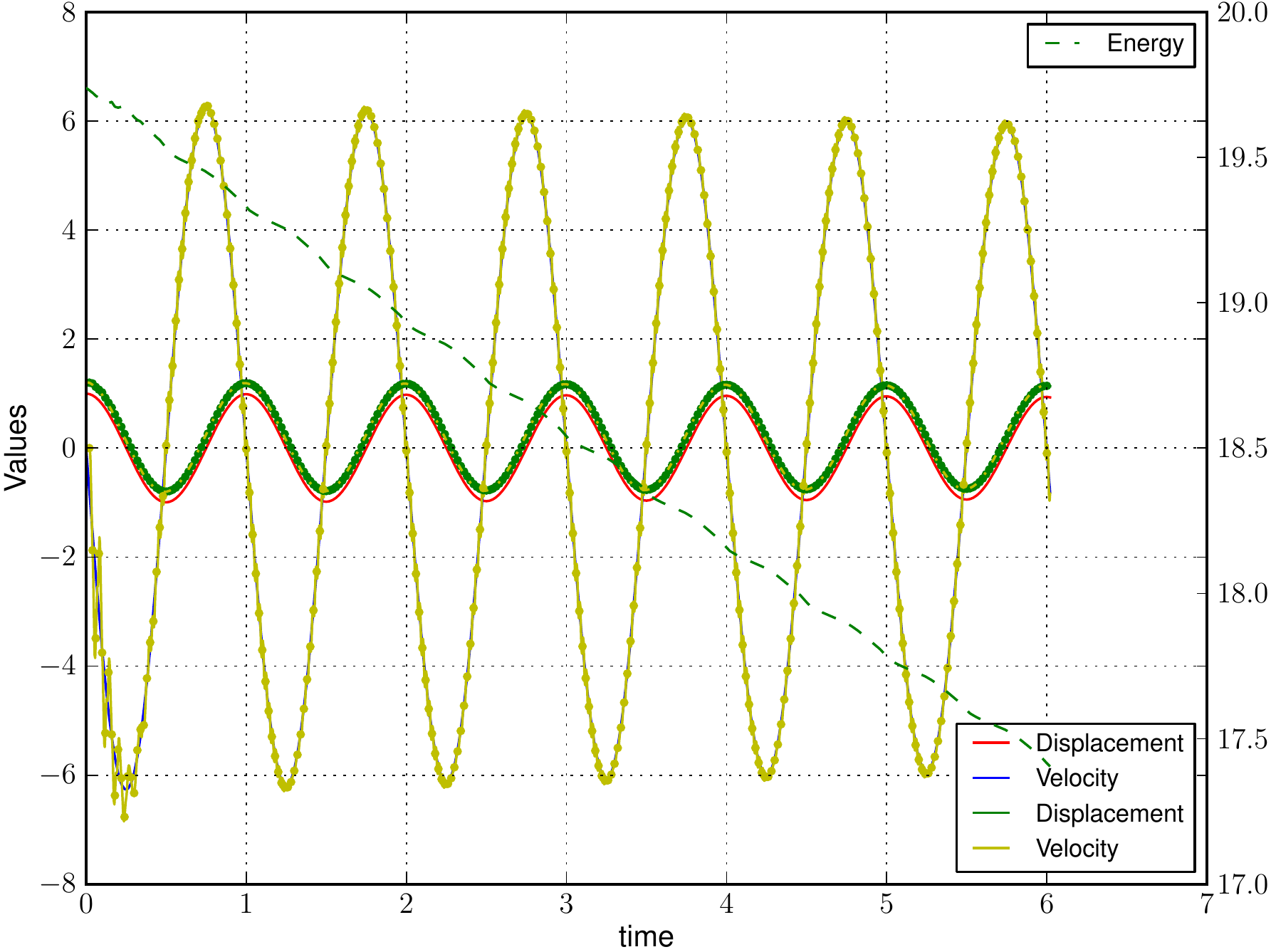}\\
\caption{\label{fig:smoothCosimStabilityDoubleSpringMass} Simulation of the system \eqref{eq:DoubleSpringMass} in the cosimulation scheme with smoothing according to \eqref{smoothSwitching} and constant (left) and linear (right) extrapolation, upper plots without, lower with balance correction.   $H = 0.02$. Smoothing, as in the problem \eqref{eq:springMass} case, negatively affects the stability.  }
\end{figure}
It is left to examine the effect of smooth switching onto the stability of the cosimulation. We have discussed that the exchange of factors of a conserved quantity leads to disbalances (see section \ref{factors}
) and further, that balance correction reliefs this effect but does not solve the problem, as corrections come with delay. If smooth switching is used, the error in input signals value is bigger than without. So the balance error of the extrapolation will also be bigger, as more quantity comes with delay, and so will be the error in the energy (the conserved quantity). This is the effect that shows up in picture \ref{fig:smoothCosimStabilityDoubleSpringMass}, and it is stated as a result of the examination that smoothing in general worsens stability.\\
To examine  the effect of smooth switching on stability in a case where smoothness matters, simulations with the system \eqref{eq:DoubleSpringMass} are considered.  The constant extrapolation here leads to a damped coupled system. Smoothing alone amplifies the damping (fig. \ref{fig:smoothCosimStabilityDoubleSpringMass}, top left), which as before is caused by the bigger error.\\
Balance correction turns this damping into an (undesired) amplification - fig. \ref{fig:smoothCosimStabilityDoubleSpringMass}, lower left. For the linear extrapolation case, the logics is analogous, but the different sign of the balance correction contibutions let the corrections have an additional damping effect. The effect is, as one expects, slightly increased by smooth switching if compared to nonsmooth switching, as fig. \ref{fig:smoothCosimStabilityDoubleSpringMass}, top, shows, same for the smooth balance corrected scheme, below.

\subsection{Splitting off balance error's early known parts}
\label{splitting}
At data exchange time $t_j$ the amount $  \int_{t_{j-1}}^{t_j}\vek u_2 dt  $  is passed on to $S_2$ along with the value $\vek u_2(t_j)$ for extrapolation. The  balance error is
$\Delta E_i^{j} = \int_{t_{j-1}}^{t_j}\vek u_i dt  - \int_{t_{j-1}}^{t_j} \overline{ \vek u_i} dt$. Mind that correction contributions alive in $\Delta_j$ concern past errors and are not part of $\overline{ \vek u_i}$.\\
The error that arises from smooth
 switching between $\operatorname{Ext}{ \vek{ u}}_{i}^{j-1}$ and $\operatorname{Ext}{ \vek{ u}}_{i}^{j}$
 obviously consists of  contributions which are already known at $t_{i-1}$.
Thus it is advantageous to start the correction of this error in interval $\left[t_{i-1},t_i\right]$, as soon as it is known. For this aim, we split the error into  an error between true $S_1$ output and  extrapolation and extrapolation and smoothly switched extrapolation.
\begin{equation}\label{splittingError}
\begin{split}
\Delta E_i^{j} = \int_{t_{j-1}}^{t_j}\vek u_i dt -
\int_{t_{j-1}}^{t_j} \overline{ \vek u_i} dt 
\\= \int_{t_{j-1}}^{t_j}\vek u_i dt -\int_{t_{j-1}}^{t_j}  \tilde{ \vek u}_i dt + \int_{t_{j-1}}^{t_j}  \tilde{ \vek u}_i dt - \int_{t_{j-1}}^{t_j} \overline{ \vek u_i} dt
\\=\underbrace{\int_{t_{j-1}}^{t_j}\vek u_i dt  -\int_{t_{j-1}}^{t_j}  \tilde{ \vek u}_i dt}_{\text{BC}} +\underbrace{\left( \int_{t_{j-1}}^{t_j}  \tilde{ \vek u}_i dt - \int_{t_{j-1}}^{t_j} \overline{  \vek u_i} dt\right)}_{\text{switching error}}
\end{split}
\end{equation}
and  the balance correction part of this is added 
beginning with $t_{j}$, while the switching error is added right away beginning at $t_{j-1}$ together with the BC contributions from  step $j-1$. 

\subsubsection{Numerical example}
\begin{figure}
\includegraphics[ width=0.5\textwidth]{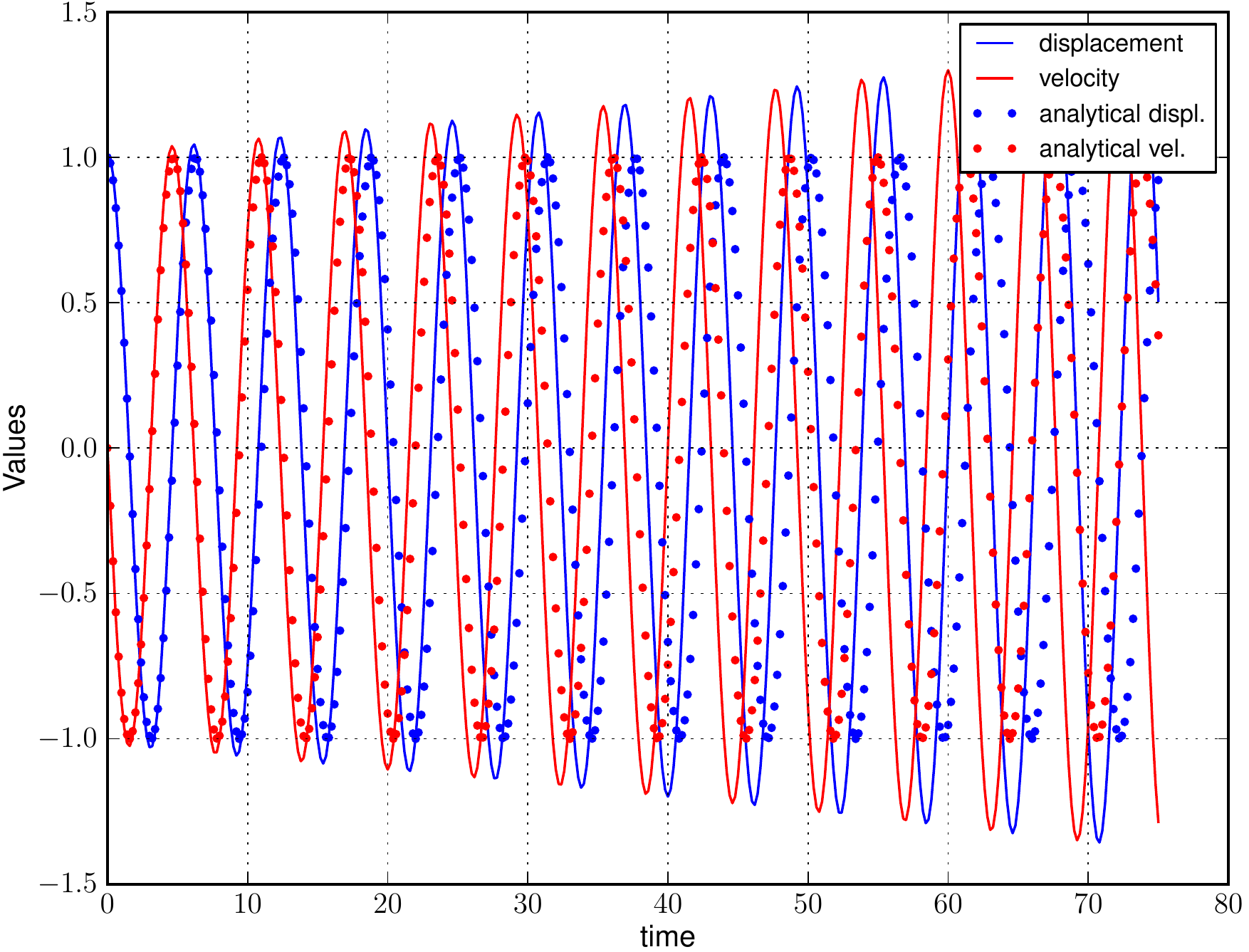}
\includegraphics[ width=0.5\textwidth]{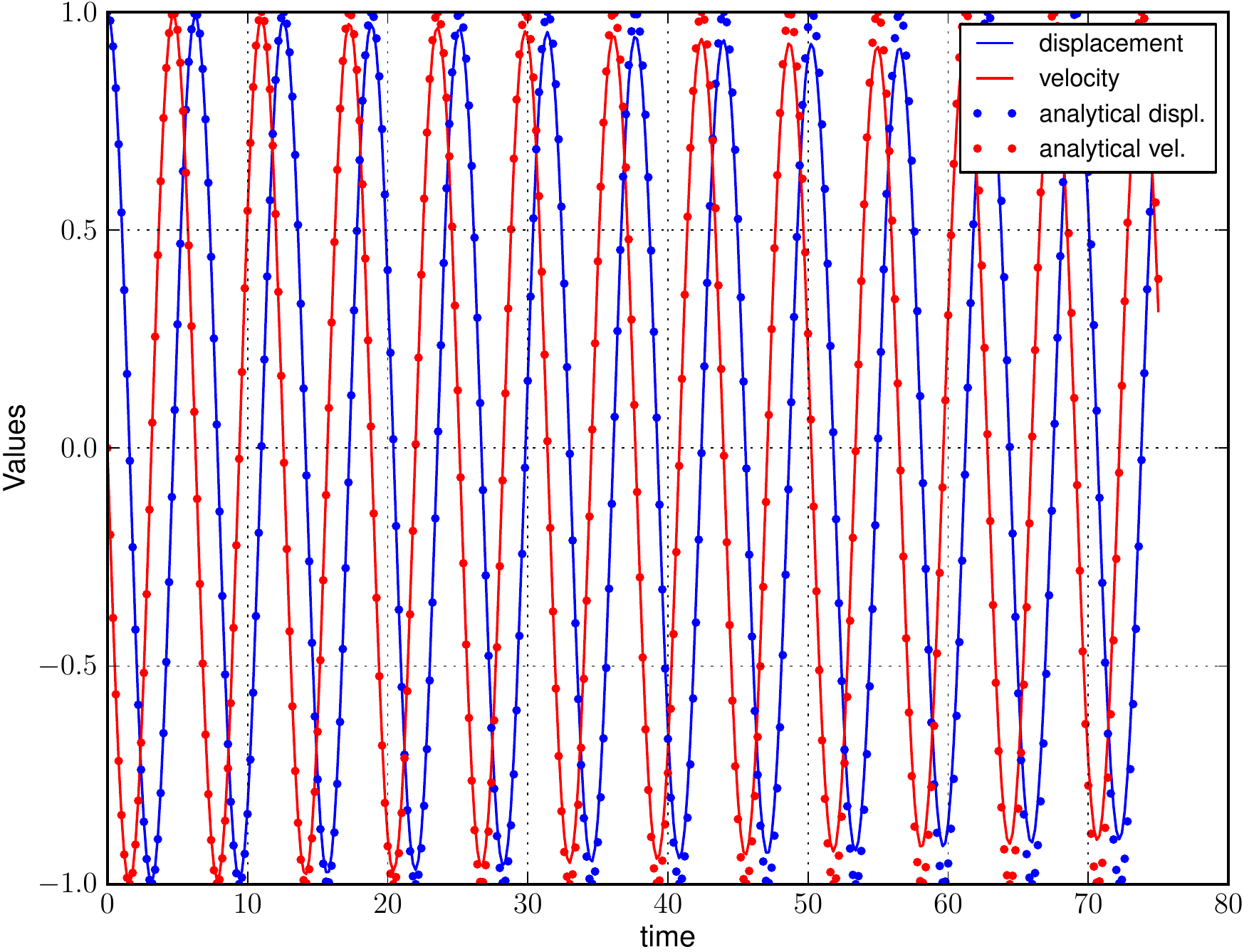}\\
\caption{\label{fig:smoothCosimStabilityQuickBCSpringMass}  Simulation of the system \eqref{eq:springMass} in the cosimulation scheme with constant (left) and linear (right) smooth extrapolation and BC with quick refeed of the switch error, $H = 0.2$. Stability properties close to, slightly worse than nonsmooth but balanced case (fig. \ref{fig:cosimStabilityReminder}), but better than in smooth and nonquickly balanced case(lower plots in \ref{fig:smoothCosimStability}).}
\end{figure}
\begin{figure}
\includegraphics[ width=0.5\textwidth]{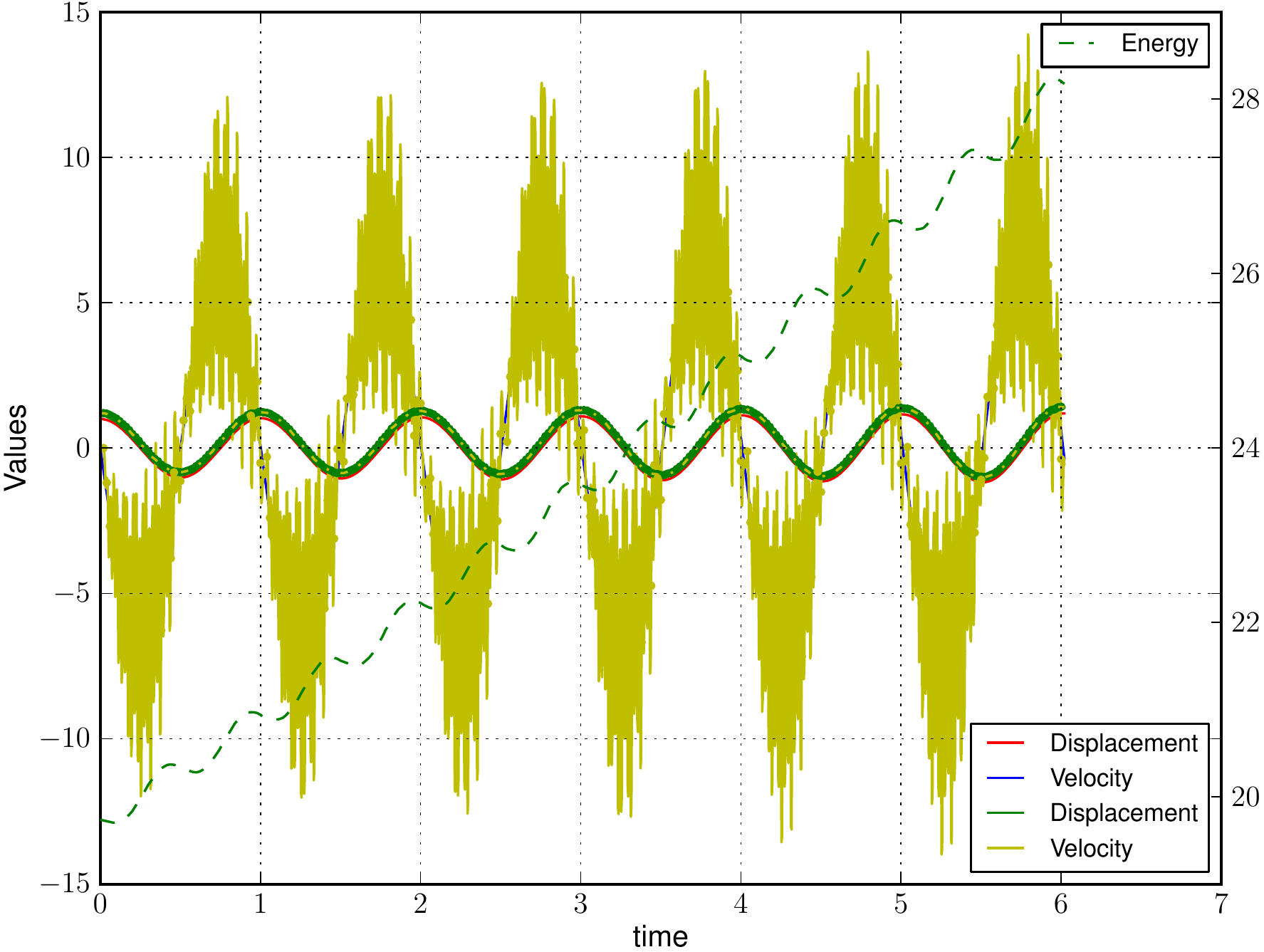}
\includegraphics[ width=0.5\textwidth]{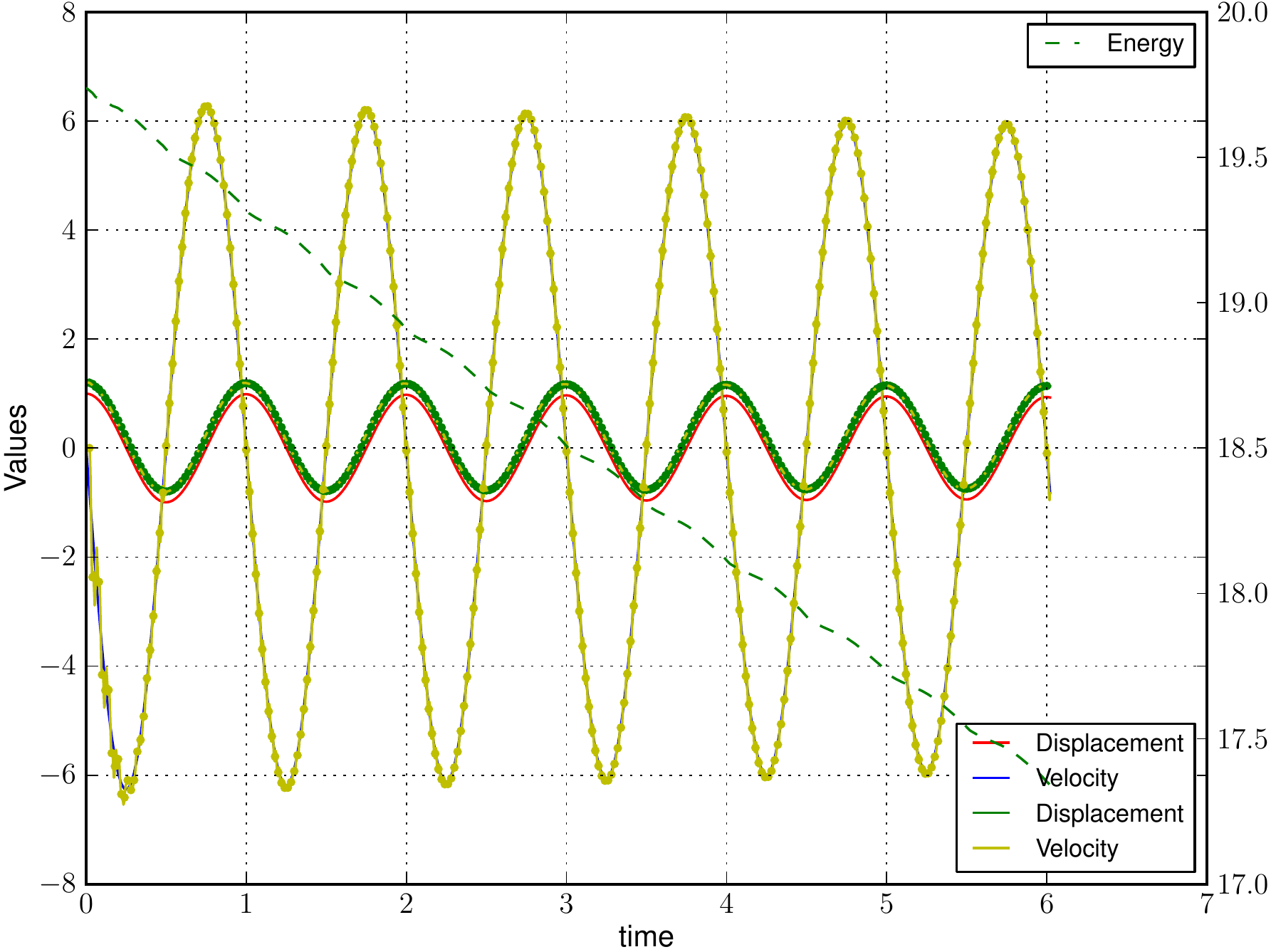}\\
\caption{\label{fig:smoothCosimStabilityQuickBCDoubleSpringMass}  Simulation of the system \eqref{eq:DoubleSpringMass} in the cosimulation scheme with constant (left) and linear (right) extrapolation $H = 0.02$ and quick refeed of the switch error. Stability properties similar to smooth, not quick refeed cas (fig. \ref{fig:smoothCosimStabilityDoubleSpringMass}), no real improvement in this setting.
    - Viell. hochfrequenter Energie"ubergang?
    - Kann man von einem steifen System reden, was an der Steifheitsgrenze aufgetrennt wird?
    - mal Eigenwertanalyse machen.
.}
\end{figure}
If one compares figures 
\ref{fig:smoothCosimStabilityQuickBCSpringMass} and
\ref{fig:smoothCosimStabilityQuickBCDoubleSpringMass} to their nonquick correspondants \ref{fig:smoothCosimStability} and 
 \ref{fig:smoothCosimStabilityDoubleSpringMass}, lower rows, one finds a positive, conservation-enhancing effect for the spring-mass example, as one would expect, but hardly a difference in the conservation behavior of the double-spring-mass example.
\todo{Look at spectral properties!}

\subsection{Smoothing and Balance Correction together with special interest on oscillation reduction}
\begin{figure}[h!bt]
\psset{xunit=3cm,yunit= 2cm,algebraic=true}
\includegraphics[ width=0.7\textwidth, trim={0 0cm  0 2.3cm},clip]{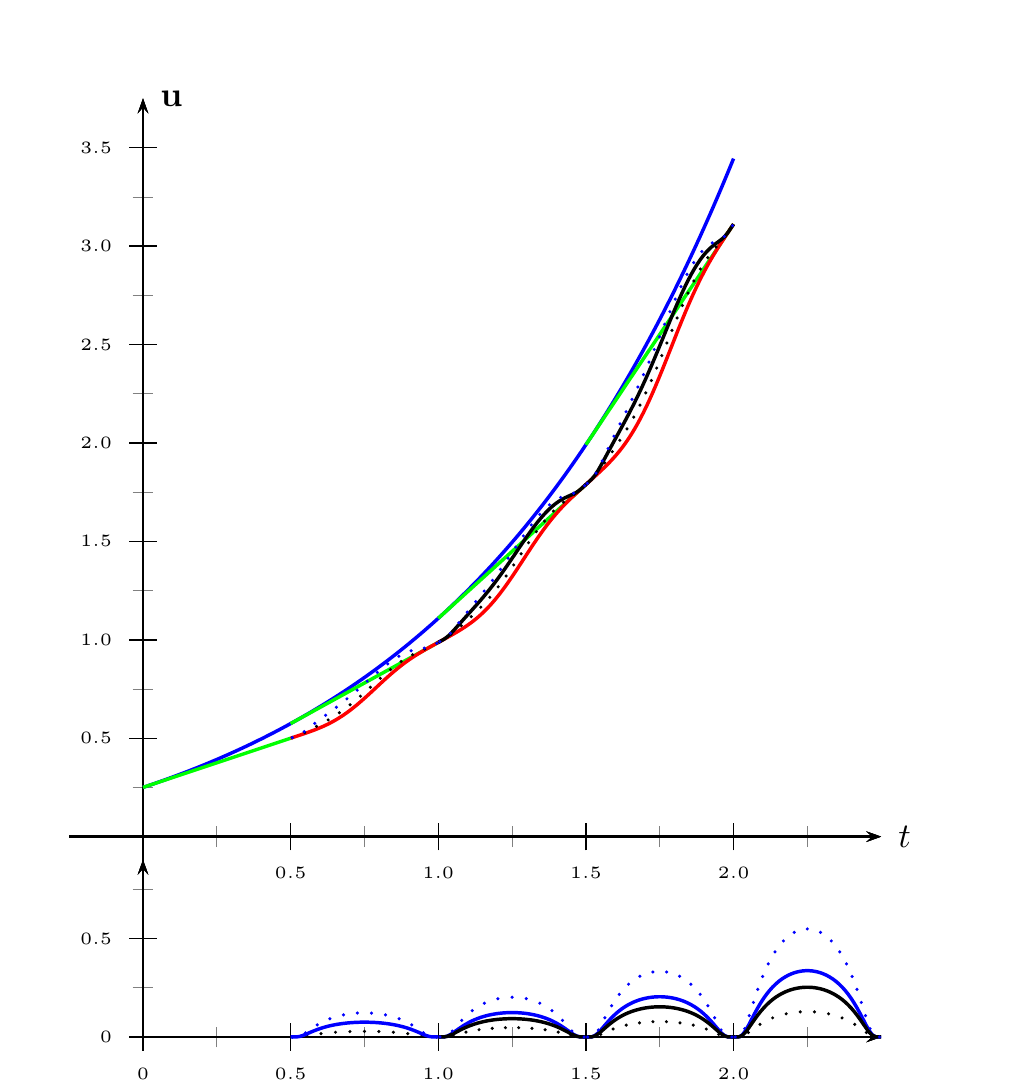}

\caption{\label{EarlyBalanceCorrection} Input signal and ways to recontribute errors by the receiving system: Early refeed of known parts: Contributions from smooth switching are recontributed earlier.  As before:  Green: linearly interpolated, red: smoothed with s-shaped $\psi$, black: error from previous time step added to smooth signal. In  the lower plot, the contributions themselves are plotted: Black: balance correction from previous timestep, blue: balance error due to smooth (slow) switching in actual timestep, black dotted: sum of both. It is obvious that switching error cannot be neglected, and that derivatives of the input signal oscillate. }
\end{figure}

\label{sec-3}
It is the aim of this work to provide a calculation of $\overline{ \vek u_i}$ and balance correction contibutions to $\tilde{ \vek u}_i$
resp. $\overline{ { \vek u}_i}$ such that $\overline{ { \vek u}_i}$
\emph{and} balance correction are smooth not only in the sense of mere
existence of derivatives, but the number of additional changes in
curvature, this is the sign of the second time derivative, is
reasonably related to number of information exchange times.  The
restriction to correct all known balance errors in the next time step,
as it is done in \cite{Scharff12}, is dropped: \emph{The balance correction contribution of $[t_j, t_{j+1})$ is added during the next $n>1$ timesteps}, thus the intervals in which the balance correction contributions are refed are now overlapping.
Adding correction
contributions immediately in the next timestep means that the hat function
$\phi$ is such that $\text{supp} \phi \subset \Delta_i t$,  which if smooth has at least one extremum
and two turning points.
 These then possibly cause local extrema and
turning points on the corrected flow input signal, making it having
unnecessarily high derivatives and look bumpy, see figure \ref{ClassicalBalanceCorrection}.\\
The unphysical derivative contributions may cause dynamics of exchange-rate frequency in the receiving subsystem and those subsequent in causality, and hide the observations one wanted to make, or worse, even influence system behavior. For example, unphysical noise may occur. All this could be seen in section \ref{sec:MovingGround}. \\
Doing cosimulation with smooth extrapolation of inputs, one has to do everything to minimize errors in derivatives of signals. 

\begin{figure}[h!bt]
\includegraphics[ width=0.7\textwidth, trim={0 0cm  0 2.3cm},clip]{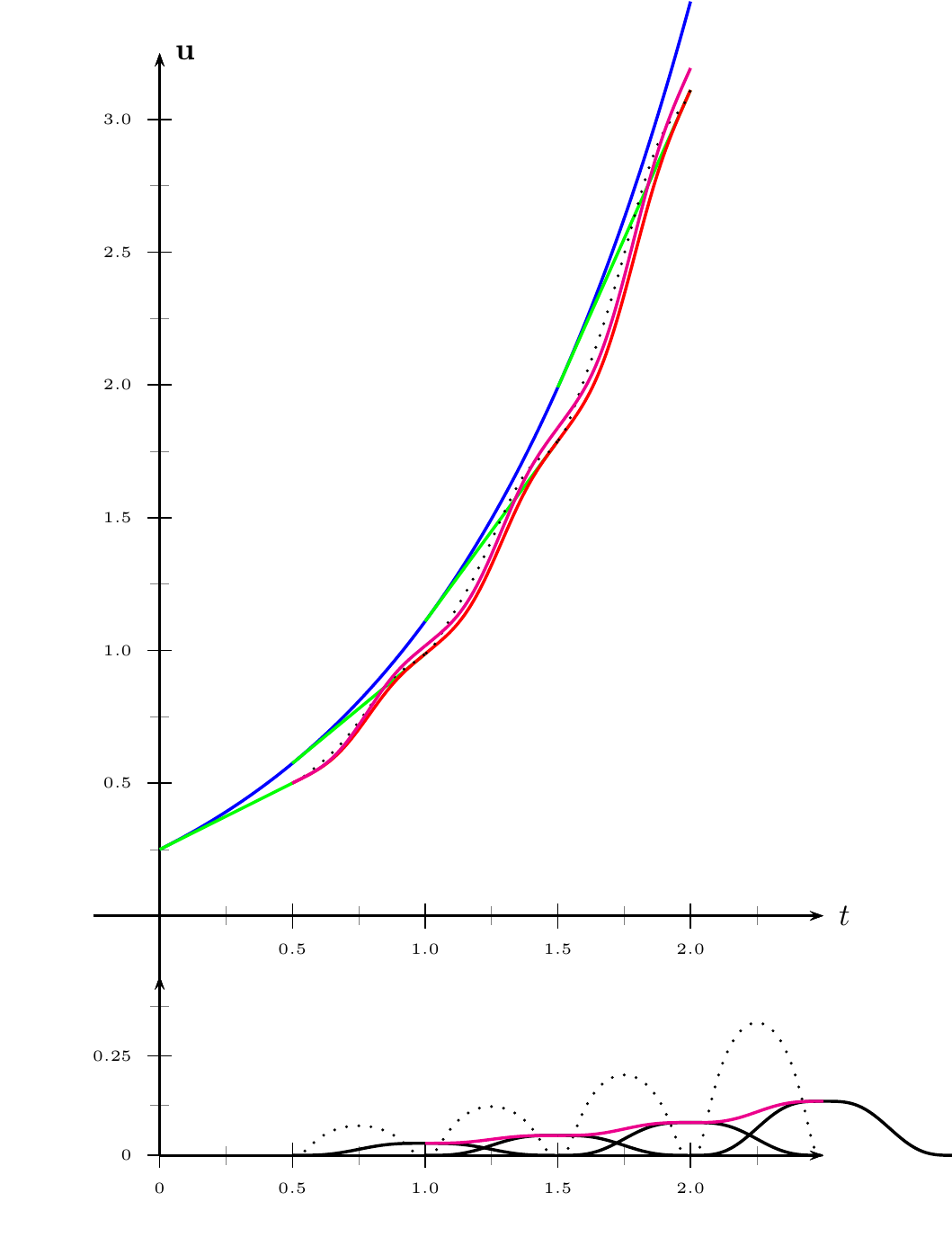}

\caption{\label{SSmoothBalanceCorrection} Early refeed of known parts, use of one-turningpoint-per-information-policy. Contributions from smooth switching are recontributed earlier.  As before:  green: linearly interpolated, red: smoothed with s-shaped $\psi$, black dotted: error from previous time step 
added to smooth signal. 
The lower plot (rescaled!) shows the correction contributions themselves: Dotted are contributions including actual switch error, using an one-interval hat function, black the same using the two-interval test function. In magenta the sum of the two-interval-test-function contributions: It shows no unnecessary  extrema.
This property is inherited to the extrapolated signal: While the signal with the one-interval hat functions (black dotted) returns to the uncorrected signal at end of time step, the magenta signal has no such bends. }
\end{figure}

Thus motivated, a more sensible $\phi$ and support is chosen. There are mainly two
ways to realize the new objective to \emph{reasonably connect the number
of changes of curvature to the number of new information}:
\begin{enumerate}
\item Let the hat function $\phi$   have  at most one convex interval and at most one concave interval per time interval, which means per change of
approximation of ${ \vek u}_i$. This means
\begin{itemize}
\item using an S-shaped function $\psi$ as in \cite{KosselDiss} for switching
between extrapolation polynomials $\tilde{ \vek u}_i^{j-1}$ and $\tilde{ \vek u}_i^{j}$:
\begin{equation}
\overline{ { \vek u}_i}^{j}= (1-\psi) \tilde{ \vek u}_i^{j-1} + \psi  \tilde{ \vek u}_i^{j}.
\end{equation}
\item and a function $\phi$ to add balance corrections as
$\Delta E_{i}^j \phi$ with the property that it is S-shaped \emph{in each
time interval} too. Straightforwardly this is done by using a hat
function consisting of a S-shaped rising branch and a mirrored falling
branch. The support of $\phi$ thus stretches over two time intervals. Such a hat function is constructed in \ref{sec:IntOfTwoHats}
\end{itemize}
\item  Let the hat function for the balance correction contribution $\phi$  have only one sign in curvature per time interval, which means that the support of it stretches over 4 time intervals, being convex in the first, concave in the second and third, having its maximum between them, and being convex again in the forth.
This method has the advantage of damped derivatives of the corrections and because usually four $\phi_i$ are  overlapping inside the time domain, they may compensate each other, bringing ${ \vek u}_i^{j-1}$ and its approximation closer together.
Stretching $\phi$ over even more time intervals can be thought of. \\
Switching between extrapolation polynomials is done as above, so in this case changes of curvature arising from this switching stay the same. However, especially the choice of a four-time-interval support of $\phi$ is charming because the change of curvature of the correction contributions  once per time interval can be seen as one reaction of the correction procedure  per information exchange.
 \end{enumerate}
The suggested method has a drawback in terms of time delay compared to classical balance correction as in \ref{sec-2-1}: In the former approach, all $\Delta E_{i}^j$ is recontributed in $[t_j, t_{j+1})$. In   the suggested way to 
distribute correction contributions
 over more  time intervals - let us  choose a support of two time intervals for the hat functions - the switching part  of $\Delta E_{i}^j$ as in equation \eqref{splittingError} is already recontributed in  $[t_{j-1}, t_{i+1}$,
 thus starting earlier and ending at the same time. But the standard BC part  arrives in $[t_j, t_{j+2})$. 
 
 As the (early) switching part amounts to roughly half of the (late) typical bilance error, 
  a time delay remains.\\
  If support of hat functions is chosen as four time intervals, the delay and its effect grows.
\begin{figure}
\includegraphics[ width=0.5\textwidth]{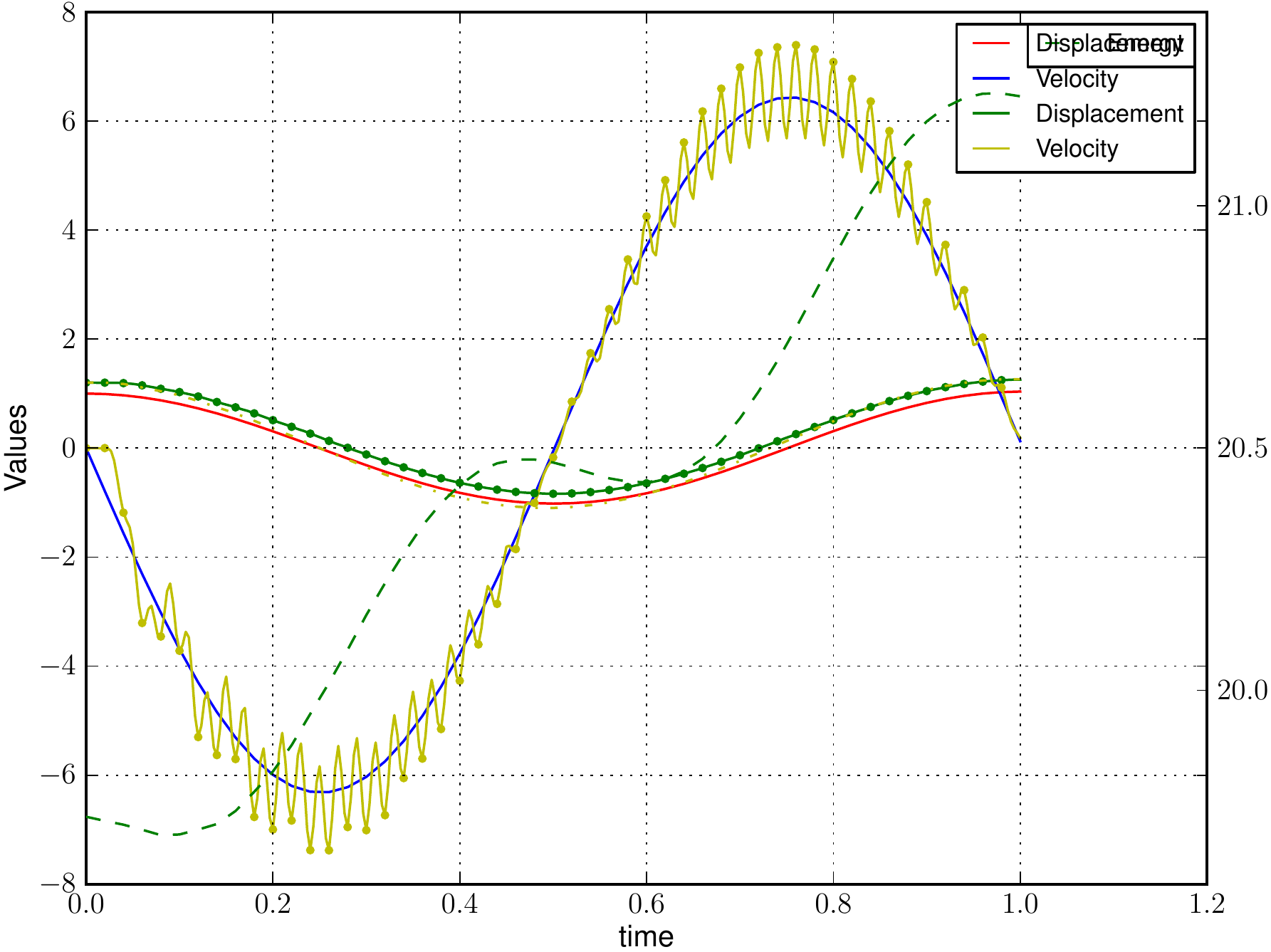}
\includegraphics[ width=0.5\textwidth]{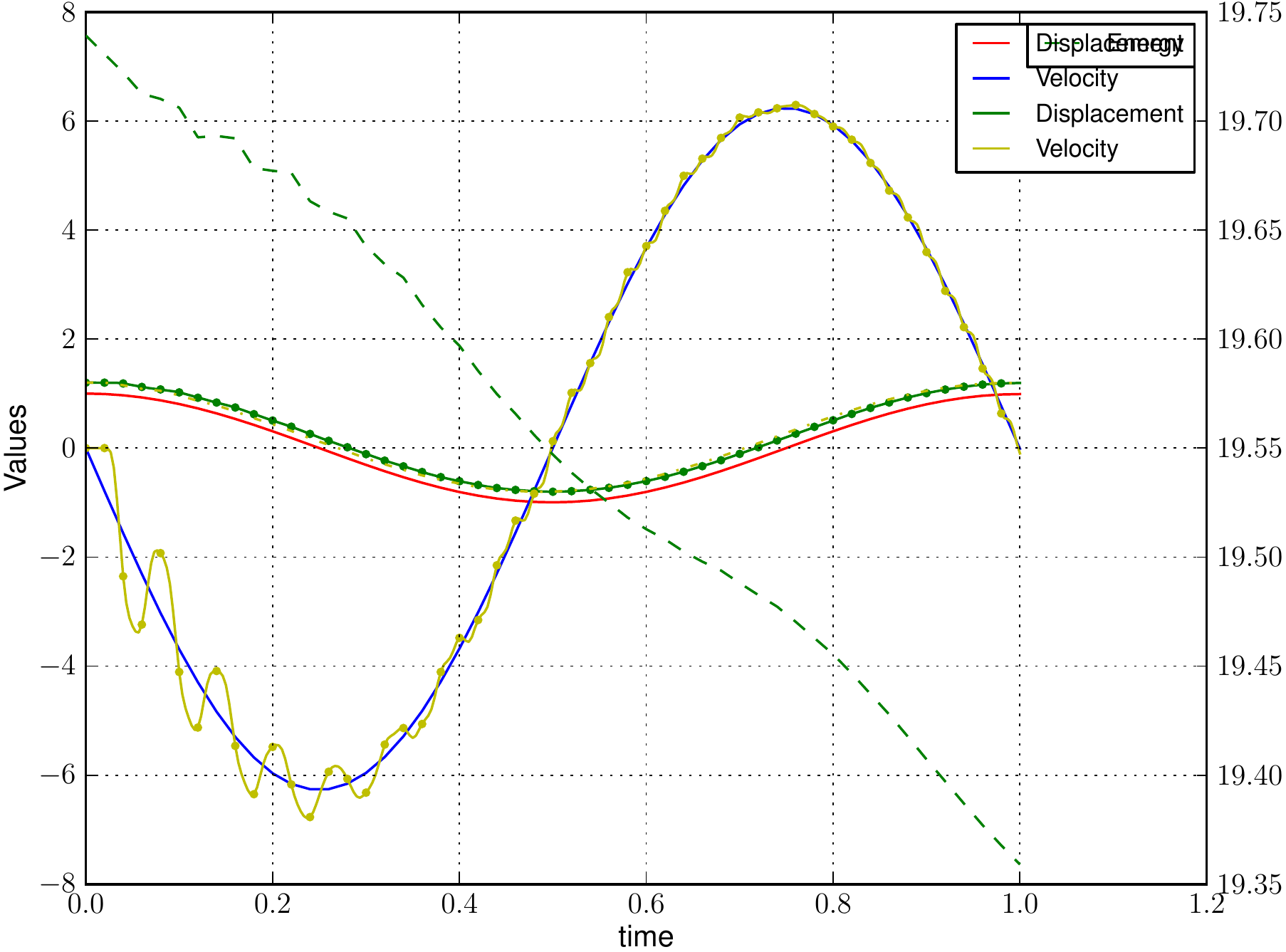}\\
\caption{\label{fig:sSmoothCosimQuickBCDoubleSpringMass} Simulation of the double spring mass system \eqref{eq:DoubleSpringMass} in the cosimulation scheme with constant (left) and linear (right) extrapolation, $H = 0.02$. Exchange induced oscillations are considerably reduced,  compare to fig.  \ref{fig:MovingGroundBC} and \ref{fig:MovingGroundSmoothBC}.}.
\end{figure}
\subsubsection{Numerical example}
\begin{figure}
\includegraphics[ width=0.5\textwidth]{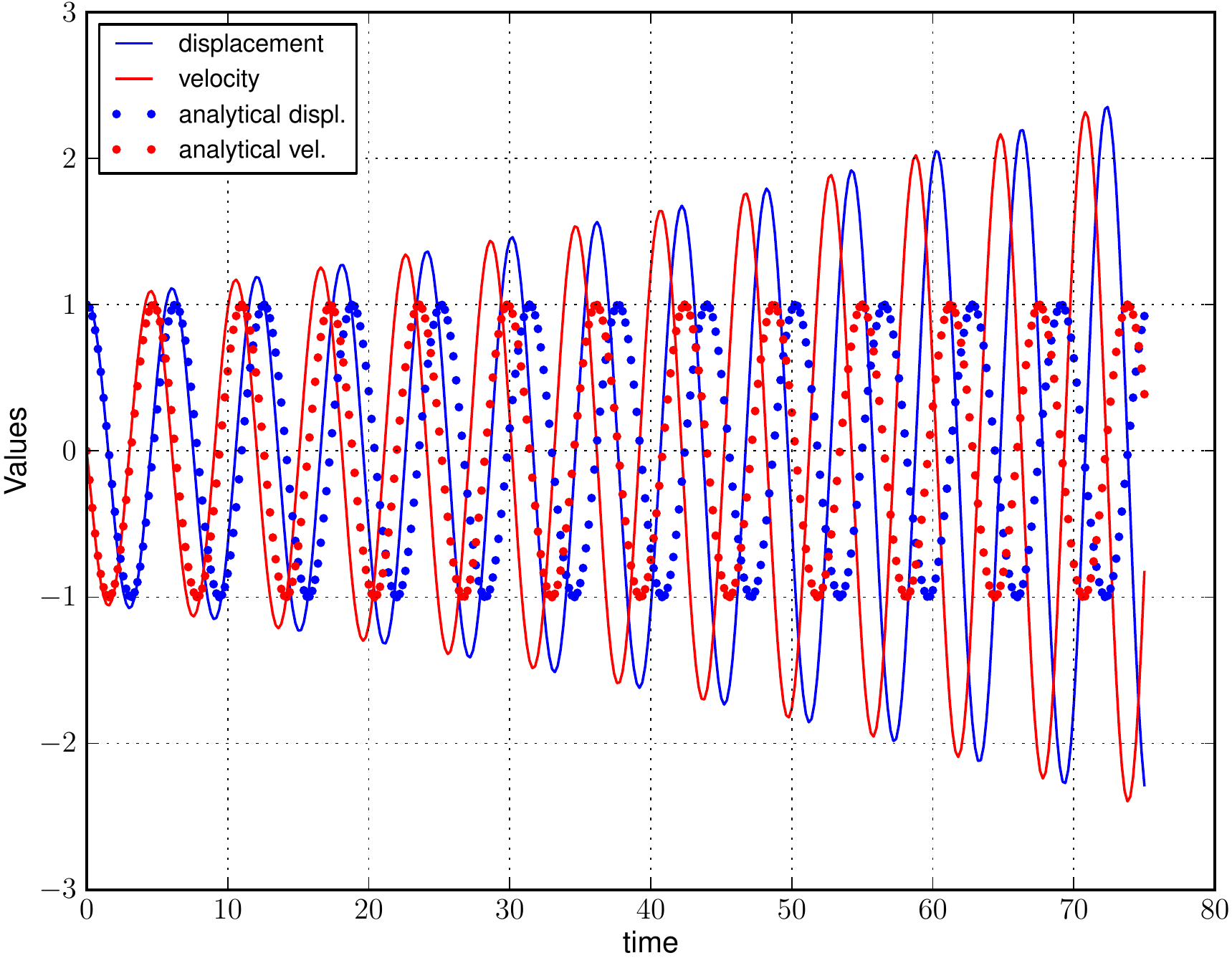}
\includegraphics[ width=0.5\textwidth]{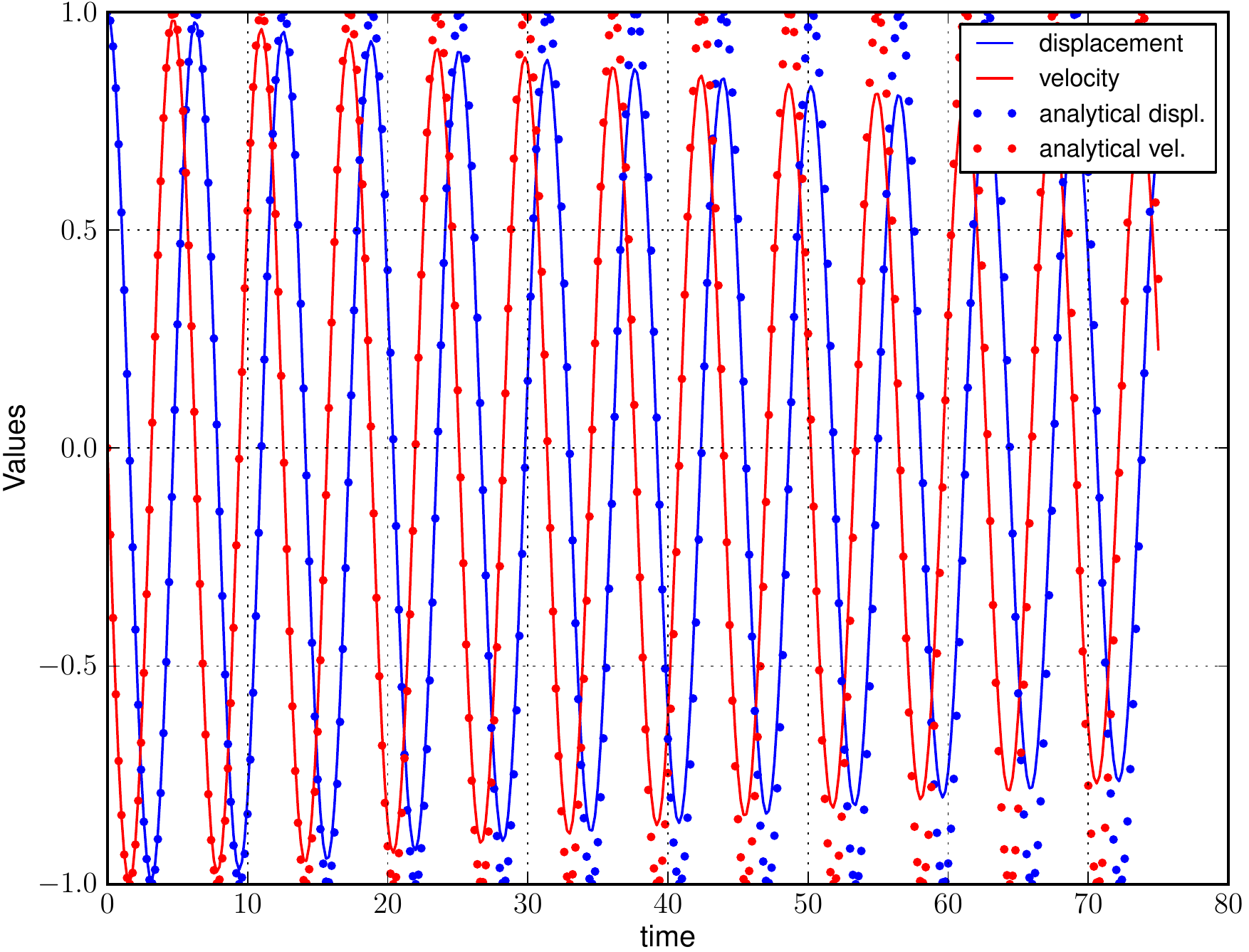}\\
\caption{\label{fig:sSmoothCosimStabilityQuickBCSpringMass}  Simulation of the system \eqref{eq:springMass} in the cosimulation scheme with constant (left) and linear (right) smooth extrapolation and BC with quick refeed of the switch error and refeed distributed over two steps, $H = 0.2$. Stability properties suffer from the time delay in refeed if compared to fig. \ref{fig:smoothCosimStabilityQuickBCSpringMass}, but are quite similar to smooth calculations without quick refeed (lower row in fig. \ref{fig:smoothCosimStability}).}
\end{figure}
\begin{figure}
\includegraphics[ width=0.5\textwidth]{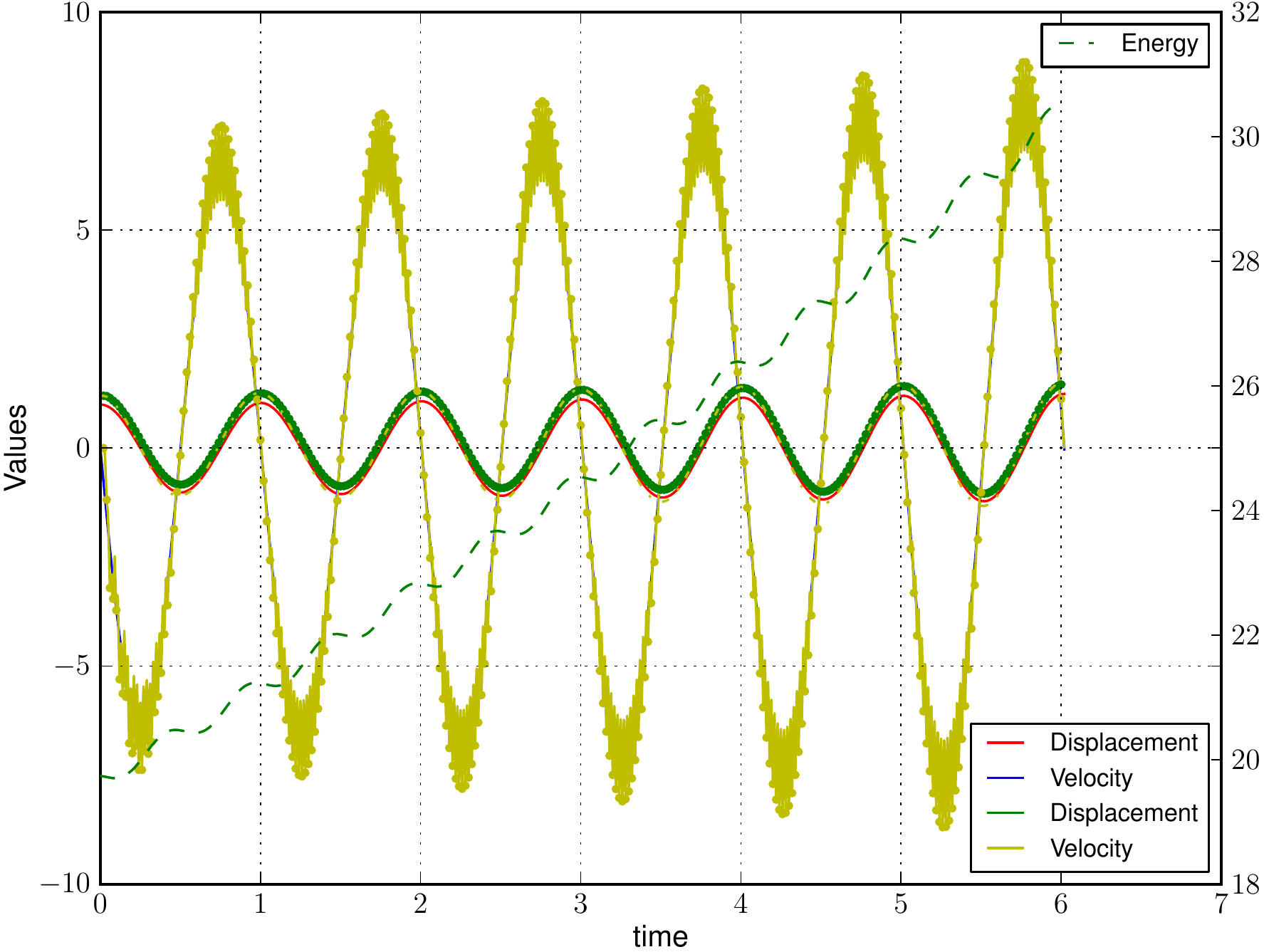}
\includegraphics[ width=0.5\textwidth]{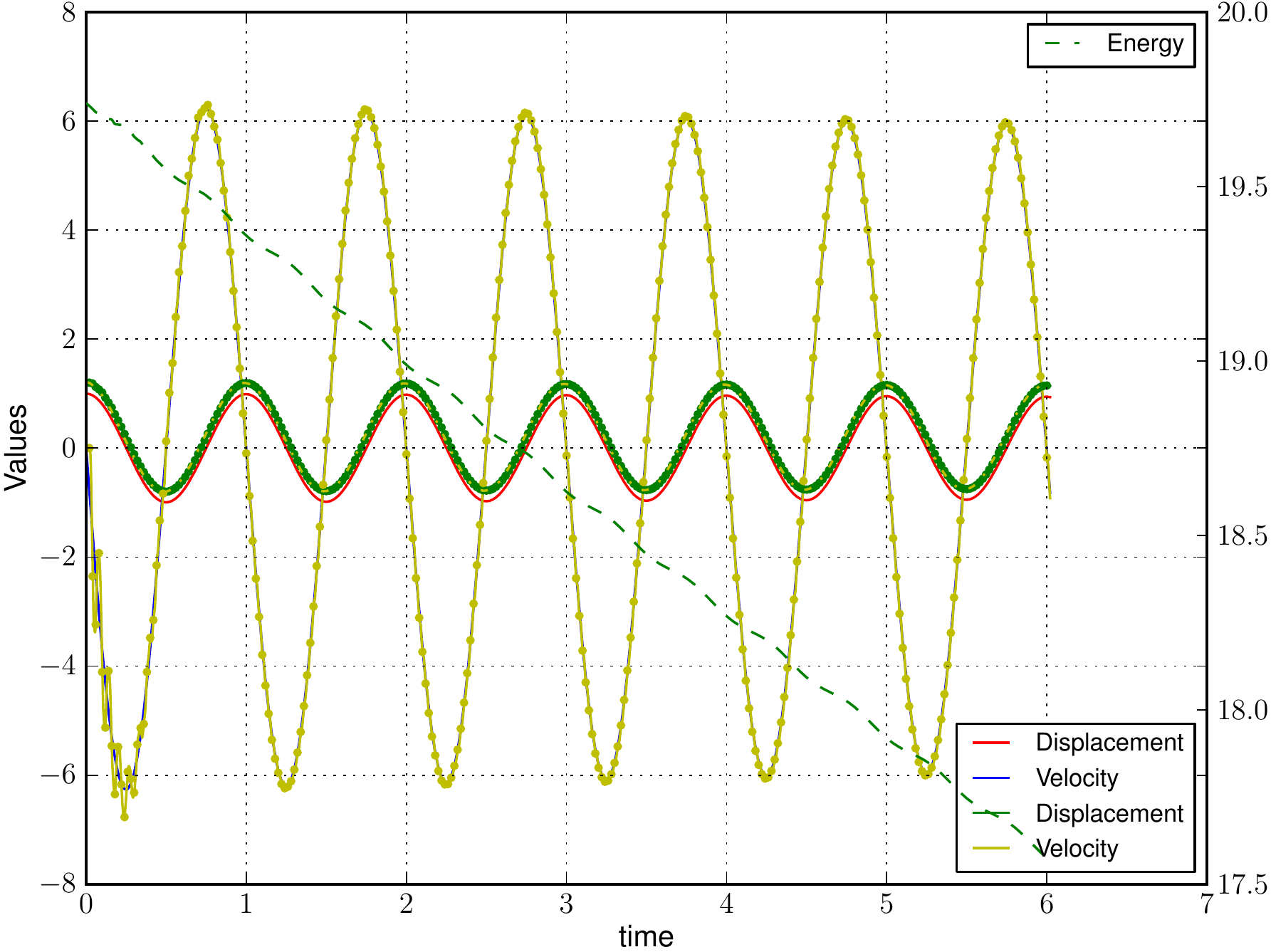}\\
\caption{\label{fig:sSmoothCosimStabilityQuickBCDoubleSpringMass} Simulation of the double spring mass system \eqref{eq:DoubleSpringMass} in the cosimulation scheme with constant (left) and linear (right) extrapolation, $H = 0.02$. Stability properties suffer from the delayed refeed, compare to fig. \ref{fig:sSmoothCosimStabilityQuickBCDoubleSpringMass}}.
\end{figure}
The aim of reducing method induced oscillations is achieved well, see figure 
\ref{fig:sSmoothCosimQuickBCDoubleSpringMass}. But the figures \ref{fig:sSmoothCosimStabilityQuickBCSpringMass} and \ref{fig:sSmoothCosimStabilityQuickBCDoubleSpringMass} show that the delayed refeed of the balance error which is inherent to the spreading of the refeed over more intervals has a negative effect on the stability, as foreseen above.

\section{Definition of  smooth switching and hat functions}
\label{sec:HatsSwitches}
We refer as \emph{hat function} to a real function with a  convex support in which it is positive and has one maximum. An example is 
\begin{equation}
d(t) = \begin{cases} e^{-\frac{1}{1-t^2}} \,\quad t\in[-1, 1]\\
0 \phantom{ e^{frac{1}/{1-t^2}}} \,\quad \text{else}
\end{cases}
\end{equation}
(e.g \cite{Walter1994}), reminding of the Gaussian standard distribution. It is not used in this context as polynomials are cheaper to evaluate. \\
All  hat functions  $\overline\phi$ are constructed here on the reference interval $\left[-1, 1\right]$ with the property that its integral is 1.
On a general interval $\left[t_{n}, t_{n+k}\right)$, $n\in\left\{1, 2,4\right\}$, we use 
\begin{equation} \label{integralTrafo}
\phi_{n}(t)= \left(\frac{ 2}{ t_{n + k } - t_k}\right)\overline\phi\left(\frac{ 2}{ t_{n + k } - t_k}\left(t -\frac{ t_{n + k }+ t_k}{ 2}\right)\right),
\end{equation}
a transformation of $\overline\phi$ which itself is a hat function on $\left[t_{n}, t_{n+k}\right)$. The property that its integral is 1  is conserved by the transformation 
such that balance correction can be added to signal $\overline{\vek u}$ as
$\Delta E_{I,m,i}\phi_{i}(t)$, %
starting at time $t_{i}$. 
\\
All switch functions $\overline\psi$ are constructed analoguously as $\overline\phi$ on the interval $\left[-1, 1\right]$ with the property that its integral is 1, and again the transformation  \eqref{integralTrafo} conserves this property.

\subsection{Definition of a polynomial smooth hat function}
\label{sec-3-1}
The function we search should be easy to calculate and implement. We
concentrate our search to the realm of polynomials. The function
should be symmetrical to its middleaxes, therefore the degree of the
polynomial has to be even.\\
Furthermore, all hat functions for our purposes should continuously vanish at the boundaries of their support, and so should at least the first and second derivative. Thus a polynomial of at least sixth degree is needed, and its first derivative  should have double roots at -1 and 1 (and of course a root at 0):
\begin{equation}
\qquad \qquad \overline\phi '= ax\left[(x-1)^2(x+1)^2\right]
=  ax(x^2-1)^2		
= a (x^5 - 2x^3 + x),
\end{equation}
the integral of which is 
\begin{equation}\label{polHat}
 \overline\phi = a \left[\frac{1}{6}x^6 - \frac{1}{2}x^4 +\frac{1}{2}x^2 + c\right].
\end{equation}
We determine $c = -\frac{1}{6}$ by using $ \overline\phi(1)=0$, and 
\begin{equation}\label{IntOfHat}
\int_{-1}^1 \overline\phi(t)dt = a \left[\frac{1}{42}x^7 - \frac{1}{10}x^5 +\frac{1}{6}x^3  -\frac{1}{6}x\right]_{-1}^1 = 1
 \end{equation} 
 yields $a = -\frac{105}{16}$.\\
No attention to turning points inside $(-1,1)$ was paid, and one easily verifies, second derivative being $ \overline\phi '' = (x-1)^2(x+1)^2\left[(x-1)(x+1)+ 2x(x+1) + 2 x(x-1)\right]$, that $\pm \sqrt{\frac{1}{5}}$ are roots of it. Turning points of this function are not equidistant, which  makes this function unsuitable for use in the second method described in \ref{sec-3}. A polynomial of 8th order  would be necessary.
%
We do not carry out these calculations because in the following a function with the desired property will be constructed.

\subsection{Construction of switch functions as integrals of hat functions and  of integral-of-two-hats type}
\label{sec:IntOfTwoHats}
If  $\sum\phi_i = \text{const}$ in some interval
, all  derivatives of $\sum\phi_i$ vanish, which means that if  corrections $\Delta E_{m,i}$ do not change, the sum $\sum\Delta E_{m,i}\phi_i $ does
 not change and thus  only a constant is added to $\overline{\tilde f_m}$. So if there shall be no contributions from the $\phi_i$ alone to derivatives, it is necessary (yet not sufficient) 
 that the
two resp. four test functions sum up to a constant on the common of  their support $\cap \text{supp} \phi_i$. \\
Nevertheless, to this objective, let  the real function $h(t)$ be one of the above defined hats, all of which have the property that they are symmetric w.r.t. $t=0$.\\
\begin{figure}[h!btp]
\includegraphics[ width=0.5\textwidth]{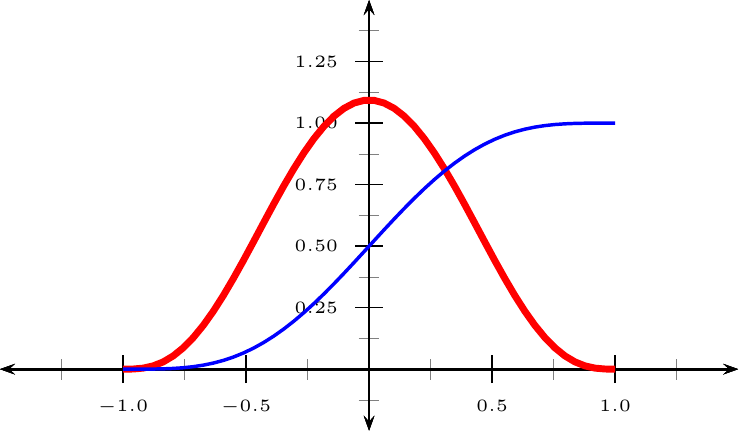}
\includegraphics[ width=0.5\textwidth]{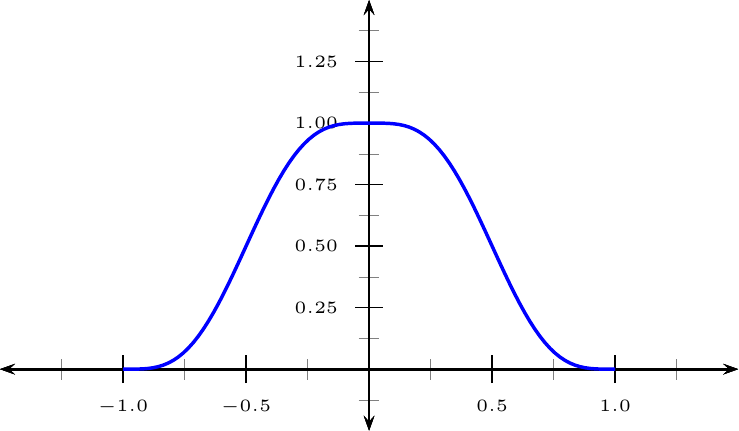}
\caption{\label{fig:PhiPsi}Left:Example for $\overline\phi$: Polynomial  \eqref{polHat} and its integral as an example for $\psi$. Right: Example for $\phi$ as Integral-of-two-hats: Polynomial \eqref{polHat} integrated, stretched and shifted and mirrored. }
\end{figure}
We define a function
\begin{equation}
\psi :=\int_{-1}^t h(\tau)d\tau.
\end{equation}
Its maximum is 1, $\psi(1)=1$, as  all hats from the former section are already normed such that  $\int_{-1}^1 h(\tau)d\tau = 1 $.  Clearly $\psi$ is a s-shaped switch function (so its name is well-chosen) with a turning point at (even rotational symmetry with respect to) $(0,\int_{-1}^1 d(\tau)d\tau/2)$. 
Moreover, as for the integral of any  function $h$ symmetric to $t_\text{symm}$ it holds that $\int_{-t}^{t_\text{symm}} h(\tau)d\tau = \int_{t_\text{symm}}^{t}h(\tau)d\tau$ is valid, $\psi(-t) = 1-\psi(t) $ follows, see fig. \ref{fig:PhiPsi}. 


\begin{figure}[h!btp]
\includegraphics[ width=0.8\textwidth]{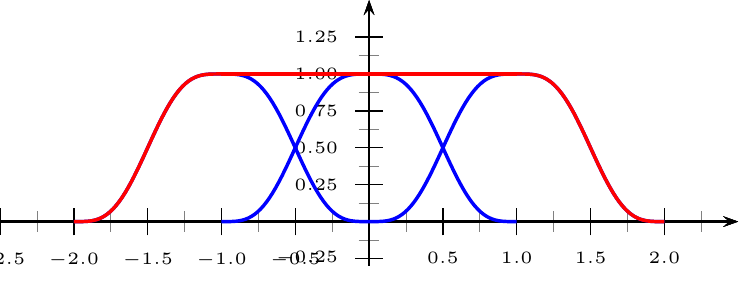}
\caption{\label{sumOfPhi}Sum of $\phi_i$ as Integral-of-two-hats adds up to 1 when $\operatorname{supp}\phi_i = [t_i,t_{i+2}]$. This fulfils the requirement of no method induced turning points according to section \ref{sec-3}. }
\end{figure}

\subsection{Construction  a of hat function of integral-of-two-hats type}

We make use of this property by defining  a hat again
\begin{equation}
\overline\phi= \begin{cases} 
	\psi\left(2\left(t + \frac{1}{2}\right)\right)\quad t<0\\
	\psi\left(2\left(-t + \frac{1}{2}\right)\right)\quad t\ge 0,
\end{cases}
\end{equation}
from a transformed   $\psi$ (compare eq. \eqref{integralTrafo}) and its mirrored counterpart. \\
According to the integral of hat \eqref{IntOfHat} and the transformation \eqref{integralTrafo} the function
\begin{multline}
\overline\phi= 
\\  \frac{-105}{16}\begin{cases}
\begin{split} \frac{(2x+1)^7}{42} - \frac{(2x+1)^5}{10} 
+ \frac{(2x+1)^3}{6} - \frac{2x+1}{6}- \frac{8}{105}\quad t<0
\end{split}\\
\begin{split}\frac{(-2x+1)^7}{42} - \frac{(-2x+1)^5}{10} 
+ \frac{(-2x+1)^3}{6} - \frac{-2x+1}{6}- \frac{8}{105}\quad t\ge 0
\end{split}
\end{cases}
\end{multline}
is a realization of an integral-of-hat type hat function, namely the one with the lowest possible degree.
%
%
%



%

\section{Discussion and conclusion}
\label{sec-4}
Unchanged by results of this examinations, problems without strong conservation properties can be well treated with cosimulation schemes, and problems with vulnerability to imbalance of exchanged conserved quantity in terms of biased system properties due to wrong quantity amount over a long time interval can be well treated with balance correction.

Considering systems with conservation properties in which factors of the conserved quantity are exchanged, the smoothing and recontibution methods presented can improve stability and smoothness at the same time, but not solve the issues.

Improving the extrapolation order is always beneficial. Besides this, all attempts to cure the drawbacks of methods that are explicit in some variables encounter new drawbacks:

If smoothing and balance correction, then of course also smooth refeed is needed. Although smooth, the recontributions may cause quite high derivatives.
The smoothing error contributes to the balance error - so to the derivatives that are caused by recontributions. 

If they matter, so if excitation of subsystems can be expected, and so one tries to tackle them by distributing the recontributions over more time intervals, balance errors depending on the exchanged quantities (energy)  are worsened.

Balance correction methods  in general remain unable to fully cure  balance violations and their consequences, even less so those that distribute recontribution   over more than one time interval.  Yet they are able to prolong the time that the simulation can deliver meaningful results. \\

\subsection{Convergence and stability considerations}
In \cite{Busch2012} convergence results and stability criteria has been derived for cosimulation of linear ODE systems. To the best of our knowledge, nobody has written down such results for cosimulation of nonlinear ODE or DAE systems. Although simple considerations make it appear likely that cosimulation methods are convergent for such systems, it remains to prove that this is the case. Once such a proof is available, it has to be shown that balance correction does not spoil convergence. 

\subsection{Conclusion}
\label{sec-5}
Using smooth contributions significantly reduces unphysical dynamics in the frequency of exchange time stepwidth. This allows to draw conclusions concerning higher-frequency dynamics from cosimulation results while using balance correction, which  would be impossible without careful choice of hat functions.\\
Finally, establishing balance of a quantity \emph{a posteriori} cannot establish consistency and thus balance in all quantities of the system. Thus, the techniques discussed in this contribution sometimes can improve stability, but it cannot make the cosimulation scheme stable. It is hardly surprising that an explicit method as the cosimulation cannot easily be turned into a stable method.\\ 
The view on cosimulation as an explicit method finally puts the results into the right context: Nobody would expect an explicit method to work  with a  stepwith close to the systems  one eigenfrequency like in the example of a slow, heavy coupled to a quick, light spring-mass system. Why should cosimulation? Dense knowledge of the simulated system remains necessary. But still, the suggested measures have the capacity to transform cosimulations Euler-Forward-like simplicity and limitedness into applicability for a big range of coupled problems - they turn cosimulation into a method with reliable results.

Anyhow, the future task will be to find methods that determine input signals such that effects of past errors, rather than the errors themselves,  on the receiving system are compensated.


\bibliography{ifacconf}
\bibliographystyle{nMCM}
\end{document}